%% file: Photospheric_spectra.tex
\documentclass[fleqn,usenatbib]{mnras}

\usepackage{newtxtext,newtxmath}

\usepackage[T1]{fontenc}

\DeclareRobustCommand{\VAN}[3]{#2}
\let\VANthebibliography\thebibliography
\def\thebibliography{\DeclareRobustCommand{\VAN}[3]{##3}\VANthebibliography}

\usepackage{graphicx}	
\usepackage{amsmath}	

\usepackage{longtable}
\usepackage{caption}
\usepackage{subcaption}
\usepackage{courier}

\usepackage{enumitem}
\usepackage{array}
\usepackage{pifont} 
\newcolumntype{P}[1]{>{\centering\arraybackslash}p{#1}}
\usepackage{orcidlink}

\title[Photospheric Spectra of SLSNe]{The Type I Superluminous Supernova Catalogue II: Spectroscopic Evolution in the Photospheric Phase, Velocity Measurements, and Constraints on Diversity}

\author[A. Aamer et al.]{Aysha Aamer$^{1}$\thanks{E-mail: aaamer01@ads.qub.ac.uk (AA)}\orcidlink{0000-0002-9085-8187},
Matt Nicholl$^{1}$\orcidlink{0000-0002-2555-3192},
Sebastian Gomez$^{2}$\orcidlink{0000-0001-6395-6702},
Edo Berger$^{2,3}$\orcidlink{0000-0002-9392-9681},
Peter Blanchard$^{2}$\orcidlink{0000-0003-0526-2248},
\newauthor
Joseph P Anderson$^{4,5}$\orcidlink{0000-0003-0227-3451},
Charlotte Angus$^{1}$\orcidlink{0000-0002-4269-7999},
Amar Aryan$^{6}$\orcidlink{0000-0002-9928-0369},
Chris Ashall$^{7}$\orcidlink{0000-0002-5221-7557},
Ting-Wan Chen$^{6}$\orcidlink{0000-0002-1066-6098},
\newauthor
Georgios Dimitriadis$^{8}$\orcidlink{0000-0001-9494-179X},
Llu\'is Galbany$^{9,10}$\orcidlink{0000-0002-1296-6887}, 
Anamaria Gkini$^{11}$\orcidlink{0009-0000-9383-2305}, 
Mariusz Gromadzki$^{12}$\orcidlink{0000-0002-1650-1518},
\newauthor
Claudia P. Guti\'errez$^{10,9}$\orcidlink{0000-0003-2375-2064}, 
Daichi Hiramatsu$^{2,3}$\orcidlink{0000-0002-1125-9187},
Griffin Hosseinzadeh$^{13}$\orcidlink{0000-0002-0832-2974}, 
Cosimo Inserra$^{14}$\orcidlink{0000-0002-3968-4409},
\newauthor
Amit Kumar$^{15,16}$\orcidlink{0000-0002-4870-9436},
Harsh Kumar$^{2}$\orcidlink{0000-0003-0871-4641},
Hanindyo Kuncarayakti$^{17,18}$\orcidlink{0000-0002-1132-1366},
Giorgos Leloudas$^{19}$\orcidlink{0000-0002-8597-0756}, 
\newauthor
Paolo Mazzali$^{20,21}$, 
Kyle Medler$^{7}$\orcidlink{0000-0001-7186-105X}, 
Tomás E. Müller-Bravo$^{22,23}$\orcidlink{0000-0003-3939-7167}, 
Mauricio Ramirez$^{24,5}$, 
\newauthor
Aiswarya Sankar.K$^{6}$\orcidlink{0009-0003-2609-3591}, 
Steve Schulze$^{25}$\orcidlink{0000-0001-6797-1889}, 
Avinash Singh$^{11}$\orcidlink{0000-0003-2091-622X}, 
Jesper Sollerman$^{11}$\orcidlink{0000-0003-1546-6615}, 
Shubham Srivastav$^{26}$\orcidlink{0000-0003-4524-6883}, 
\newauthor
Jacco H. Terwel$^{22}$\orcidlink{0000-0001-9834-3439},
David R. Young$^{1}$\orcidlink{0000–0002–1229–2499}
\\
\\
\textit{Affiliations are listed at the end of the paper}}

\date{Accepted XXX. Received YYY; in original form ZZZ}

\pubyear{\the\year{}}

\begin{document}
\label{firstpage}
\pagerange{\pageref{firstpage}--\pageref{lastpage}}

\maketitle

\begin{abstract}
Hydrogen-poor superluminous supernovae (SLSNe) are among the most energetic explosions in the universe, reaching luminosities up to 100 times greater than those of normal supernovae. Although there is debate in the literature about the powering mechanisms of H-poor SLSNe, detailed spectral analysis holds the potential to reveal their progenitors and underlying energy sources. This paper presents the largest compilation of SLSN photospheric spectra to date, encompassing data from the advanced Public ESO Spectroscopic Survey of Transient Objects (ePESSTO+), the Finding Luminous and Exotic Extragalactic Transients (FLEET) search, and all published spectra up to December 2022. The dataset includes a total of 974 spectra of 234 SLSNe. By constructing average phase binned spectra, we find SLSNe initially exhibit high temperatures (10000$-$11000\,K), with blue continua and weak lines. A rapid transformation follows, as temperatures drop to 5000$-$6000\,K by 40 days post peak, leading to stronger P-Cygni features. These averages also suggest that a fraction of SLSNe may contain some He at explosion.
Variance within the dataset is slightly reduced when defining the phase of spectra relative to explosion, rather than peak, and normalising to the population’s median e-folding decline time. Principal Component Analysis (PCA) supports this, requiring fewer components to explain the same level of variation when binning data by scaled days from explosion, suggesting a more homogeneous grouping. Using PCA and K-means clustering, we identify outlying objects with unusual spectroscopic evolution and evidence for energy input from interaction, but find no support for groupings of two or more statistically significant subpopulations. We find Fe II $\lambda$5169 line velocities closely track the radius implied from blackbody fits, indicating formation near the photosphere. We also confirm a correlation between velocity and velocity gradient, which can be explained if all SLSNe are in homologous expansion but with different scale velocities. This behaviour aligns with expectations for an internal powering mechanism.

\end{abstract}




\section{Introduction}

Hydrogen-poor superluminous supernovae (SLSNe) are some of the brightest explosions in the universe, releasing 10$-$100 times more energy than normal supernovae, and spanning over hundreds of days \citep[e.g.][]{Gal-Yam2009, Barbary2009, Quimby2011, Chomiuk2011, Gal-Yam2019b, Inserra2019, Nicholl2021, Gomez2024, Moriya2024}. Understanding the physical mechanisms behind these extraordinarily bright explosions is one of the major unsolved problems in stellar astrophysics. Despite significant observational efforts, the origins and powering mechanisms of SLSNe remain elusive, signalling that we lack a complete understanding of stellar life cycles, particularly at the upper end of the powering scale.

Spectroscopically, H-poor SLSNe-I (hereafter SLSN) are characterised by a steep blue continuum, often alongside a series of singly ionised oxygen (O II) absorption lines, indicating extremely high temperatures and non-thermal excitation at early times \citep{Dessart2012, Mazzali2016}. These lines found between $3737-4650$\,\AA\ \citep{Quimby2018, Gal-Yam2019a} produce a characteristic W-shape that weakens with time as the temperature of the photosphere drops \citep{Quimby2011}. However, some SLSNe do not show these O II absorption lines \citep[e.g.][]{Gutierrez2022}. During this photospheric phase, the spectrum is dominated by blackbody radiation from the hot, dense photosphere of the SN, with absorption features from elements in the outer ejecta \citep[e.g.][]{Pastorello2010, Inserra2013b, Howell2013, Nicholl2016b}. As the temperature drops, the spectra increasingly resemble those of traditional stripped envelope SNe (SE-SNe) at the same temperatures, likely indicating similar progenitors or environments between the two classes \citep{Pastorello2010}. As the ejecta expand and cool, the SN transitions to the nebular phase where the photosphere becomes optically thin and stops producing continuous blackbody emission. Instead, the spectrum is dominated by emission lines from elements such as oxygen and magnesium \citep{Gal-Yam2009, Jerkstrand2017a, Nicholl2016a, Nicholl2019c}. 

Although their star-forming environments \citep{Neill2011, Chen2013, Lunnan2014, Leloudas2015b, Angus2016, Perley2016, Schulze2018} point to the explosions of massive stars, there are still many questions regarding their powering mechanisms. The light curve evolution for typical SE-SNe is dictated by the decay of $^{56}$Ni produced in the explosion \citep{Colgate1969}. The mass of $^{56}$Ni produced by typical SE-SNe is $\lesssim 0.2\,M_{\odot}$ \citep[e.g.][]{Drout2011, Anderson2019, Rodriguez2023}. For SLSNe, several solar masses of $^{56}$Ni would be needed to power such a long duration light curve, as well as power the peak luminosities observed. Observations of SLSNe suggest total ejecta masses ranging from a few up to a few tens of solar masses \citep{Blanchard2020b, Gomez2024}, which would often result in ejecta compositions dominated by nickel. This is not supported by SLSN spectra, which are dominated by intermediate mass elements such as carbon and oxygen \citep{Dessart2012}. The amount of nickel needed could be synthesised in pair-instability SNe (PISNe), which are theorised to occur in main sequence stars with masses $140-260\,M_{\odot}$. In these stars, spontaneous pair production of electron and positron pairs can reduce the internal radiation pressure, resulting in explosion of a PISN with over $100\,M_{\odot}$ of ejecta \citep{Heger2002, Barkat1967}. These explosions could have been more common in the very early universe when the metallicity was lower and stars could retain more mass \citep{Bromm2004, Woosley2007}. So far the only promising candidate for this mechanism is SN\,2018ibb \citep{Schulze2024}. Instead another mechanism is needed to power most SLSNe. Current theories propose a range of both internal and external powering mechanisms. 

A possible explanation for the powering mechanism of SLSNe is interaction with circumstellar material (CSM). This theory also proposes a SE-SN that has large amounts of CSM created through stellar winds and ejections throughout the stars life \citep{Chevalier2011, Chatzopoulos2012a, Chatzopoulos2013}. Pulsational pair instability (PPI) is a mechanism thought to occur in lower mass stars ($95-130\,M_{\odot}$) than required for PISNe where the star ejects material towards the end of its life due to pair instability; this has been suggested as a way to get massive CSM in interaction models \citep{Woosley2007}. The SN ejecta are able to catch up to this material because they have much higher velocities, and are rapidly decelerated if the CSM is massive enough. This creates a shock that deposits energy in the ejecta and CSM, the cooling of which can produce a bright and long-lived light curve \citep{Smith2007b, Chevalier2011, Ginzburg2012, Moriya2013, Sorokina2016}. Interaction with CSM is the mechanism thought to power H-rich SLSNe-II \citep{Kangas2022, Hiramatsu2024}, but is less easily reconciled with H-poor SLSNe-I. Several PPI candidates exist showing a variety of different observables such as multi-peaked light curves \citep[e.g.][]{Wang2022, Bose2018, Hosseinzadeh2022, Zhu2023, Gomez2019a}, early emission of "nebular" lines \citep[e.g.][]{Aamer2024, Angus2024}, discovery of late time interaction with H \citep[e.g.][]{Yan2017a}, or the discovery of CSM shells through Mg II absorption lines \citep[e.g.][]{Lunnan2018a, Schulze2024, Gkini2025}.

The most popular theory, with observational support particularly from late-time observations, is the magnetar central engine model. In this picture, a high-mass star undergoes a SN explosion at the end of its life much like a typical core collapse SN (CCSN). However, in this case the remnant is a fast rotating neutron star with a very strong magnetic (B) field (around $10^{13}-10^{14}$ G) called a magnetar \citep{Kasen2010, Woosley2010, Dexter2013, Metzger2015}. Although these are very strong B-fields, nearly 10\% of newly born neutron stars have B-fields in the range $10^{13}-10^{15}$ G lasting over 1000 years after their birth, so it is plausible that these could exist to produce SLSNe  \citep{Kouveliotou1998, Woods2006}. As these magnetars spin down, they inject energy into the system, powering the observed light curves. If the initial spin period is close to maximal, around a few milliseconds, the energy released is sufficient to boost the luminosity to the levels observed in SLSNe \citep{Ostriker1971, Kasen2010}. There is also growing evidence that some CCSNe at peak luminosity are not solely powered by radioactive $^{56}$Ni decay, and instead require the presence of a central engine such as a magnetar \citep{Rodriguez2024}. If true, this would support the presence of magnetars in SLSNe. Despite the magnetar model being widely applied to SLSNe \citep[e.g.][]{Inserra2013b, Nicholl2017a}, multiple powering mechanisms may be at play \citep{Chen2023b}, depending on the progenitor star and its environment.

SLSNe show striking diversity in their observed properties with some events reaching peak absolute magnitudes as bright as $M=-22.5$ \citep{Gomez2024}. Similarly, their light curves can vary vastly in duration from around 50 to over 600 days \citep{Nicholl2015a, DeCia2018, Lunnan2018b, Angus2019, Chen2023a}. This variability extends to the mass of material ejected in the explosion, which can range from a few solar masses, up to around 40 solar masses in most SLSNe \citep{Blanchard2020b, Chen2023a, Gomez2024}. However, this number can reach up to 100\,$M_{\odot}$ like in the case of SN\,2018ibb \citep{Schulze2024}. Understanding this diversity is key to determining whether SLSNe represent a single class of explosion or whether they encompass multiple sub-types. This observed diversity may also point to the possibility of multiple progenitor channels.

A key question is whether SLSNe represent an extension of normal SNe Ic \citep{Pastorello2010, Inserra2013b, Metzger2015, Liu2017, Gomez2022a}. These are stripped envelope SNe without H or He lines present. Spectroscopically, H-poor SLSNe share some similarities with SNe Ic, especially in the late-time nebular phase \citep{Nicholl2019c, Gutierrez2022} when the ejecta becomes optically thin and individual element lines dominate the spectrum. However, the extreme luminosity and extended light curves of SLSNe require an additional, long-term heating source compared to the $^{56}$Ni that powers normal SNe Ic. Moreover, unlike typical SNe Ic, SLSNe occur almost exclusively in low-metallicity galaxies \citep{Chen2013, Lunnan2013, Lunnan2014, Leloudas2015b, Perley2016, Schulze2018, Cleland2023}. This likely influences the distribution of natal stellar masses, and the loss of mass and angular momentum prior to explosion.


The spectroscopic evolution of SLSNe is key to understanding their origins, and the diversity of this class necessitates a statistical approach with large samples of objects. For example, recent work from the Zwicky Transient Facility \cite[ZTF;][]{Bellm2019} contains a sample of 78 SLSNe, focusing on the photometry and light curves of these events, with spectra used to derive photospheric velocities \citep{Chen2023a, Chen2023b}. A study by \citet{Lunnan2018b} analyses the light curves and spectra of 17 SLSNe from the Panoramic Survey Telescope and Rapid Response System \cite[Pan-STARRS1;][]{Chambers2016} Medium Deep Survey. However, the spectral analysis in that work was again limited to expansion velocities derived from the spectra. A more comprehensive study was conducted by \citet{Quimby2018} on spectra of 23 objects from the Palomar Transient Facility \cite[PTF;][]{Rau2009}. That work looked at specific features within the spectra such as the ionised Mn II and O II lines, how they evolved with time and analysed the velocity distributions of different element absorption lines. A sample of 12 events with 41 spectra from \citet{Nicholl2019c} focussed only on the nebular phase. Other studies have looked at the possibility of subpopulations within the class such as \citet{Inserra2018b} who analysed the photometric evolution in combination with spectroscopic properties of 22 SLSNe. Work by \citet{Konyves-Toth2021} claimed differences between fast and slow events for 28 SLSNe based on photospheric velocities and the appearance of O II absorption lines. Another study of note is the work by \citet{Liu2017} looking at spectral similarities between 32 SLSNe, 17 normal SNe Ic, and 21 broad-lined SNe Ic (Ic-BL), with particular focus on the Fe II $\lambda$5169 line. That work found similarities between SLSNe and SNe Ic-BL which displayed similar absorption velocities and line widths, compared to SNe Ic which had systematically lower velocities and narrower lines. This raises the question of whether these classes are distinct populations or if they exist on a smooth continuum.

In this paper, we present a sample of 974 spectra from 234 H-poor SLSNe. This is the second paper in our series analysing the largest dataset to date of SLSNe, and constitutes the second data release, focusing on the photospheric spectra. In the first data release, \citet{Gomez2024} present photometry and light curve modelling of 262 SLSNe. The final data release, featuring late-time nebular spectra (Blanchard, P. et al., in prep.), will serve as the third instalment. This data will be publicly available on Github\footnote{https://github.com/gmzsebastian/SLSNe}, and in the supplementary material of this journal. Previously unpublished raw spectra will also be publicly uploaded to the Weizmann Interactive Supernova Data Repository (WISeREP\footnote{https://www.wiserep.org}).

The paper is structured as follows. Section \ref{sec:sample} describes where the data was collected from, how it was processed, and key properties of the sample. Section \ref{sec:average_spectra} analyses the average spectra in different time bins, as well as general line identifications. In Section \ref{sec:pca}, we discuss potential subpopulations of SLSNe based on their spectra. Section \ref{sec:fe_line} presents our measurements and analysis of the key spectral line Fe II $\lambda$5169, and how it compares to the photospheric velocity. We present our conclusions in Section \ref{sec:conclusions}.

\section{SLSN Sample}
\label{sec:sample}
\subsection{Data Collection}
\label{sec:datacoll} 

In total this collective study contains data from 234 events with 1252 spectra collected and reduced from public and private sources detailed below. The focus of this paper is on the photospheric phase of the SN where the spectra are dominated by blackbody emission and absorption lines. For the scope of this paper we limit this to spectra with a phase of up to 200 days post explosion, or corresponding phases up to +160 days relative to the r-band peak of the light curve, both defined in \citet{Gomez2024}. This results in 974 photospheric spectra used in the analysis of this paper after applying the phase cuts and coadding spectra from the same instrument on the same day. These spectra will be made available in the data release alongside this paper.

This sample contains fewer objects than \citet{Gomez2024} as some objects in that sample do not have publicly available spectra or are due to be published in future works. The majority of the public data consisting of 1064 spectra were obtained from WISeREP, and the Transient Name Server (TNS\footnote{https://www.wis-tns.org}), with 32 obtained through private communications such as spectra of SN\,2002gh, SN\,2011kl, SN\,1999as, and those from \cite{Lunnan2018b}. 82 private spectra were obtained through the extended-Public ESO Spectroscopic Survey for Transient Objects (ePESSTO+) \citep{Smartt2015}, and 74 spectra from the Finding Luminous and Exotic Extragalactic Transients (FLEET) search \citep{Gomez2020b, Gomez2023a}.

All spectra from PESSTO, ePESSTO and ePESSTO+ were obtained using the 3.56m European Southern Observatory New Technology Telescope (ESO NTT) at La Silla Observatory in Chile, with Grisms 11, 13 and 16. All spectra were reduced using the PESSTO pipeline\footnote{https://github.com/svalenti/pessto} \citep{Smartt2015}. Spectra dating up to April 2019 have been obtained from PESSTO Spectroscopic Data Releases 1-4. We also include in our analysis previously unpublished spectra from ePESSTO+ shown in Table \ref{tab:unpub_spec}.


Spectra collected through the FLEET follow-up programme were obtained using both the Blue Channel \citep{Schmidt1989} and Binospec \citep{Fabricant2019} on the MMT 6.5m Telescope, the Low Dispersion Survey Spectrograph \citep[LDSS3c;][]{Stevenson2016} and the Inamori-Magellan Areal Camera and Spectrograph \citep[IMACS;][]{Dressler2011} on the Magellan 6.5m telescopes, as well as the FAst Spectrograph for the Tillinghast Telescope \citep[FAST;][]{Fabricant1998} on the 1.5m Tillinghast Telescope. The objects with unpublished spectra are shown in Table \ref{tab:unpub_spec}


\begin{center}
\begin{table}
    \begin{tabular}{|P{2.5cm}|P{1.2cm}|P{3.5cm}|} 
    \hline
    Object & Spectra & Source \\ 
    \hline\hline
        SN\,2018bym & 2 & FLEET \\
        SN\,2018cxa & 2 & FLEET \\
        SN\,2018fcg & 2 & FLEET \\
        SN\,2018fd & 1 & FLEET \\
        SN\,2018ffs & 1 & FLEET \\
        SN\,2018lfe & 4 & FLEET \\

        SN\,2019gfm & 5 & PESSTO \\
        SN\,2019gqi & 3, 1 & PESSTO, FLEET \\
        SN\,2019itq & 4 & FLEET \\
        SN\,2019kcy & 5 & PESSTO \\
        SN\,2019nhs & 4 & PESSTO \\
        SN\,2019otl & 2 & FLEET \\
        SN\,2019pvs & 2 & FLEET \\
        SN\,2019sgh & 1 & FLEET \\
        SN\,2019ujb & 2 & FLEET \\
        SN\,2019unb & 11 & PESSTO \\
        SN\,2019une & 1 & FLEET \\
        
        SN\,2020abjc & 1 & FLEET \\
        SN\,2020abjx & 2 & PESSTO \\
        SN\,2020adkm & 1, 2 & PESSTO, FLEET \\
        SN\,2020ank & 7 & PESSTO \\
        SN\,2020myh & 1 & FLEET \\
        SN\,2020qlb & 1 & FLEET \\
        SN\,2020rmv & 1 & FLEET \\
        SN\,2020uew & 16 & PESSTO \\
        SN\,2020xgd & 4 & FLEET \\
        SN\,2020wnt & 4 & FLEET \\
        SN\,2020znr & 7 & FLEET \\
        SN\,2020zzb & 1 & FLEET \\
        
        SN\,2021een & 3, 1 & PESSTO, FLEET \\
        SN\,2021ek & 4 & PESSTO \\
        SN\,2021fpl & 2, 2 & PESSTO, FLEET \\
        SN\,2021hpx & 1 & PESSTO \\
        SN\,2021txk & 2 & FLEET \\
        SN\,2021uvy & 11, 3 & FLEET \\
        SN\,2021vuw & 4 & FLEET \\
        SN\,2021xfu & 2 & FLEET \\
        SN\,2021ybf & 1, 1 & PESSTO, FLEET \\
        SN\,2021ynn & 2 & FLEET \\
        SN\,2021zcl & 13, 3 & PESSTO, FLEET \\
        
        SN\,2022le & 2 & FLEET \\
        SN\,2022lxd & 1 & FLEET \\
        SN\,2022npq & 1 & PESSTO \\
        SN\,2022pjq & 1 & FLEET \\
        SN\,2022vxc & 3, 1 & PESSTO, FLEET \\
        SN\,2022ued & 1 & FLEET \\

    \hline
    \end{tabular}

    \caption{Objects with previously unpublished spectra included in this sample, the number of spectra included, and the data source.}
    \label{tab:unpub_spec}
\end{table}
\end{center}

\begin{figure}
	\begin{center}
		\includegraphics[width=\columnwidth]{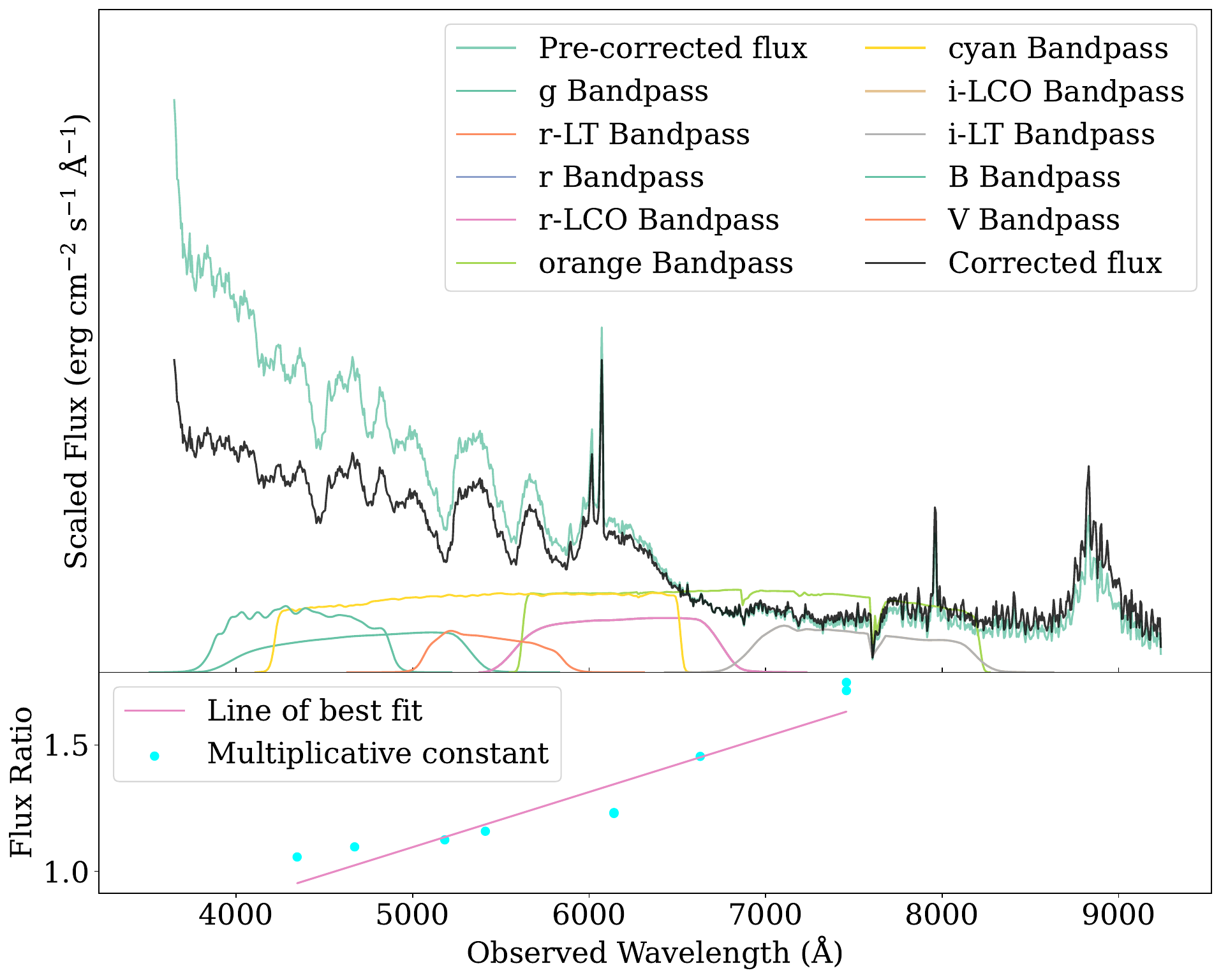}
		\caption{Flux correction (mangling) of an example spectrum of SN\,2019szu \citep{Aamer2024}. 
        \textit{Top:} In green is the pre-corrected spectrum of SN\,2019szu. The filter passbands used for the correction are overlaid. The black line represents the flux corrected spectrum. Both spectra are normalised to their median values and remain uncorrected for redshift at this stage of processing.
        \textit{Bottom:} The cyan points indicate the multiplicative constants required in each filter passband to align the observed spectrum with the photometry from \citet{Gomez2024}, plotted against the effective wavelength of each filter. These constants correspond to the ratio of the spectral derived flux and model fluxes in each passband. The pink line shows the line of best fit, which is used to scale the spectrum for flux correction by extending it over the full wavelength range of the spectrum.}
            \label{fig:flux_slope}
	\end{center}
\end{figure}

\begin{figure}
	\begin{center}
		\includegraphics[width=\columnwidth, trim={1.5cm 0 3cm 2cm},clip]{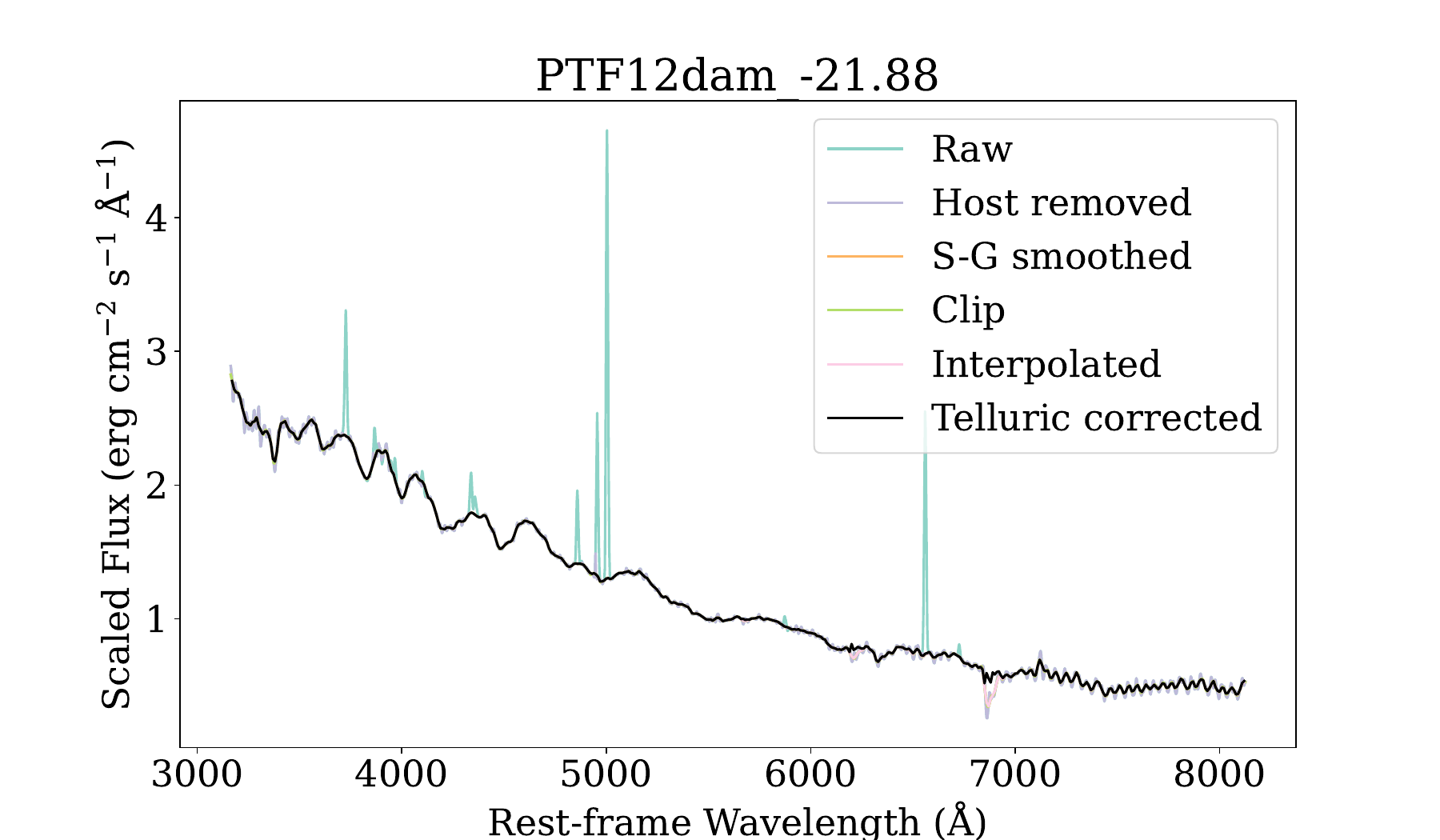}
		\caption{Spectrum of PTF12dam at a phase of $-$21.88 days with respect to peak. The figure illustrates the steps applied at the post processing stage to all spectra with host galaxy line removal, Savitzky-Golay smoothing, clipping the edges of the spectra if needed, telluric removal, and interpolating and remapping onto the same wavelength grid. The legend indicates the order these steps were performed in.}
            \label{fig:post_processing}
	\end{center}
\end{figure}

\subsection{Data Processing}
\label{sec:data_processing}

After obtaining all the available 1D spectra for this sample, the data were required to be standardised and processed into a homogeneous dataset.

The first step was to remove telluric absorption lines due to the Earth's atmosphere; this was done using the python package tellurics-begone\footnote{https://github.com/EJRidley/tellurics-begone}. This package works by convolving model spectra of H$_2$O and O$_2$ with each spectrum and subtracting them off. If a spectral resolution was specified in the original file headers, this value was used, otherwise a default value of 10$\,\Angstrom$ was used. This process was also only applied to spectra with any flux between 4000$-$10000$\,\Angstrom$ as the model spectra of H$_2$O and O$_2$ only covered this wavelength range.

The data were then corrected for Galactic extinction with the host galaxy extinction assumed to be negligible. This is typically a reasonable assumption for SLSNe, as the host galaxies are very low mass \citep{Lunnan2014} and light curve studies find extinctions consistent with zero \citep{Nicholl2017a}. This was done using values from \citet{Schlafly2011} and using the implementation of the \citet{Gordon2023} extinction law in Astropy \citep{AstropyCollaboration2018}. This uses four representative samples of extinction curves and combines them into an extinction relation from 912\,\AA\ in the far UV, up to 32$\mu$m in the MIR \citep{Gordon2009, Fitzpatrick2019, Decleir2022, Gordon2021}. The visual extinction to reddening ratio was assumed to be $R_{\rm{V}}=3.1$.

The spectra were then flux calibrated in a process called "mangling" \citep{Hsiao2007} to model photometry from \cite{Gomez2024}. The real photometry is host subtracted as described in \citet{Gomez2024}. This data was first vetted and cleaned by removing points with errors larger than 0.2\,mag. For flux calibration, we only used bands that had at least one data point meeting this criterion. Filter passbands were obtained for each band from the SVO filter profile server\footnote{http://svo2.cab.inta-csic.es/theory/fps/}. We caveat that some of the filters obtained from SVO do not account for atmospheric absorption. The integrated flux in each of the bandpasses with a full overlap with the target spectrum was measured and this was compared with the observed flux in each band. The ratio of these values is plotted in Figure \ref{fig:flux_slope} against the effective wavelength in each band for an example spectrum of SN\,2019szu. Calculating a linear fit to these ratios indicated the correction needed to be applied to the entire spectrum to calibrate to the photometry. If only one band fit the criteria then the multiplicative constant for that band was applied to the entire target spectrum. Figure \ref{fig:flux_slope} shows in green the original, uncorrected spectrum of SN\,2019szu, and in black the flux-corrected spectrum. The wavelengths of the spectra were then redshift corrected, using the redshifts from \citet{Gomez2024}. These values were verified spectroscopically where possible.

The last step was to co-add spectra obtained from the same telescope and instrument in the same night if they were obtained with the same observing set-up (i.e the same grism, grating or arm). This final version of the spectra will be available in the data release on the SLSNe repository\footnote{https://github.com/gmzsebastian/SLSNe}.

For the spectra used in further analysis in this paper, post processing was applied to remove unwanted artefacts. The spectra were all smoothed using a Savitzky-Golay filter to reduce noise. Spectra that contained any flux between 3000$-$9000\,\AA\ were remapped onto a constant wavelength grid running between these wavelengths in 10\,\AA\ intervals to create a uniform dataset. Spectra that did not cover this entire wavelength range were padded with NaN values. Lastly, all spectra were normalised to their median value. An example of this post-processing is shown in Figure \ref{fig:post_processing}, where these steps are applied to a spectrum of PTF12dam.

Host galaxy lines were removed by fitting a Gaussian around the central wavelengths of prominent host galaxy lines within a 50$\,\Angstrom$ window and with a 20$\,\Angstrom$ leeway in central wavelength. These lines were identified by having narrow widths ($\lesssim 200$\,km\,s$^{-1}$)and, in the case of hydrogen, by exhibiting similar line widths to other host lines in the spectrum. The lines used were [O II] $\lambda\lambda$3726, 3728, H I $\lambda$3835, H I $\lambda$3889, H I $\lambda$3969, H$\delta$ $\lambda$4101, H$\gamma$ $\lambda$4340, [O III] $\lambda$4363, H$\beta$ $\lambda$4861, [O III] $\lambda$4959, [O III] $\lambda$5007, He I $\lambda$5876, H$\alpha$ $\lambda$6563, [S II] $\lambda$6716, and [S II] $\lambda$6731. In most cases, [N II] $\lambda$6548 and $\lambda$6583 were blended with H$\alpha$. These Gaussians were then subtracted off from the target spectra. A further check was made by choosing a 20\,\AA\ buffer on either side of a 40\,\AA\ region centred on each line. The median and standard deviation ($\sigma$) was calculated in this buffer region. The central region was then checked and any flux that was more than 1$\sigma$ above or below the median of the buffer region was clipped. Finally, some prominent noise spikes remained in the data. These were removed by clipping any flux above 10 times the median, or below $-$5 times the median. This process also accounted for any residuals from imperfect single-Gaussian fits to the H$\alpha$+[N II] blends. Further telluric removal was applied to a small subset of spectra after smoothing and rebinning where the original removal was unsuccessful, these are denoted in Appendix \ref{tab:all_spec}.

There were 101 spectra with either noisy edges or chip gaps that could not be removed. In these cases, these wavelength regions were clipped to eliminate as much noise as possible. Noisy spectra with no discernable SN flux were omitted from further analysis and are denoted in Appendix \ref{tab:all_spec}.

We label each spectrum with an observed phase, the number of rest-frame days from maximum light in the r-band, at which the spectrum was obtained. The time of maximum is taken from \citet{Gomez2024}. Because SLSNe evolve over very different timescales, with different rise times, this observation time may not be the most physically-motivated label. We therefore also define a "scaled phase from explosion", where the explosion time is measured by fitting a magnetar plus radioactive decay model to the light curves in \citet{Gomez2024}. This was done by normalising the phases to the decay time ($\tau_e$), defined as the time taken to reach 1/e times the maximum bolometric luminosity, and scaling up by the median value of $\tau_e = 44$ days for the entire sample determined in \citet{Gomez2024}. For the phases from explosion, this resulted in the phase range extending up to 500 scaled days. For one spectrum this resulted in a negative phase (SN\,2017egm at $-$3.2 days before explosion). This is likely due to an uncertain explosion time due to the unusual bumpy light curve. For this spectrum, the time was instead given as time of explosion.


\begin{figure}
	\begin{center}
		\includegraphics[width=\columnwidth]{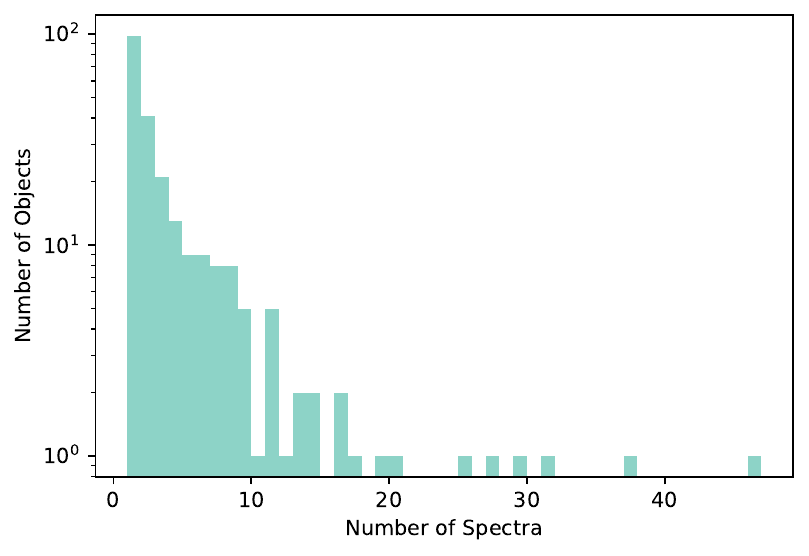}
		\caption{The total number of spectra per event in this sample. This is calculated after processing the spectra and co-adding spectra taken on the same night using the same instrument and grisms.}
            \label{fig:number_spectra}
	\end{center}
\end{figure}

\begin{figure}
	\begin{center}
		\includegraphics[width=\columnwidth]{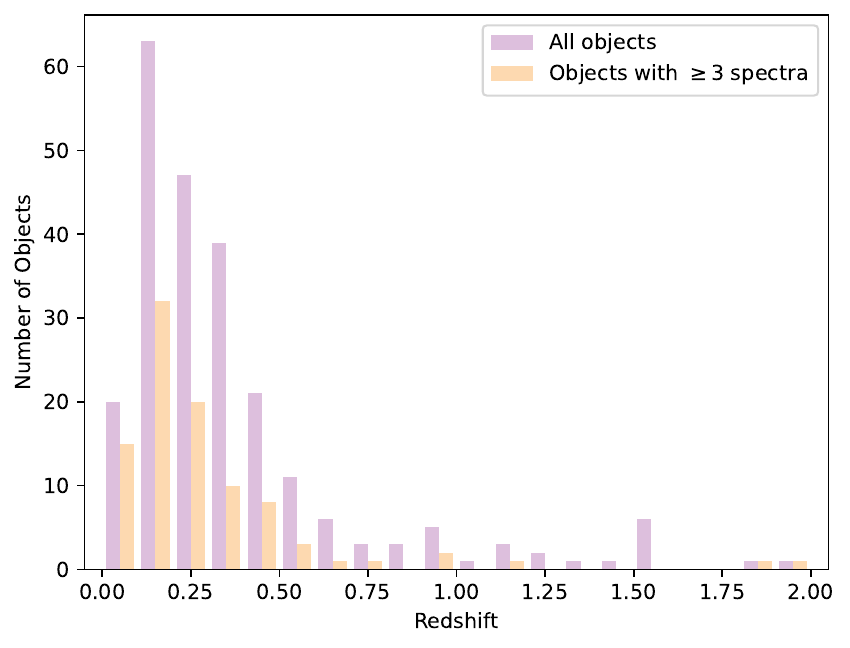}
		\caption{Distribution of redshifts of objects in this sample in bins of 0.1. In purple are all objects within this sample, and orange are objects with at least three spectra.}
            \label{fig:redshfit_dist}
	\end{center}
\end{figure}

\begin{figure}
	\begin{center}
		\includegraphics[width=\columnwidth]{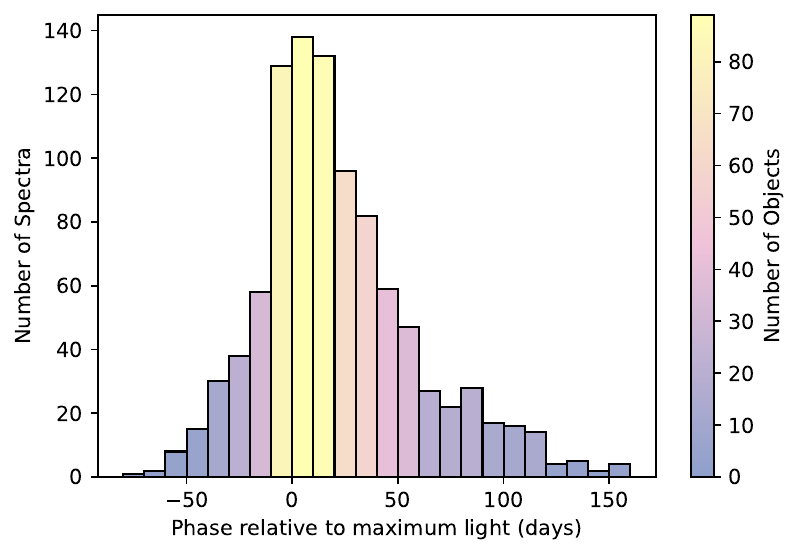}
		\caption{Phases of spectra within this sample relative to maximum light in rest frame. Colourmap indicates the number of objects within each bin.}
            \label{fig:phase_dist}
	\end{center}
\end{figure}

\begin{figure}
	\begin{center}
		\includegraphics[width=\columnwidth]{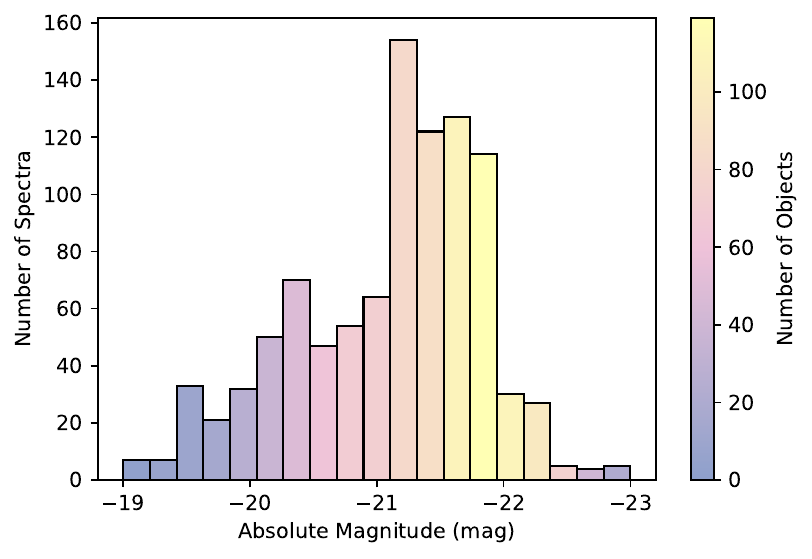}
		\caption{Absolute magnitude distribution. Absolute magnitudes are in the r band and determined based on the model fits from \citet{Gomez2024}.}
            \label{fig:mag_dist}
	\end{center}
\end{figure}

\subsection{Sample Properties}

Figure \ref{fig:number_spectra} shows the distribution of the number of events versus the number of spectra after processing. There are 98 objects in this sample with only a single spectrum, typically the publicly uploaded classification spectrum. This does not rule out the possibility that there exists further follow-up for these objects.

Figure \ref{fig:redshfit_dist} shows the redshift distribution of all of the events in our sample, and also all the events with at least three spectra, in bins of $\Delta z =0.1$. At higher redshifts there is a dearth of events with multiple spectra, likely owing to the dimmer apparent magnitude and therefore difficulty obtaining spectra over a longer timeframe. This is especially apparent between redshifts of $z=1-1.8$. Performing a two-sample Kolmogorov-Smirnov (KS) test on the two samples, we find a p-value of 0.03, indicating the samples are not drawn from the same distribution and the lack of multiple spectra affects the objects at higher redshifts disproportionately. However, as these spectra are collated from both public and private data, this may have introduced biases into the sample. It is also important to note that events at higher redshifts will only contribute to the bluer part of the wavelength range once shifted to the rest frame. This is dealt with in further analysis by imposing minimum wavelength ranges for the data to be included in the analysis.

Figure \ref{fig:mag_dist} shows the distribution of absolute magnitude ($M$) of the spectra within this sample, with the colourbar representing the number of objects in each bin. From this we can see there is not a bias towards brighter objects as may be expected. Instead the imposed nominal cut-off of $M \leq -21$ is apparent with the the spike of spectra below this threshold, and the increase in number of objects per bin at brighter magnitudes.


The distribution in Figure \ref{fig:number_spectra} shows a long tail above 20 spectra per object. These are typically the most nearby events: the eight events within this tail have an average redshift of $\bar{z}_{\rm{tail}} = 0.08$, much lower than the overall average of $\bar{z}_{\rm{all}} = 0.23$. The events with the most spectra, SN\,2018hti and SN\,2017egm, have 50 and 51 spectra respectively. These events had peak observed magnitudes of $m_{\rm{V}}=14.74 (\pm 0.06)$ and $m_{\rm{r}}=15.26 (\pm 0.01)$, at redshifts of $z=0.0612$ and $z=0.0307$ \citep{Lin2020, Nicholl2017c}. The event with the lowest redshift, SN\,2018bsz, had a redshift of $z=0.0267$ \citep{Anderson2018}, but a comparable observed peak magnitude to these other two events. SN\,2018bsz has 23 spectra within this sample.

Figure \ref{fig:phase_dist} shows the distribution of phases relative to maximum light of all of the spectra within this sample. The colourmap indicates the number of objects that fall into each 10 day bin. As expected the majority of spectra are taken soon after peak where the SN is the brightest. 

\begin{figure*}
	\begin{center}
            \large{Phases from Peak}\par
		\includegraphics[width=2\columnwidth, trim={4cm 3cm 3.5cm 4cm},clip]{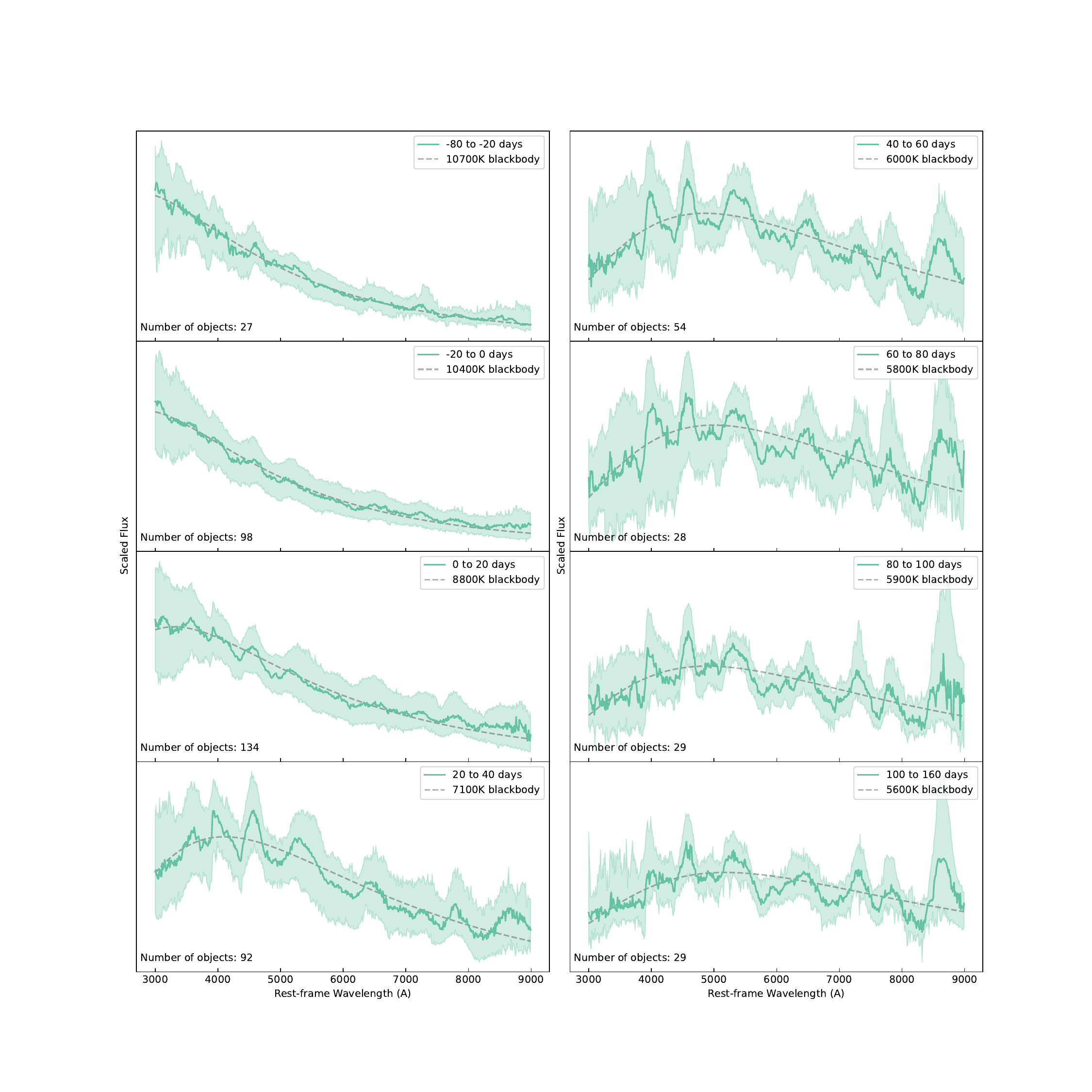}
		\caption{Composite average spectra in time bins relative to maximum light. The light green shaded region denotes the 1$\sigma$ uncertainty. Number of objects contained in each bin are tagged, with a minimum of 20 events imposed per bin. Best fit blackbody are indicated in grey dashed lines.}
            \label{fig:av_specs}
	\end{center}
\end{figure*}

\begin{figure*}
	\begin{center}
            \large{Scaled Phases from Explosion}\par
		\includegraphics[width=2\columnwidth, trim={4cm 3cm 3.5cm 4cm},clip]{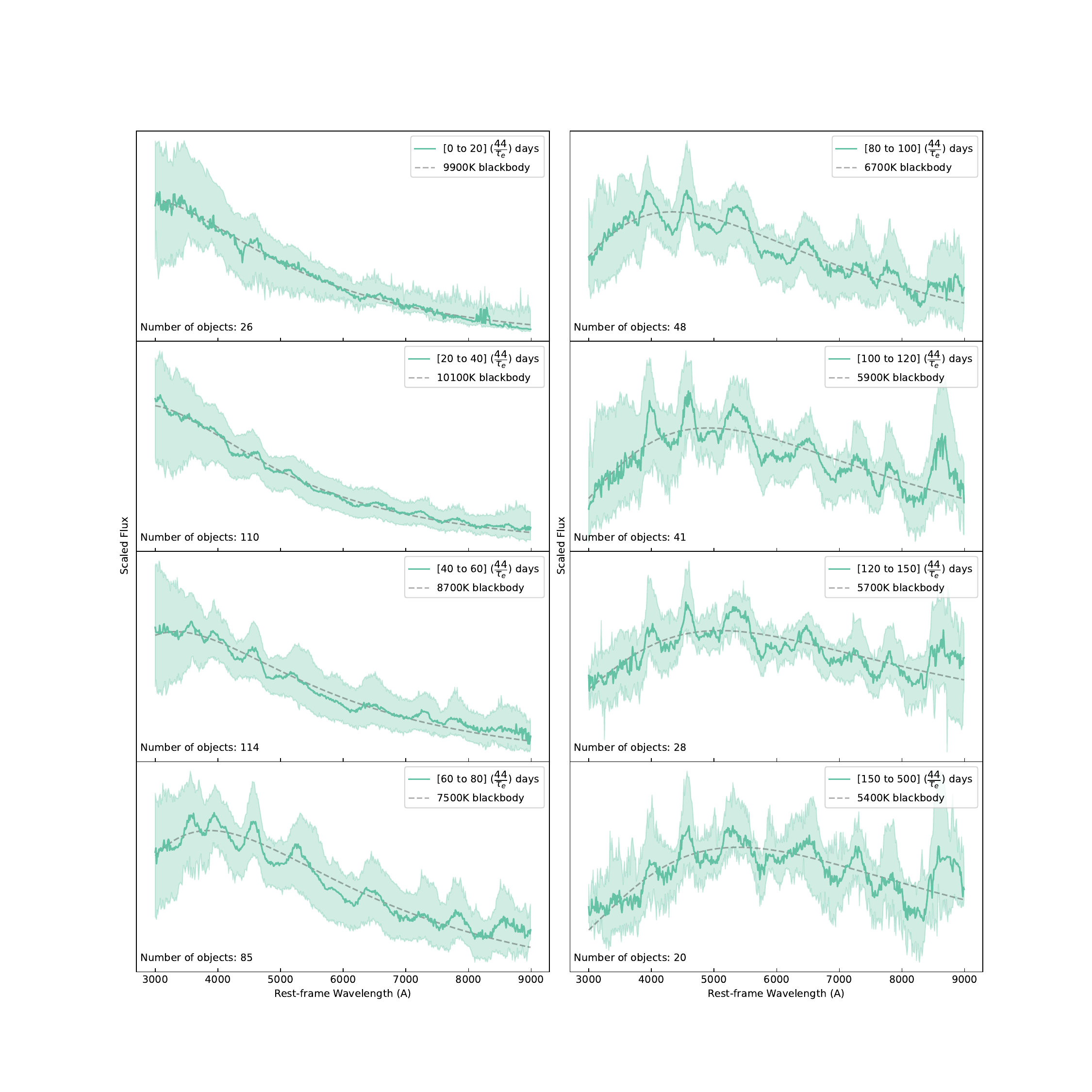}
		\caption{Composite average spectra in time bins relative to explosion, and scaled . The light green shaded region denotes the 1$\sigma$ uncertainty. Number of objects contained in each bin are tagged, with a minimum of 20 events imposed per bin. Best fit blackbody are indicated in grey dashed lines.}
            \label{fig:av_specs_exp}
	\end{center}
\end{figure*}

\begin{figure}
	\begin{center}
		\includegraphics[width=1\columnwidth]{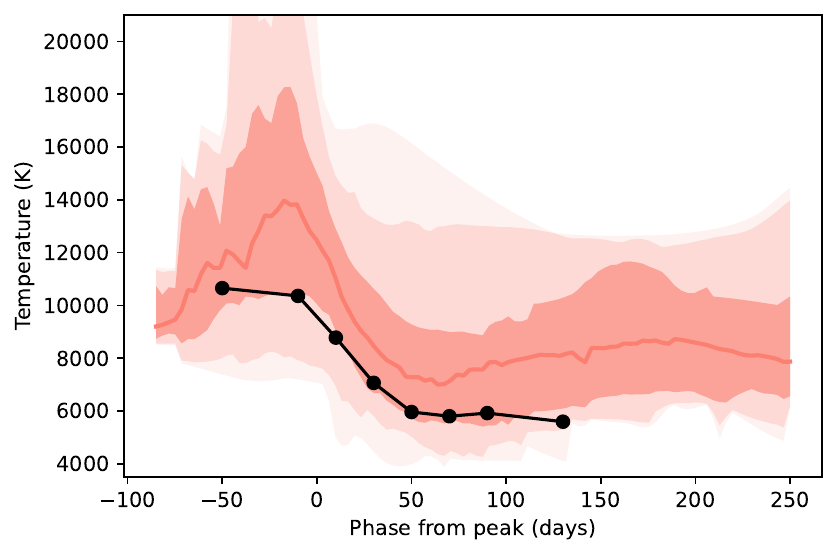}
		\caption{The temperature evolution of average spectra from Figure \ref{fig:av_specs}. These are grouped in phase from peak and plotted in black. In comparison, the temperature evolution derived from the photometry in \citet{Gomez2024} is plotted in red. The shades of red denote the 1, 2, and 3$\sigma$ ranges. The spectroscopic temperatures fall within the 2$\sigma$ spread of photometric values but are systematically lower, likely due to the spectral coverage not extending as far into the UV.}
            \label{fig:temp_ev}
	\end{center}
\end{figure}

\section{Average Spectra}
\label{sec:average_spectra}

To investigate the typical behaviour of SLSNe as a function of time, we constructed average spectra in different time bins. This was also done so features that may be faint in some spectra could become more evident in the averages. Similar techniques have been applied to other astrophysical transients such as SNe Ic, GRB afterglows and quasars \citep[e.g.][]{Modjaz2016, Christensen2011, Selsing2016}. The evolution of the average spectra in this analysis was examined using both the phases from peak and the scaled phases from explosion. Bin sizes were chosen by starting at the earliest phase, and incrementing the sizes by 10 days until a minimum of 20 objects were included, with a minimum bin size of 20 days. These average spectra will be included in this data release.

\subsection{Method of Averaging}
\label{sec:av_spec_method}

The average spectra were created by taking the smoothed and remapped spectra described in Section \ref{sec:data_processing}. Each spectrum was first normalised by its mean value to 1. Then, for each wavelength point, the median flux across all spectra was calculated. In order to not bias bins with multiple spectra from the same objects, this process was first applied to each object to create a single median spectrum per object per bin, which were then used to compute the overall statistics for that time bin.


The average spectra created are displayed in Figure \ref{fig:av_specs} for phases from peak, and in Figure \ref{fig:av_specs_exp} for scaled phases from explosion as described in Section \ref{sec:data_processing}. The 1$\sigma$ uncertainty is measured by calculating the standard deviation of the flux measurements at each wavelength. Each panel has been fit with a single blackbody, the temperature of which is indicated in the legend. This is a very good approximation at early times where the equivalent widths of absorption/emission lines are weak compared to the flux from the blackbody. At these earliest epochs, the blackbody temperature is $\sim$10000\,K. However, SLSNe are often described using a modified blackbody, where the flux below a cutoff wavelength has a linear suppression applied to it \citep{Nicholl2017a, Yan2018}. This is due to the significant absorption in the UV from lines such as  C II $\lambda\lambda$2325, 2328, Si III $\lambda$2543, and Mg II $\lambda$2800 \citep{Quimby2011, Chomiuk2011, Howell2013, Vreeswijk2014}. In SLSNe, this cutoff wavelength has been found to be $\lambda = 3400^{+1000}_{-700}$\,\AA\ \citep{Gomez2024}. However, our spectral analysis does not extend beyond 3000\,\AA\ into the UV and so we cannot comment on any trends regarding the use of a modified blackbody. 

Over the next few months, the spectra begin to transition to more emission lines, and the underlying blackbody temperature decreases to 7000$-$8000\,K. We can see the blackbody still captures the general shape of the spectral energy distribution (SED) very well. Popular light curve fitting codes for SLSNe such as \texttt{MOSFiT} and \textsc{Redback} assume a modified blackbody SED \citep{Nicholl2017c, Sarin2024}, which is supported by our analysis down to 3000\,\AA.

\texttt{MOSFiT} and \textsc{Redback} also introduce an additional free parameter to fit of the final plateau temperature ($T_{\rm{min}}$) in their magnetar models. This parameter is important for extending these fits to late times and simply introduces a minimum temperature to which the fits converge. In \texttt{MOSFiT}, the default prior for this parameter is a Gaussian with a mean of 6000\,K and a standard deviation of 1000\,K, whereas in \textsc{Redback} the default is a uniform prior between 3000$-$10,000\,K. Applying \texttt{MOSFiT} to a sample of 238 SLSNe, \citet{Gomez2024} obtain a value of $T_{\rm{min}} = 6500^{+1700}_{-1400}$\,K. This is consistent with previous work by \citet{Nicholl2017c}, who used a sample of 38 SLSNe and obtained an average of $\sim$6400\,K. We can see the blackbody temperature in both Figure \ref{fig:av_specs} and Figure \ref{fig:av_specs_exp} plateauing to about 5000-6000\,K which supports the use of $T_{\rm{min}}$ in the fitting.

We can also compare the average temperature evolution derived from the spectra to the temperature derived from fitting \texttt{MOSFiT} models to the light curves. This is shown in Figure \ref{fig:temp_ev} which displays the mean, 1$\sigma$, 2$\sigma$, and 3$\sigma$ ranges for the population. The temperatures derived from the spectra fall within the 2$\sigma$ spread and are therefore consistent with those obtained from photometry. However, they are systematically lower, which may be attributed to our spectra not extending as far into the UV as the photometry. This sample is limited to 3000$-$9000\,\AA\ in the rest frame (the range covered by the vast majority of spectra). This restriction likely prevents us from capturing the peak of the blackbody fits, particularly at early times.

\subsection{Time From Peak vs Time From Explosion}
\label{sec:binning_scheme}
 
Looking at the evolution of the spectra in Figure \ref{fig:av_specs} and Figure \ref{fig:av_specs_exp}, we can see the earliest two bins corresponding to the time before maximum light show high and roughly constant temperatures, between 10000$-$11000\,K. At these phases we see a hot blue continuum with weak absorption lines in comparison. The next three bins show a fairly rapid change in properties, with the emergence of stronger P-Cygni and potentially emission lines which will be discussed further in Section \ref{sec:line_IDs}. This is accompanied by a drop to $\sim$6000\,K. The final three bins then show a very slow evolution, with the continuum emission decreasing in strength relative to the lines, but no evidence for further cooling. We note that we see very similar behaviour using both binning schemes.

We investigated whether the behaviour of SLSNe is more homogeneous when grouped by phase from peak (convenient observationally), or scaled phase from explosion (perhaps better motivated physically). The fits from these methods were quantified by calculating the relative scatter (standard deviation, $\sigma_{F}$, divided by the median flux, $F_{\rm med}$) in each group, averaged over wavelength. Averaging over the eight time bins in each binning scheme, we find only a marginal change in the scatter using the two methods. The averages using phases from peak showed $\langle \sigma_{F, {\rm peak}} / F_{\rm med,peak} \rangle = 0.66$, whereas using the scaled phases from explosion showed $\langle \sigma_{F, {\rm exp}} / F_{\rm med,exp} \rangle = 0.64$. This quantifies the visual observation that both Figure \ref{fig:av_specs} and Figure \ref{fig:av_specs_exp}, and therefore both methods of averaging produce reasonable results. However, the averages obtained when binning in scaled phases from explosion exhibit slightly tighter fits, but only marginally.

As the fractional standard deviations were so close, we cannot say for certain if either binning scheme provides a more homogenous grouping. The reason for this may be that SLSNe seem to show a constant temperature and very little spectral evolution up to maximum light. After this, the spectra only evolve significantly during the rapid photospheric cooling phase, usually immediately after peak. Since we scale the explosion phases based on the light curve evolution rate, so that all scaled spectral series effectively peak in the same bin, this largely erases any differences in temperature in the bins between measuring from peak or measuring from explosion. This is also apparent in the blackbody fits, which are very similar for the two binning schemes. Physically, this suggests that a characteristic temperature evolution drives the spectral evolution we see in SLSNe.

\begin{figure}
	\begin{center}
		\includegraphics[width=1\columnwidth]{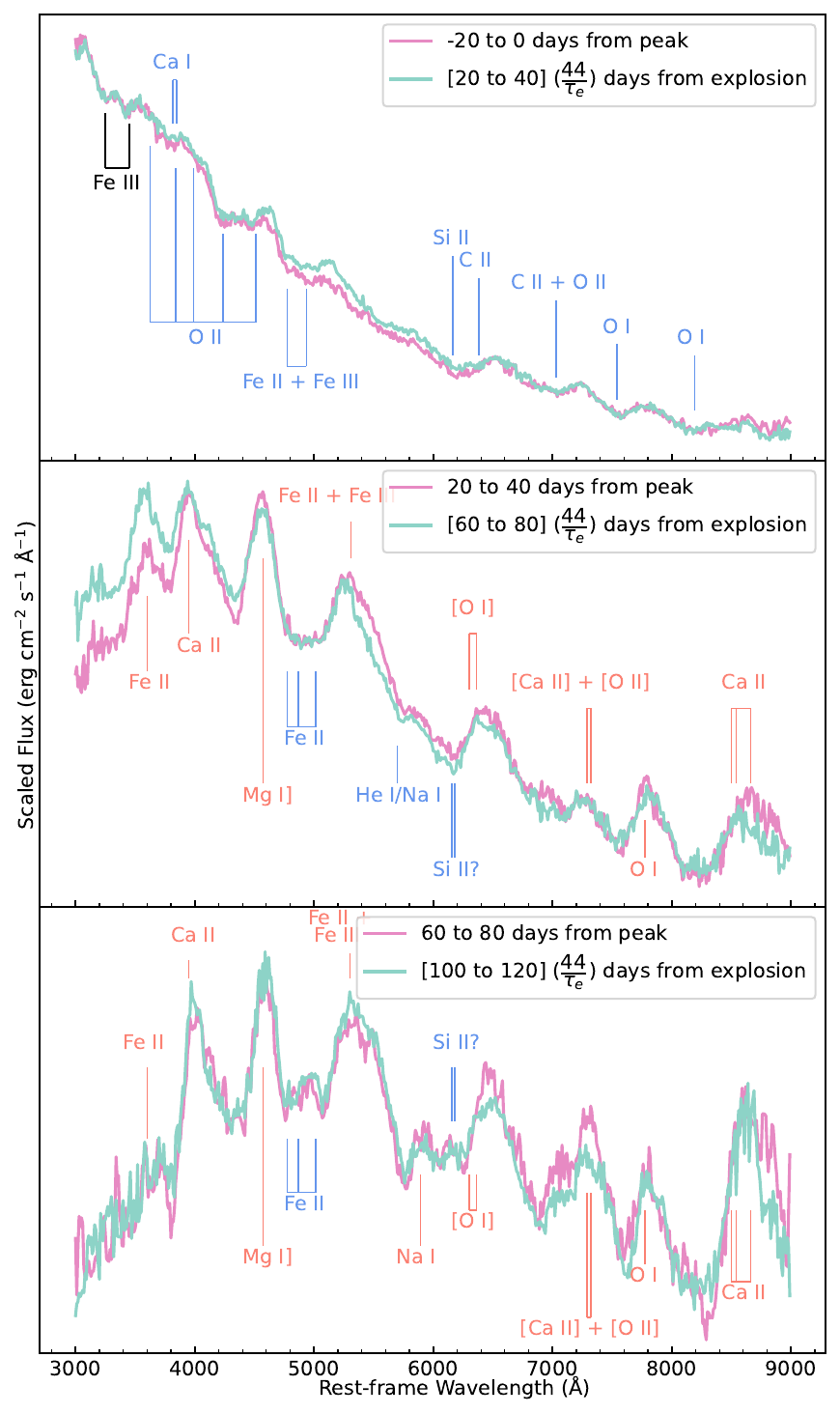}
		\caption{Subset of average spectra. Plotted both in days from peak, as well as scaled days from explosion to match the median e-folding decline time of SLSNe (44 days). $\tau_{e}$ refers to the time taken for the light curve to reach 1/e times the maximum luminosity. Blue labels indicate absorption lines which have all been shifted by 9000\,km\,s$^{-1}$. Red labels are for emission lines which are shown at rest.}
            \label{fig:av_specs_lines}
	\end{center}
\end{figure}

\begin{figure}
	\begin{center}
		\includegraphics[width=1\columnwidth]{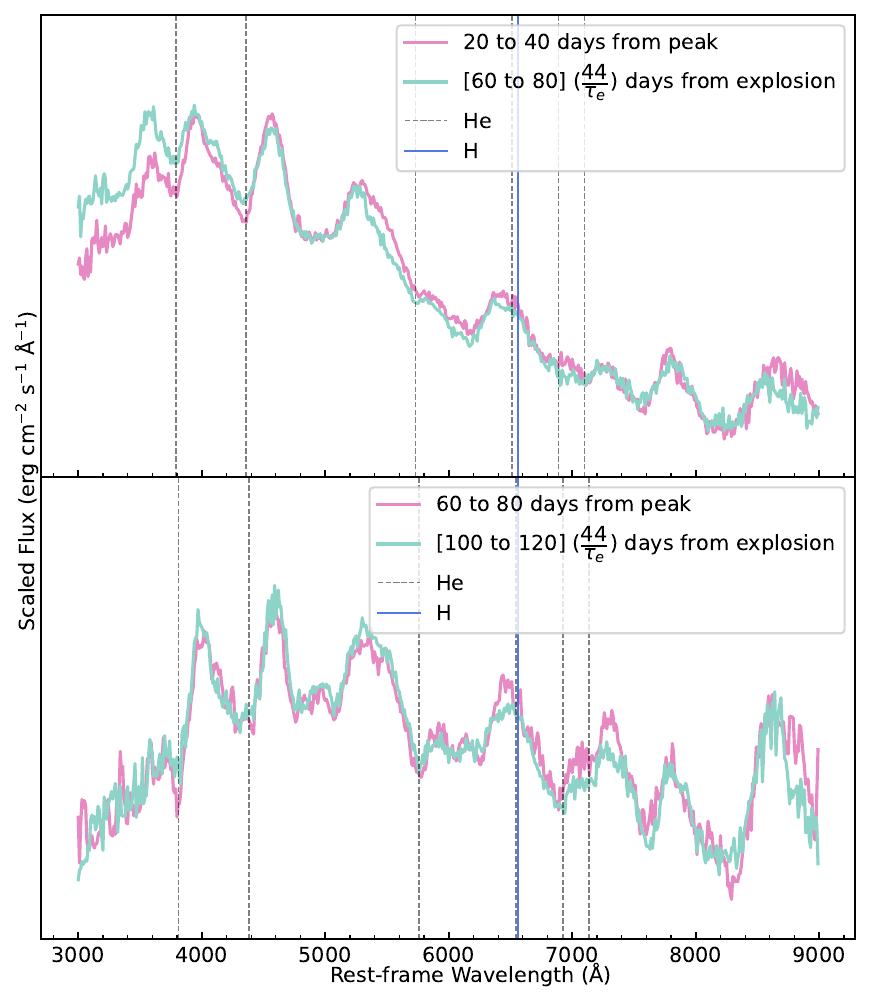}
		\caption{Subset of average spectra. Plotted both in days from peak, as well as scaled days from explosion to match the median e-folding decline time of SLSNe (44 days). Dashed vertical lines indicate He absorption lines which have been blueshifted by 7500\,km\,s$^{-1}$ in the top panel, and by 6000\,km\,s$^{-1}$ in the bottom panel in order to match observed troughs in the spectra. The solid blue line corresponds to rest-frame H$\alpha$.}
            \label{fig:he_check}
	\end{center}
\end{figure}

\subsection{Line Identifications}
\label{sec:line_IDs}

Figure \ref{fig:av_specs_lines} shows a subset of the average spectra at representative epochs. Averages from both explosion and peak are plotted together in corresponding bins from each, with both having the same bin sizes. As discussed in Section \ref{sec:binning_scheme}, both methods result in a similar spectral evolution, and therefore it is not surprising to see that the corresponding bins look similar. There are only minor discrepancies in the shape of the continua and line strengths but these fall into the 1$\sigma$ uncertainties (not plotted in this figure for visibility). All absorption lines are labelled in blue and have been blueshifted by $\sim$9000\,km\,s$^{-1}$ to match the observed troughs of features. Emission lines are labelled in red and are plotted at rest.

The top panel shows the spectra between $-$20 to 0 days relative to peak, and 20 to 40 scaled days from explosion. The blackbody temperature is about 10000\,K. We note that although we can see the series of characteristic O II absorption lines, these spectra do not show a strong "W" shape. Instead this shape appears to be more of a broad trough due to dilution when averaging over spectra with a range of velocities. The effective wavelengths of these O II blends are 4650.71\,\AA, 4357.97\,\AA, 4115.17\,\AA, 3959.83\,\AA, and 3737.59\,\AA\ \citep{Quimby2018}. At redder wavelengths we see lines from Si II $\lambda$6355, C II $\lambda$6580, and O I $\lambda$7774 and O I $\lambda$8446. There are also potential Fe III lines at around 3200\,\AA\ \citep{Leloudas2012, Nicholl2013}. This is supported by spectral models created by \citet{Dessart2019} around peak, where Fe III is the only ion to show significant features in the form of P-Cygni lines.

Looking at Figure \ref{fig:av_specs} and Figure \ref{fig:av_specs_exp}, at the very early spectra from $-$80 days before peak, and 20 scaled days from explosion, we can see a small bump indicating excess flux around $\sim$7300\,\AA. This is coincident with a known nebular line due to emission from forbidden lines of [Ca II] and [O II]. This shows early emission of nebular lines may be more prevalent than previously thought. Only a handful of SLSNe have shown this line before or at peak, including SN\,2019szu \citep{Aamer2024}, and SN\,2018ibb \citep{Schulze2024}. Some normal SNe Ic have also shown early appearances of this line. This includes SN\,2022xxf, an unusual event with a double peaked light curve and appearance of forbidden lines before the second peak \cite{Kuncarayakti2023}. There does not seem to be any evidence at these phases for [O III] $\lambda\lambda$4959, 5007, which could imply that in the majority of these events the early emission is dominated by calcium. This does not rule out the excess flux being attributed to [O II] $\lambda\lambda$7320,7330 as this line is expected to be stronger than both [O III] lines based on models due to the higher temperatures required for [O III] \citep{Jerkstrand2017a, Omand2023}. In the case of SN\,2019szu, the identification of [O II] at peak is secure \citep{Aamer2024}. We therefore performed a visual search through all objects for other events that show evidence of other [O II] and [O III] lines, which would suggest that [O II] dominates the early emission at 7300\,\AA. We find one additional case similar to SN\,2019szu, SN\,2022le which also displays strong forbidden lines of oxygen at early phases. This object will be the subject of a future paper (Blanchard, P. et al., in prep.).

The middle panel shows the spectra between 20 to 40 days relative to peak, and 60 to 80 scaled days from explosion. The blackbody temperature is about 7000\,K at these phases. By this stage the hot blue continuum has decreased and the line profiles are mostly P-Cygni shaped. These include a blend of Fe II at $\sim$3600\,\AA\ and $\sim$5300\,\AA, Ca II $\lambda\lambda$3934, 3968, [Ca II] $\lambda\lambda$7291, 7323, and [O II] $\lambda\lambda$7320, 7330. Some of the lines look more symmetrical and therefore may consist of pure emission such as Mg I] $\lambda$4571, and O I $\lambda$7774. The feature around 6000-6500\,\AA\ is difficult to identify, with a complex profile and an emission peak wavelength that does not match any common line. We suggest this could be caused by [O I] $\lambda\lambda$6300,6364 in emission combined with Si II $\lambda\lambda$6347, 6371 in absorption. It is interesting to note the appearance of more forbidden and semi-forbidden transitions which require lower density material in order to form. At this stage in the evolution, the spectral lines look very similar to those of typical SNe Ic. The one exception to this is the region around 5700-6500\,\AA, where SNe Ic often do not display the trio of lines seen in SLSNe \citep{Nicholl2016b}. However, it is difficult to reconcile the hotter underlying continua for SLSNe, compared to the cooler temperature of the line forming region if the SN ejecta is in radiative equilibrium. \citet{Chen2017} explain this by the presence of porous or clumpy, H- and He poor material, which the ejecta can interact with and introduces an additional hot black-body component to the flux. Alternatively, the ejecta can become clumped due to engine-driven turbulence, leaving low-density regions capable of producing the emission lines and dense regions that provide the continuum \citep{Chen2020}.

The final panel in Figure \ref{fig:av_specs_lines} presents the spectra from 60 to 80 days relative to peak and 100 to 120 scaled days from explosion. While many of the same lines visible in the middle panel remain, the profiles have transitioned from P-Cygni-like features to being more emission dominated. Additionally, the underlying blackbody continuum has cooled significantly to approximately 6000\,K. In this phase, we observe the emergence of the Na I $\lambda\lambda$5890, 5896 doublet and a more pronounced Fe II absorption feature. A prominent feature appears around 7100\,\AA\ in the averages relative to peak, though it remains unidentified and is not labelled in this panel. The profile shape could be a result of absorption from He (see next section). Notably, the Fe II emission near 3800\,\AA\ is weaker compared to the previous panel.

\subsection{Degree of envelope stripping}

The post-maximum spectra exhibit a small absorption feature around 5700\,\AA\ which could be attributed to either He I $\lambda$5876 or Na I $\lambda\lambda$5890, 5896, both of which would need to be blueshifted by $\sim$9000\,km\,s$^{-1}$. A He-rich sub-class of SLSNe has been proposed by \citet{Yan2020}. To investigate whether this feature is indeed due to helium, we searched for evidence of other He lines in the average spectra. This is illustrated in Figure \ref{fig:he_check}, where prominent He lines at 3888\,\AA, 4471\,\AA, 5876\,\AA, 6678\,\AA, 7065\,\AA, and 7281\,\AA\ are plotted for the bottom two panels of Figure \ref{fig:av_specs_lines}. By blueshifting the lines in the top panel by approximately 7500\,km\,s$^{-1}$, we observe that several lines align with troughs in the spectra. Notably, the 7065\,\AA\ and 7281\,\AA\ lines match up to small dips in the spectra which are not attributed to other features. This trend continues in the bottom panel with the lines blueshifted by $\sim$6000\,km\,s$^{-1}$. Determining with certainty whether all these lines are present is difficult, particularly since the 3888\,\AA\ and 4471\,\AA\ lines coincide with troughs caused by other emission lines, and the 6678\,\AA\ line overlaps with an emission peak. However, it could be argued that a slight dip is visible at the location of the 6678\,\AA\ line, which is especially noticeable in the spectra grouped by phase from peak. This shows creating average spectra helps identify features which may not be obvious in individual spectra, but are strengthened by combining them. The possible presence of He in the averages suggests that many SLSNe may retain some residual He at the time of explosion.

Checks were also performed to look for signatures of H and C, however we did not find evidence for these elements in the averages. Rest-frame H$\alpha$ is plotted in Figure \ref{fig:he_check} with a solid blue line. The lack of H is to be expected as this sample looks at Type I SLSNe which are inherently H poor. There have been some events with evidence for H at late times, in the form of emission lines from late-onset interaction with CSM \citep{Yan2015, Yan2017a}, however we do not find strong indicators for this in our sample showing this is not ubiquitous in the population. The lack of C shows the presence of this may also be a unique and rare feature \citep[e.g.][]{Anderson2018, Gutierrez2022, Gkini2024}.

\begin{figure*}
\centering
\begin{minipage}{.48\textwidth}
\begin{center}
        \begin{subfigure}{1\columnwidth}
            \centering
            \includegraphics[width=1\columnwidth]{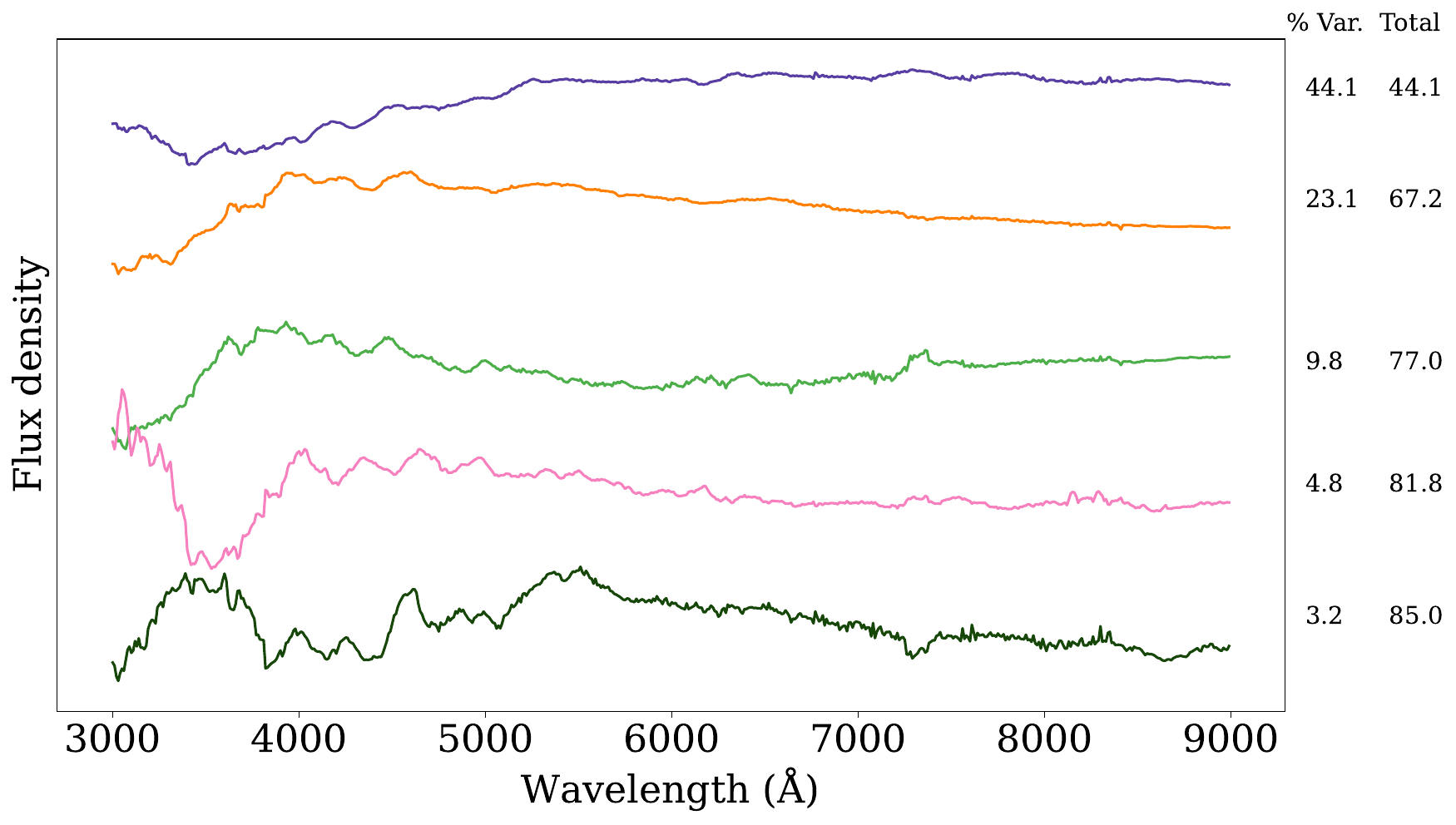}
            \caption{Days -80 - -5}
            \label{fig:-80--5_peak}
        \end{subfigure}
     \hfill
        \begin{subfigure}{1\columnwidth}
            \centering
            \includegraphics[width=1\columnwidth]{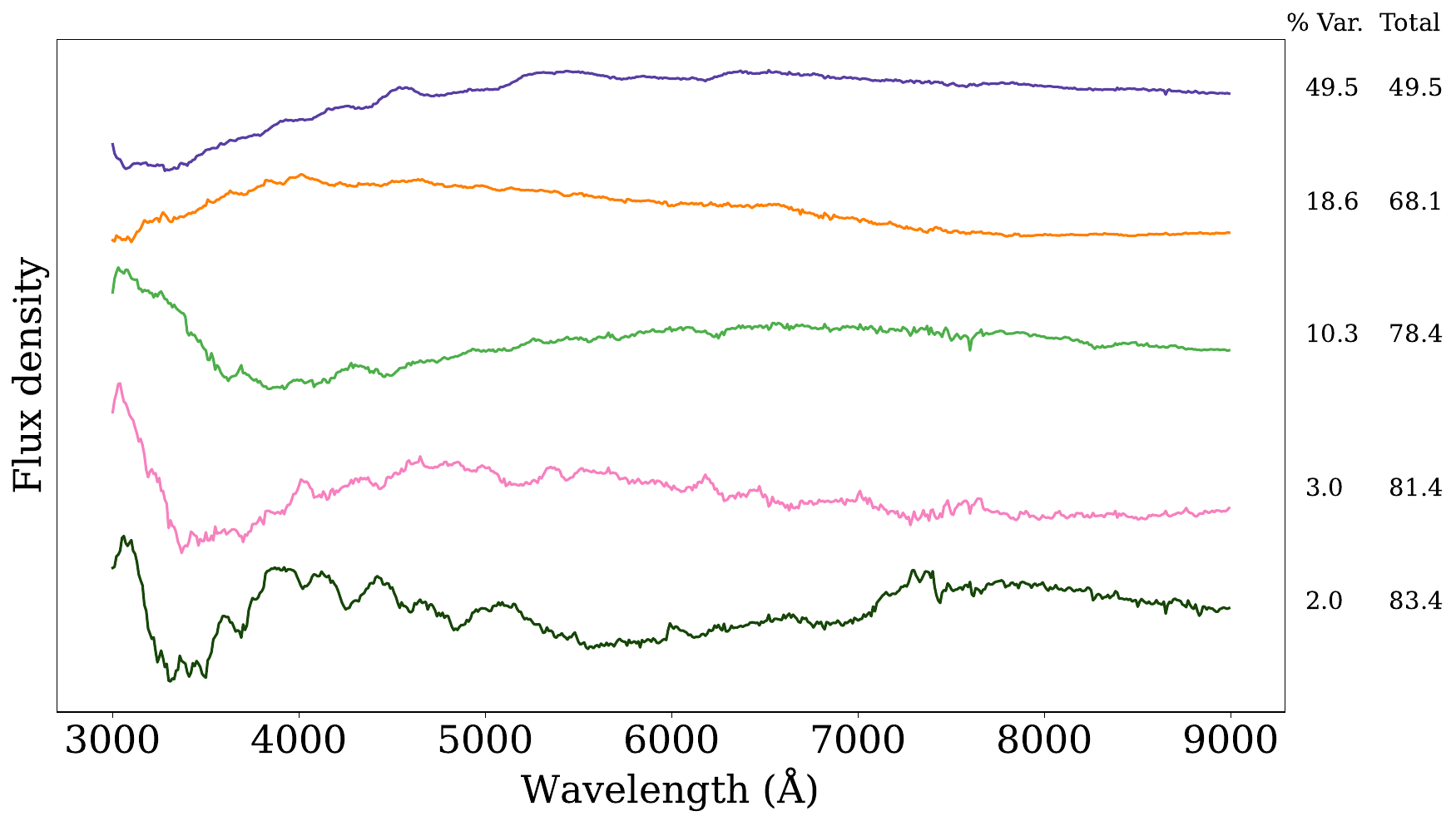}
            \caption{Days -5 - 15}
            \label{fig:-5-15_peak}
        \end{subfigure}
     \hfill
        \begin{subfigure}{1\columnwidth}
            \centering
            \includegraphics[width=1\columnwidth]{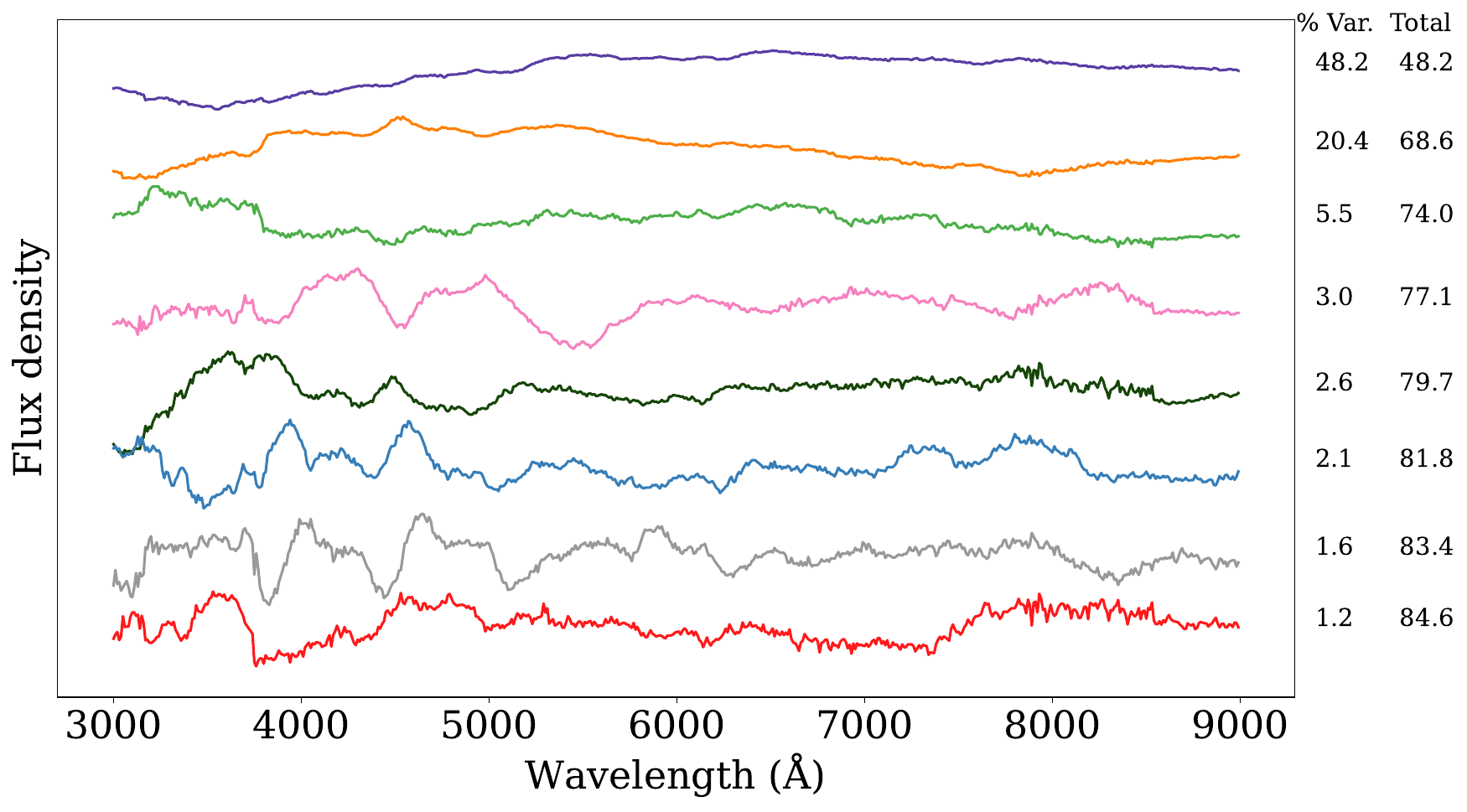}
            \caption{Days 15 - 40}
            \label{fig:15-35_peak}
        \end{subfigure}
     \hfill
        \begin{subfigure}{1\columnwidth}
            \centering
            \includegraphics[width=1\columnwidth]{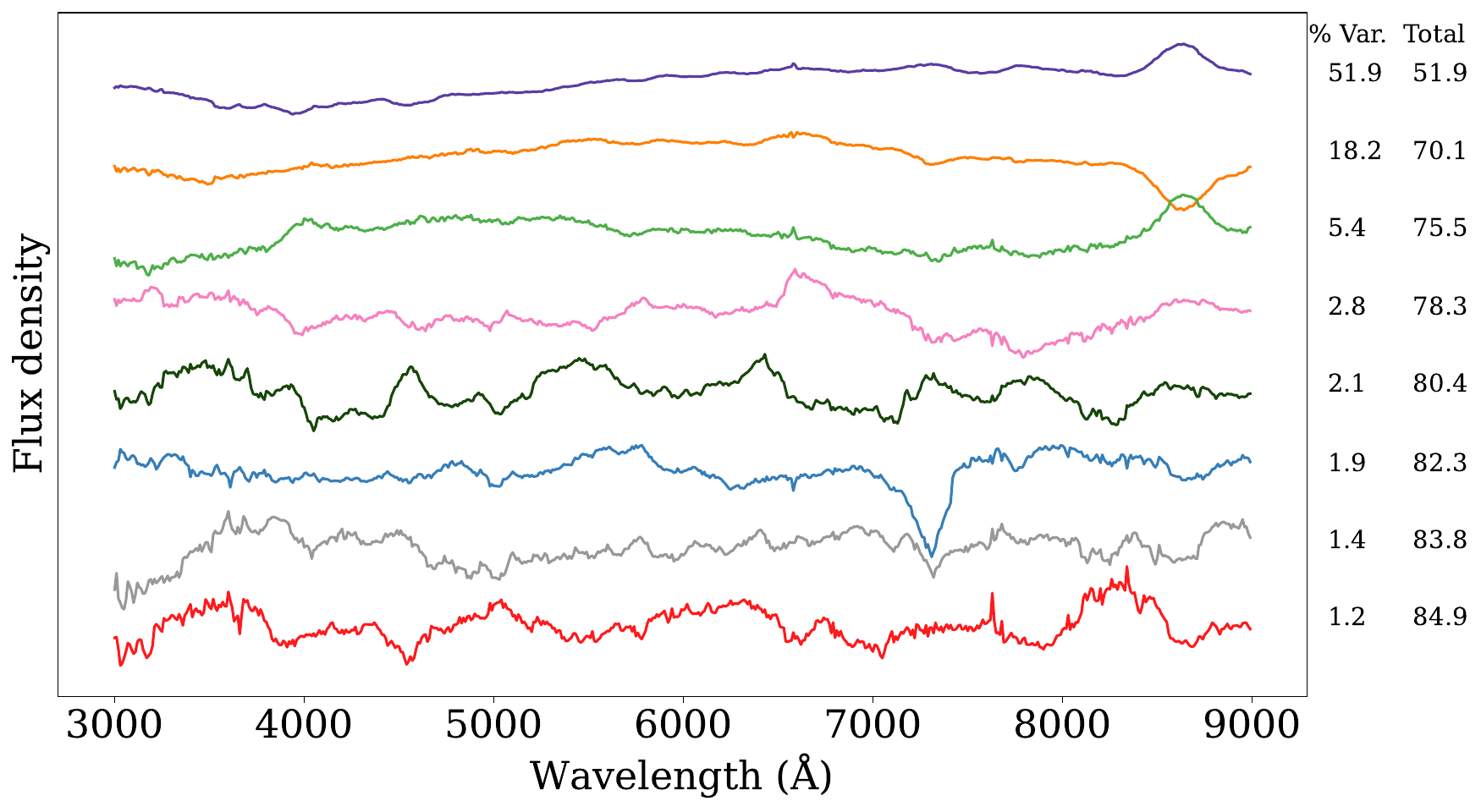}
            \caption{Days 40 - 160}
            \label{fig:35-160_peak}
        \end{subfigure}
     \hfill
        \caption{PCA decomposition of SLSN spectra in different time bins relative to peak. Explaining variations up to 85\% of the sample. The Var column indicates the percentage variation explained by each component, and the Total percentage column shows the cumulative percentage explained by the addition of each new component. More components are needed at later times meaning greater variation in the spectra at late times.}
        \label{fig:PCA_peak}
    \end{center}
\end{minipage}%
\hfill
\begin{minipage}{.48\textwidth}
    \vspace{4.6mm}
    \begin{center}
        \begin{subfigure}{1\columnwidth}
            \centering
            \includegraphics[width=1\columnwidth]{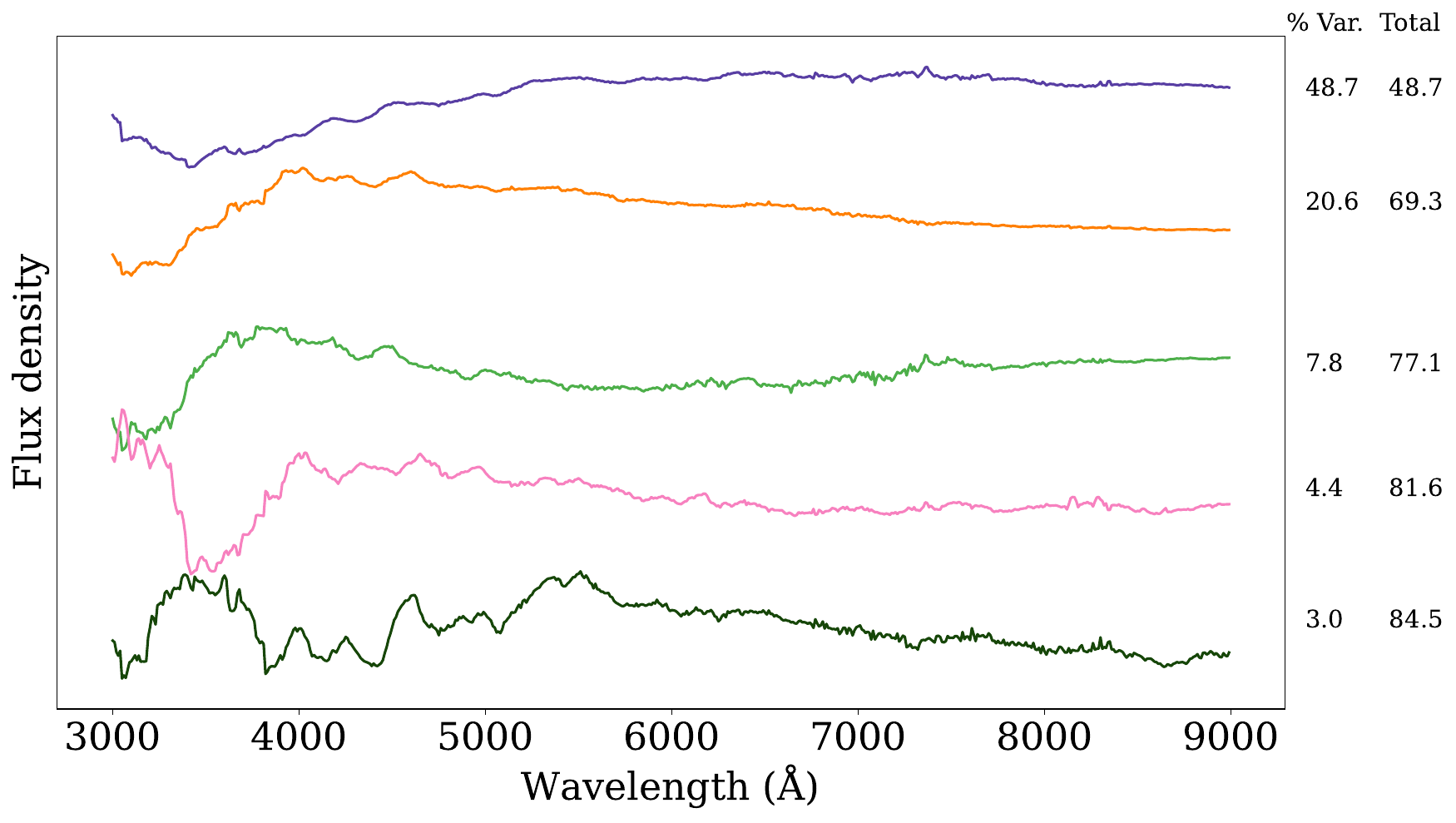}
            \caption{Days 0 - 35}
            \label{fig:-5-35_exp}
        \end{subfigure}
     \hfill
        \begin{subfigure}{1\columnwidth}
            \centering
            \includegraphics[width=1\columnwidth]{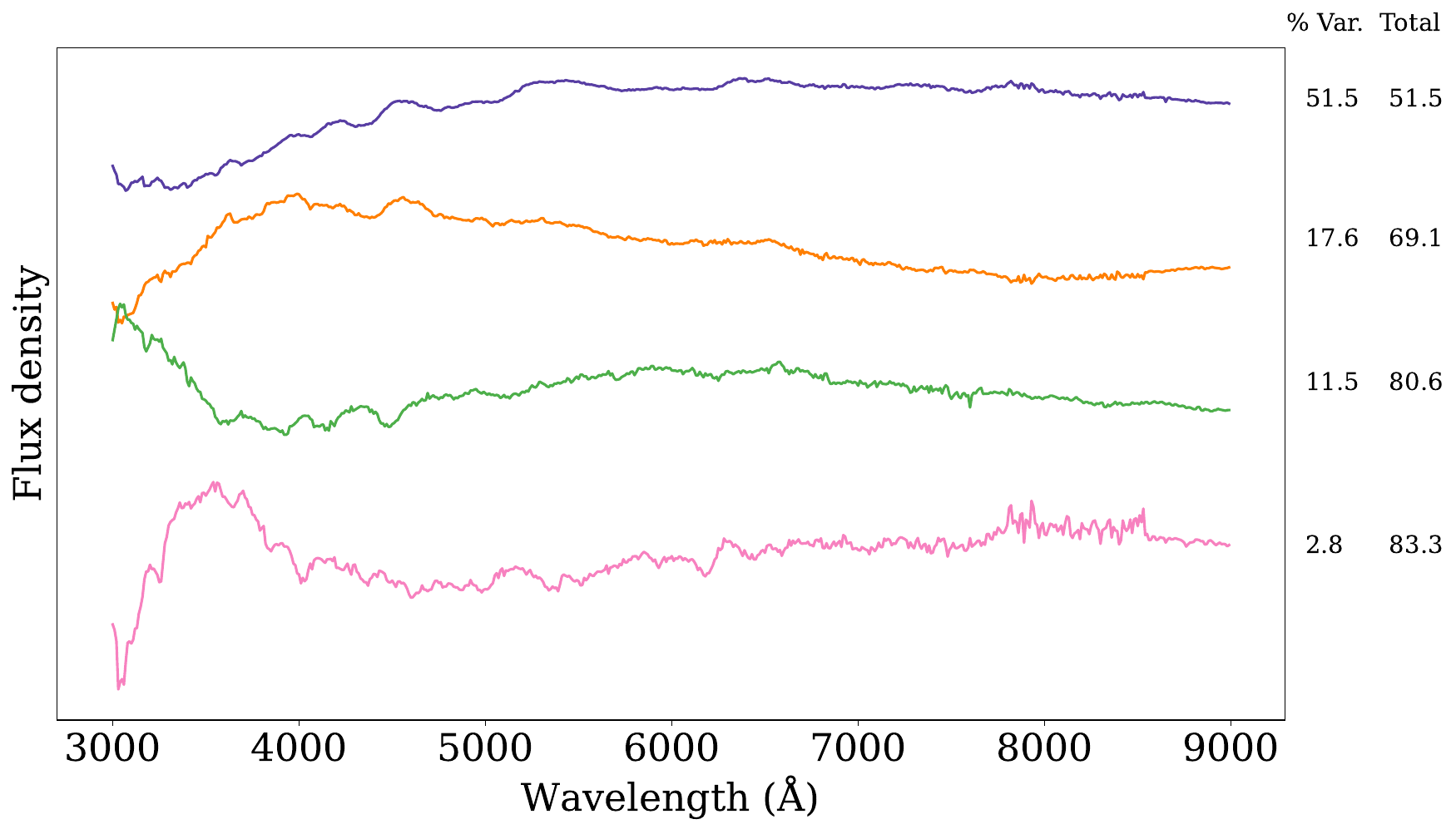}
            \caption{Days 35 - 55}
            \label{fig:35-55_exp}
        \end{subfigure}
     \hfill
        \begin{subfigure}{1\columnwidth}
            \centering
            \includegraphics[width=1\columnwidth]{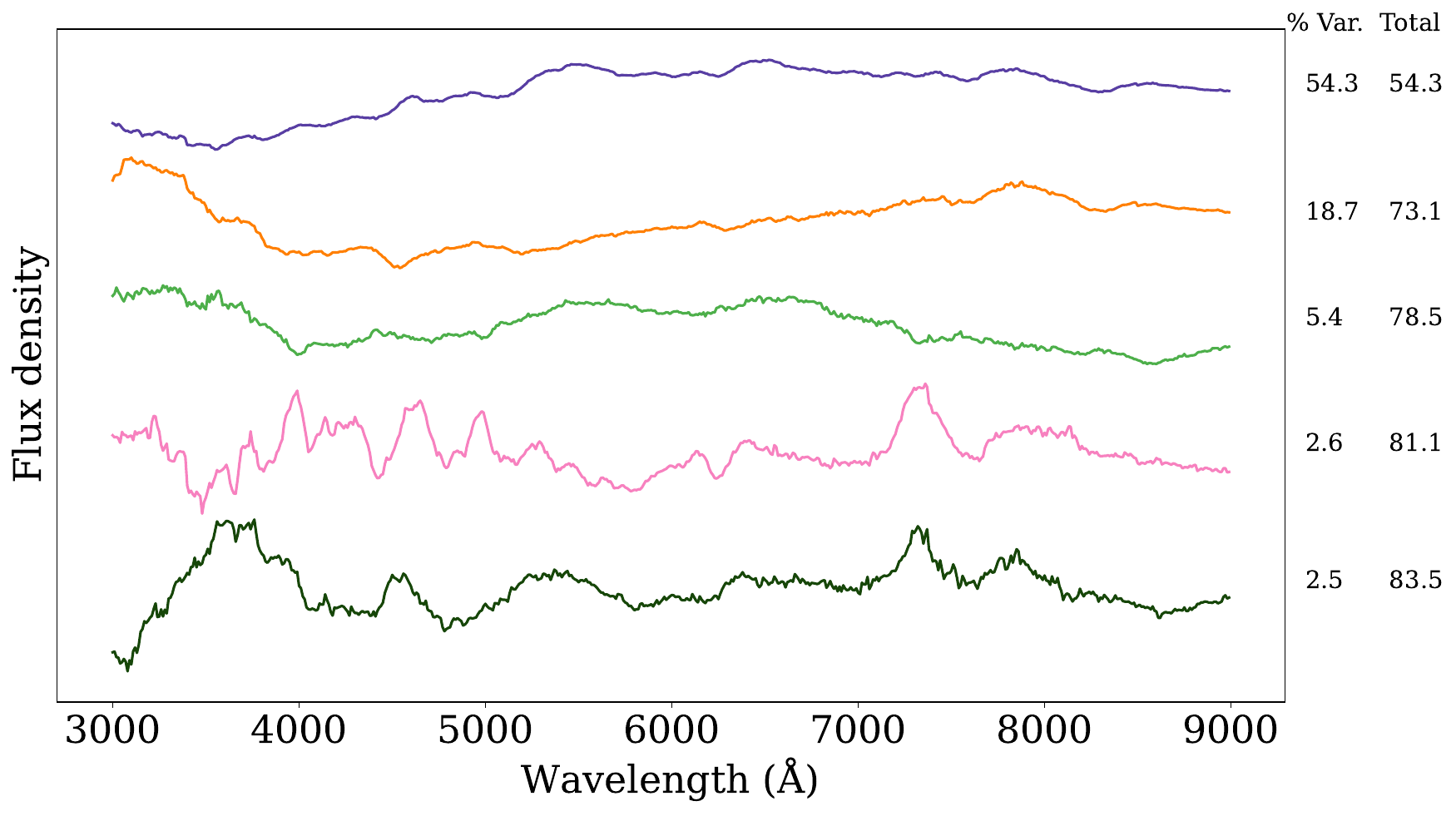}
            \caption{Days 55 - 85}
            \label{fig:55-85_exp}
        \end{subfigure}
     \hfill
        \begin{subfigure}{1\columnwidth}
            \centering
            \includegraphics[width=1\columnwidth]{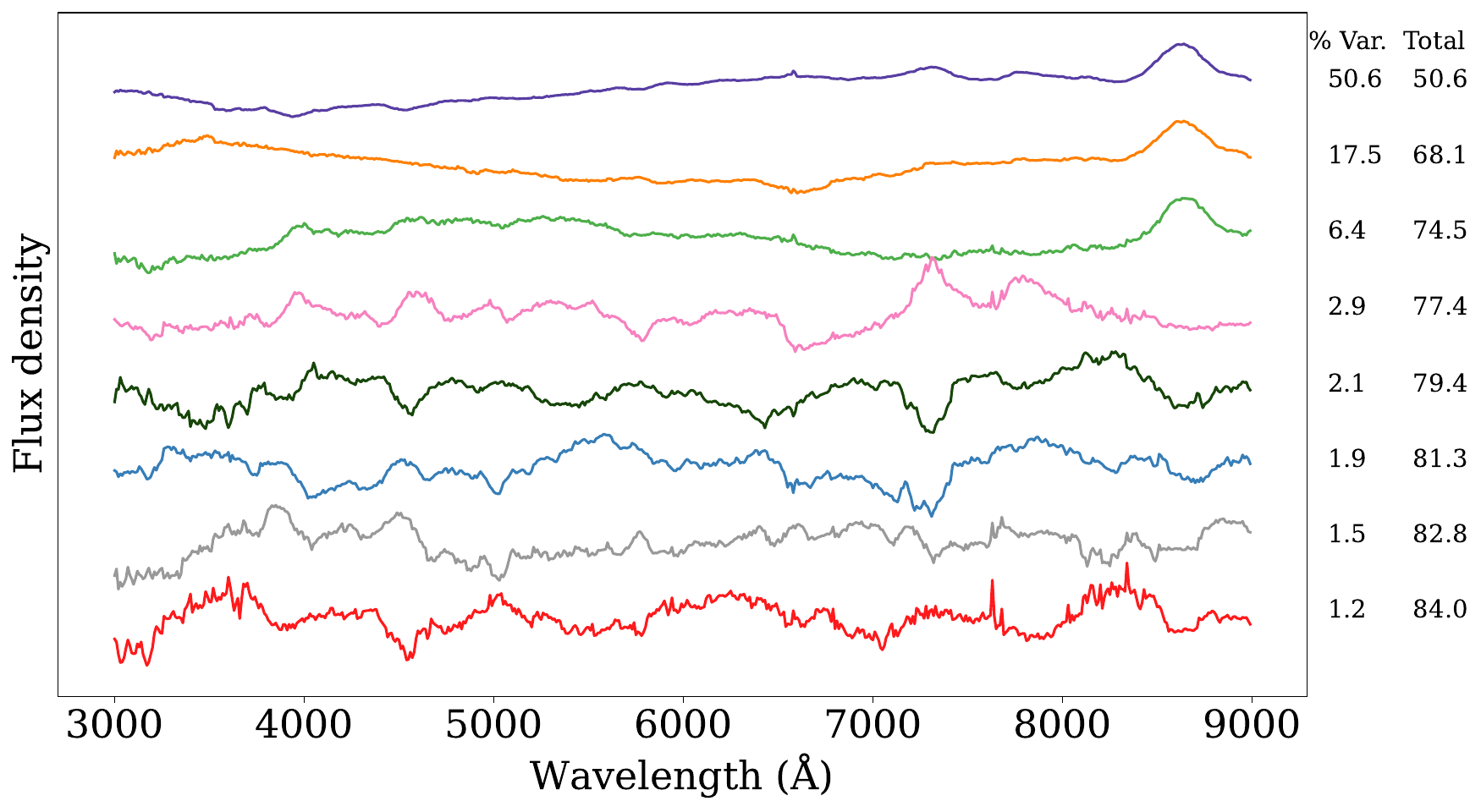}
            \caption{Days 80 - 500}
            \label{fig:80-500_exp}
        \end{subfigure}
     \hfill
        \caption{PCA decomposition of SLSN spectra in different time bins relative to explosion, with the phase scaled to match the median e-folding decline time of SLSNe (44 days). Explaining variations up to 85\% of the sample. The Var column indicates the percentage variation explained by each component, and the Total percentage column shows the cumulative percentage explained by the addition of each new component. More components are needed at later times meaning greater variation in the spectra at late times.}
        \label{fig:PCA_exp}
    \end{center}
\end{minipage}
\end{figure*}

\section{Principal Component Analysis}
\label{sec:pca}

In the Section \ref{sec:average_spectra} we investigated the evolution for "representative" SLSNe and searched for similarities between the population as a whole. Now, we will investigate the diversity of the population and which spectral features account for this.

Principal Component Analysis (PCA) is a technique used to reduce the dimensionality of a dataset, whilst retaining its key information. This decomposition creates a set of principal components, from which all the spectra in that set can be recreated. The first principal component explains the most variance, and each subsequent one explains the next largest portion of variation, with the key requirement that each new component is uncorrelated with the ones before it. The resulting set of principal components form an orthogonal basis in the space of the data \citep{Pearson1901, Hotelling1933}. Spectral data often contains a lot of redundant information because flux from nearby wavelengths tend to be highly correlated. PCA can summarise the data by identifying a few key components that describe most of the variability in the sample. These components can then be used to group samples, identify outliers, or understand underlying patterns in the data. 

The data in this sample was split by phase into four bins. Using the same bin sizes as in Section \ref{sec:average_spectra} would result in too few spectra per time bin and so new boundaries were chosen. Each bin was chosen to contain roughly 150 objects. We note that we see little evolution over the phase ranges included in these broader bins. Spectra that did not cover at least 60\% of the wavelength range between 3000-9000\,\AA\ were not included in this analysis. Cuts were also applied to objects that were too noisy to contribute meaningfully to the analysis. The breakdown of these bins is given in Table \ref{tab:pca_peak} and Table \ref{tab:pca_exp}. The data was also standardised by normalising to the median. This was to ensure all spectra were contributing equally to the analysis.

\begin{center}
\begin{table}
    \begin{tabular}{|P{4.5cm}|P{1.2cm}|P{1.5cm}|} 
    \hline\hline
    Phases from Peak (days) & Spectra & Components \\ 
    \hline
    -80 \hspace{0.1em} -- \hspace{0.1em} -5 & 154 & 5 \\ 
    -5 \hspace{0.1em} -- \hspace{0.1em} 15 & 168 & 5 \\ 
    15 \hspace{0.1em} -- \hspace{0.1em} 40 & 155 & 8 \\ 
    40 \hspace{0.1em} -- \hspace{0.1em} 160 & 185 & 8 \\ 
    \hline\hline
    \end{tabular}

    \caption{Bin sizes used for the PCA analysis using phases from peak, the number of spectra in each bin after removing noisy spectra, and the number of components needed to explain 85\% of the variation in the sample.}
    \label{tab:pca_peak}
\end{table}
\end{center}

\begin{center}
\begin{table}
    \begin{tabular}{|P{4.5cm}|P{1.2cm}|P{1.5cm}|} 
    \hline\hline
    Scaled Phases from Explosion (days) & Spectra & Components \\ 
    \hline
    0 \hspace{0.1em} -- \hspace{0.1em} 35 & 160 & 5 \\ 
    35 \hspace{0.1em} -- \hspace{0.1em} 55 & 154 & 4 \\ 
    55 \hspace{0.1em} -- \hspace{0.1em} 80 & 162 & 5 \\ 
    80 \hspace{0.1em} -- \hspace{0.1em} 500 & 186 & 8 \\
    \hline\hline
    \end{tabular}

    \caption{Bin sizes used for the PCA analysis using scaled phases from explosion, the number of spectra in each bin after removing noisy spectra, and the number of components needed to explain 85\% of the variation in the sample.}
    \label{tab:pca_exp}
\end{table}
\end{center}

\subsection{Component Spectra}
\label{sec:component_specs}

Figure \ref{fig:PCA_peak}, and Figure \ref{fig:PCA_exp} show the component spectra from the PCA decomposition for phases from peak, and scaled phases from explosion respectively. The columns on the right indicate the percentage that each component contributes to the total variation, and the cumulative total when adding these components sequentially. These components form the basis of the set from which all spectra in the sample can be created. A threshold of 85\% was used, meaning only components that can explain up to 85\% of the total variation are included. We can see that as the phase increases, the number of components needed to explain the same amount of variation in the data increases. This means the spectra at early times tend to be more similar to one another than at late times. This can be seen in both Figure \ref{fig:PCA_peak}, and Figure \ref{fig:PCA_exp} and summarised in Table \ref{tab:pca_peak} and Table \ref{tab:pca_exp}. 

In Figure \ref{fig:-80--5_peak} and Figure \ref{fig:-5-35_exp}, we see that only five components are needed to explain 85\% of the variation within the sample. The first couple of components in both Figure \ref{fig:-80--5_peak}  and Figure \ref{fig:-5-35_exp} show different continuum shapes implying the majority of the variation at early times is due to the underlying blackbody temperature. This implies at early times the spectra are relatively simple, likely due to the flux being dominated by the blackbody continuum and the lines being comparatively weak. 


The final component in Figure \ref{fig:-80--5_peak} and Figure \ref{fig:-5-35_exp} show the characteristic O II "W"-shape just below 4500\,\AA\ meaning there is significant variation in these lines at these early phases. This feature is also clearly visible in Figure \ref{fig:-5-15_peak}, versus Figure \ref{fig:35-55_exp} which only marginally shows variation at the location of this feature. Sample studies of SLSNe by \citet{Konyves-Toth2022} have suggested separate sub-classes of SLSNe where the O II absorption lines at peak either display a strong "W"-shape, or a "15bn"-like shape with no discernable double trough. Here we see that if we remove the dependence on observing relative to peak, the diversity in O II lines decreases, indicating that the separate sub-classes may just be a reflection of the different times the objects are observed relative to explosion. \citet{Quimby2018} suggested a similar spectroscopic division into "2011ke"-like or "PTF12dam"-like events. Spectra were assigned a spectral phase $\phi$ which ranged from $\phi = -1$ for early-phase spectra with strong O II lines, $\phi = 0$ when the O II lines disappeared, and $\phi = 1$ when the spectra resembled SNe Ic near maximum light. Spectra grouped into the "2011ke"-like category exhibited broader absorption features between $-0.3 < \phi < 0.3$, suggesting high ejecta velocities.

The middle panels of Figure \ref{fig:PCA_peak} and Figure \ref{fig:PCA_exp} correspond to the rapid cooling stages seen in Section \ref{sec:average_spectra}. Here many more components are required to explain the same level of variation, especially when grouping the spectra by phases from peak. This is especially apparent in Figure \ref{fig:15-35_peak} which corresponds to the latter stages of this cooling period and requires eight components to reach the threshold. This suggests that at least some of the diversity in SLSN spectra results from comparing objects evolving with different timescales. By contrast, Figure \ref{fig:55-85_exp} only requires five components, which is the largest discrepancy we find in terms of numbers of components needed for corresponding bins. This may indicate that grouping spectra by scaled phases from explosion could group together more similar spectra, offering a more effective way of describing the spectral evolution of SLSNe.

The middle panels of Figure \ref{fig:PCA_peak} and Figure \ref{fig:PCA_exp} begin to show more individual lines and the differences in these line strengths. There appears to be a line at 7300\,\AA\ in the final components of Figure \ref{fig:-80--5_peak} which persists at later epochs. This is coincident with a forbidden emission line either from [Ca II] or [O II]. This line also shows up in the average spectra created in Section \ref{sec:average_spectra} as early as $-80$ days before peak, and can be seen in the 1$\sigma$ spread. Early appearance of this line (such as at the phases in Figure \ref{fig:-5-15_peak}) has been linked to interaction with low-density circumstellar material which, as discussed, may be a sign of PPI in SLSNe. By contrast, the panels in Figure \ref{fig:PCA_exp} do not show as clear variation at the location of the feature until 55 days post explosion (Figure \ref{fig:55-85_exp}). This reinforces the idea that these lines are formed outside the SN ejecta, as it takes time for the ejecta to reach these line forming regions and interact with them.


The last set of panels show the spectra once they have started transitioning to the pseudo-nebular phase, but still with an underlying blackbody component. We once again see the most variation explained by the underlying continuum. These panels correspond to the plateau in temperature seen in Section \ref{sec:average_spectra}. We can see that at these later times as the SLSNe are evolving more slowly, the diversity is less dependent on binning scheme and therefore the number of components is the same regardless of whether we normalise the timescales. This is because the evolution timescale at this phase is much longer and is therefore less sensitive to the bin edges. We also begin to see the emergence of different lines to explain the variation. At these stages, more of the diversity is attributable to the strengths of identifiable emission lines. These include Fe II at $\sim$3500\,\AA\ , Ca II $\lambda\lambda$3934, 3968, Mg I] 4571, Fe II at around 5100\,\AA\, [O I] $\lambda\lambda$6300, 6364, [Ca II] $\lambda\lambda$7291, 7323, [O II] $\lambda\lambda$7320, 7330, and Ca II $\lambda\lambda\lambda$8498, 8542, 8662. Once the spectra fully evolve into the nebular phase, the diversity can largely be explained by a few strong emission lines \citep{Nicholl2019c}. However, we caveat this with the fact that at these later phases, the SLSNe are much fainter and noise plays a significant role in the variation seen.

\begin{figure*}
	\begin{center}
		\includegraphics[width=2\columnwidth]{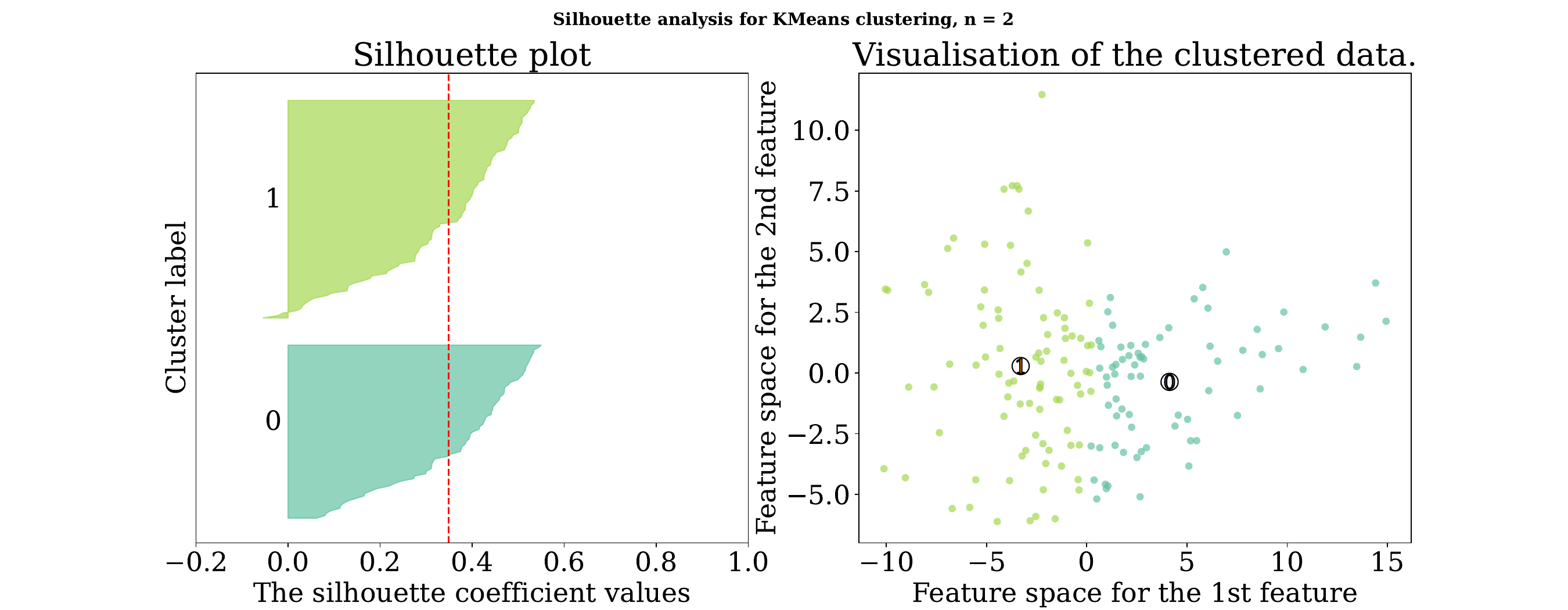}
		\caption{Silhouette plot using a cluster size of two for the first two components from the PCA decomposition in the time bin $0 - 35$ scaled days post explosion. These plots show the quality of clustering by displaying the silhouette coefficient for each point, which measures how well the point fits within its assigned cluster compared to the nearest neighbouring cluster. Values range from $-1$ to 1, with higher values indicating well-separated clusters, while negative values suggest possible misclassification. The mean silhouette score, shown as a red vertical line, summarises clustering performance. With a score below 0.5, this suggests no evidence of clustering. The right panel displays these clusters in the PCA parameter space.}
        \label{fig:sihouette_2}
	\end{center}
\end{figure*}

\begin{figure*}
	\begin{center}
		\includegraphics[width=2\columnwidth]{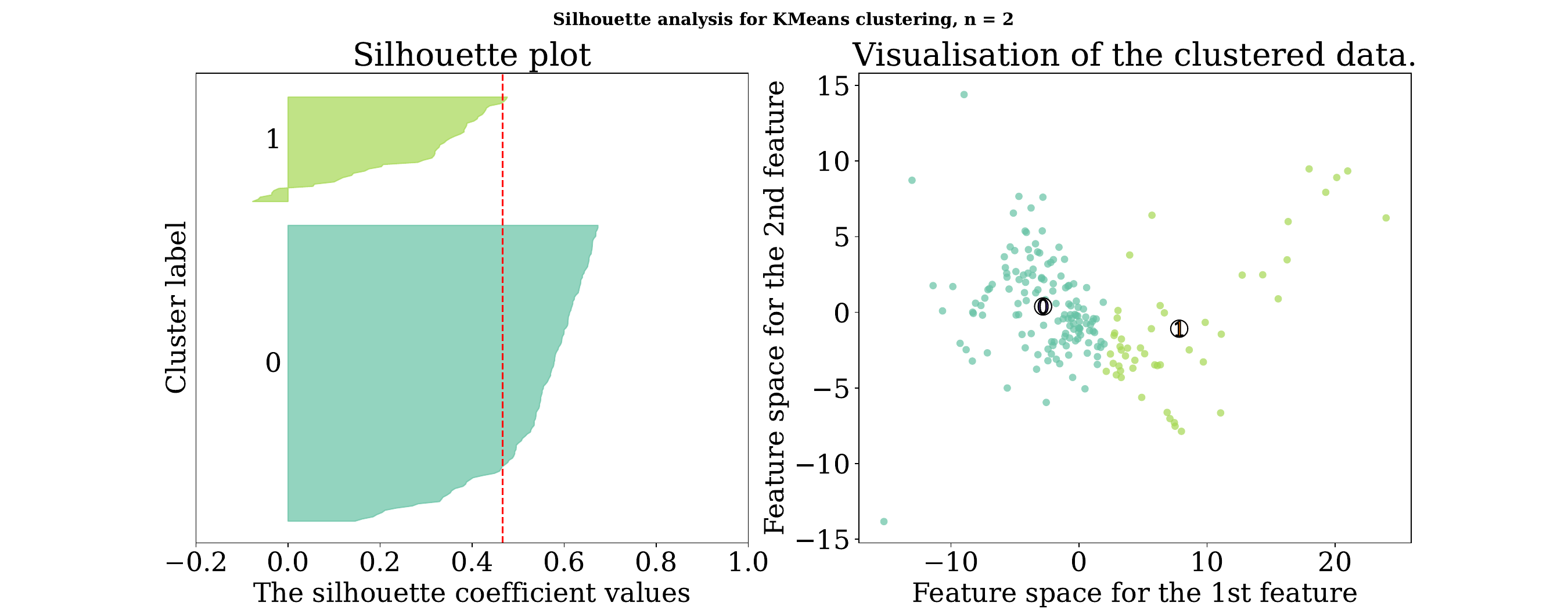}
		\caption{Silhouette plot using a cluster size of two for the first two components from the PCA decomposition in the time bin 80 - 500 scaled days post explosion. These plots show the quality of clustering by displaying the silhouette coefficient for each point, which measures how well the point fits within its assigned cluster compared to the nearest neighbouring cluster. Values range from $-1$ to 1, with higher values indicating well-separated clusters, while negative values suggest possible misclassification. The mean silhouette score, shown as a red vertical line, summarises clustering performance. With a score just below 0.5, this suggests a slight preference for two distinct clusters. However, the second cluster is primarily composed of spectra from SN\,2017egm which we show to be spectrally unique in Section \ref{sec:clustering}. The right panel displays these clusters in the PCA parameter space.}
            \label{fig:sihouette_3}
	\end{center}
\end{figure*}

\begin{figure*}
\centering
\begin{minipage}{0.48\textwidth}
\begin{center}
        \begin{subfigure}{1\columnwidth}
            \centering
            \includegraphics[width=1\columnwidth]{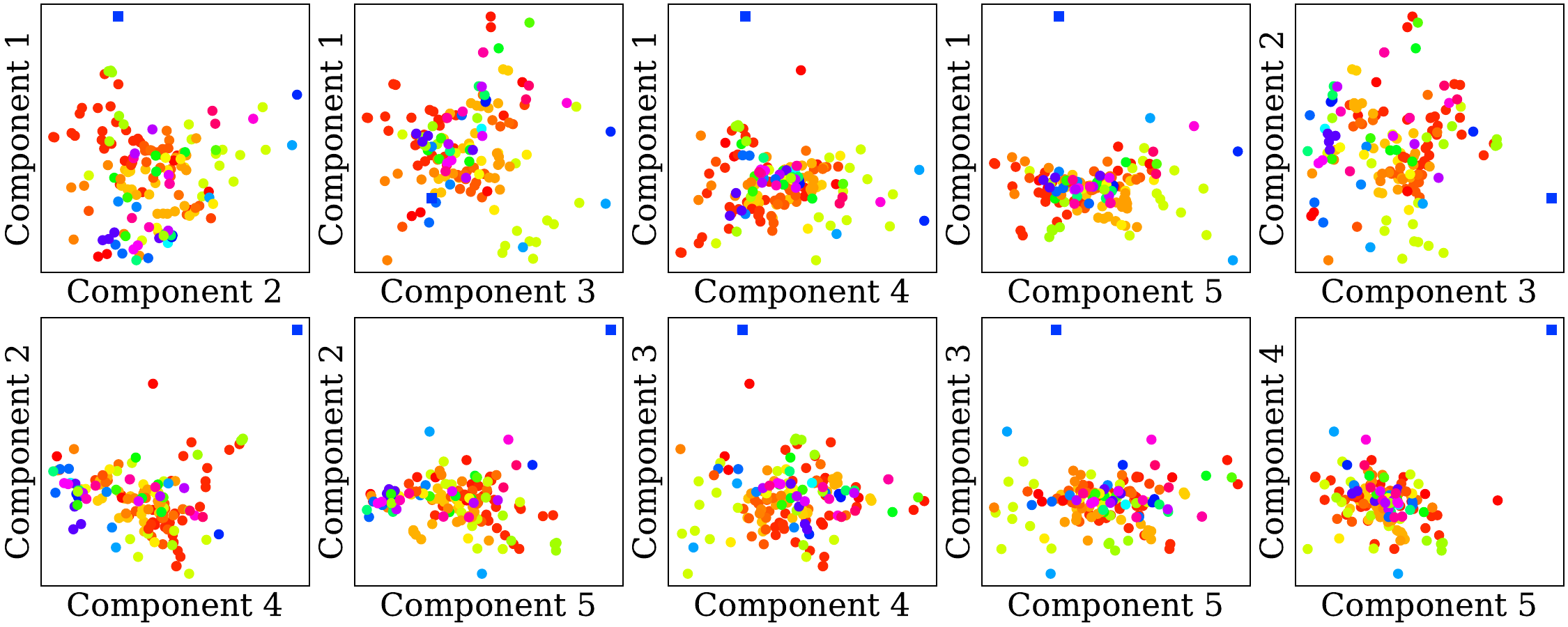}
            \caption{Days -80 - -5}
            \label{fig:-80--5_comp_peak}
        \end{subfigure}
    \par\bigskip
    \hfill
        \begin{subfigure}{1\columnwidth}
            \centering
            \includegraphics[width=1\columnwidth]{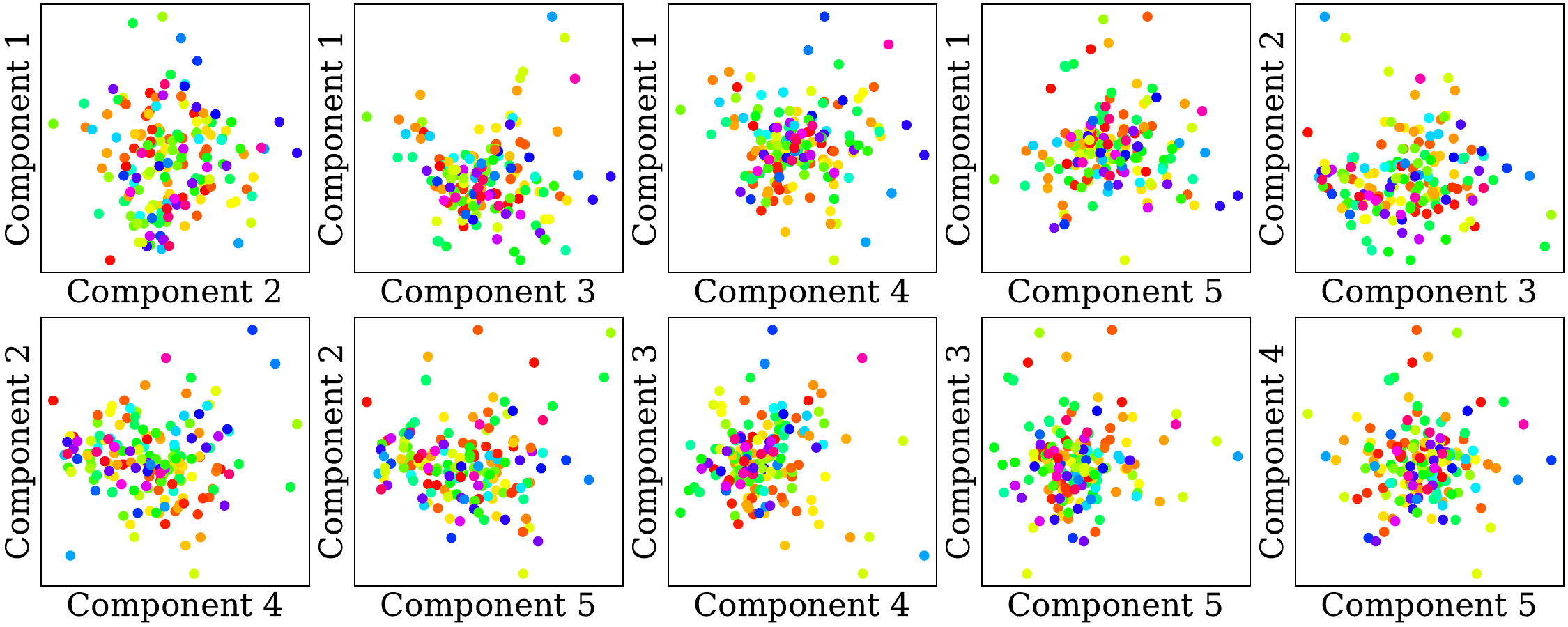}
            \caption{Days -5 - 15}
            \label{fig:-5-15_comp_peak}
        \end{subfigure}
    \par\bigskip
    \hfill
        \begin{subfigure}{1\columnwidth}
            \centering
            \includegraphics[width=1\columnwidth]{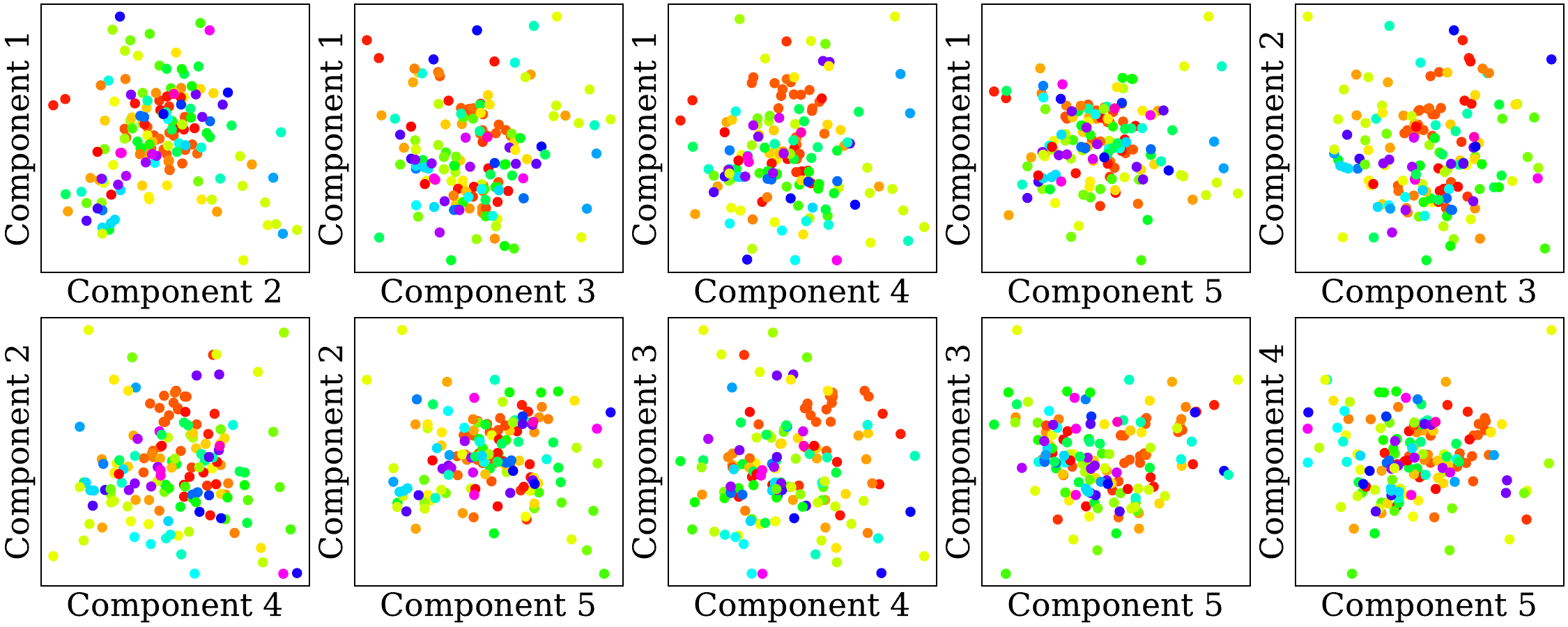}
            \caption{Days 15 - 40}
            \label{fig:15-35_comp_peak}
        \end{subfigure}
    \par\bigskip
    \hfill
        \begin{subfigure}{1\columnwidth}
            \centering
            \includegraphics[width=1\columnwidth]{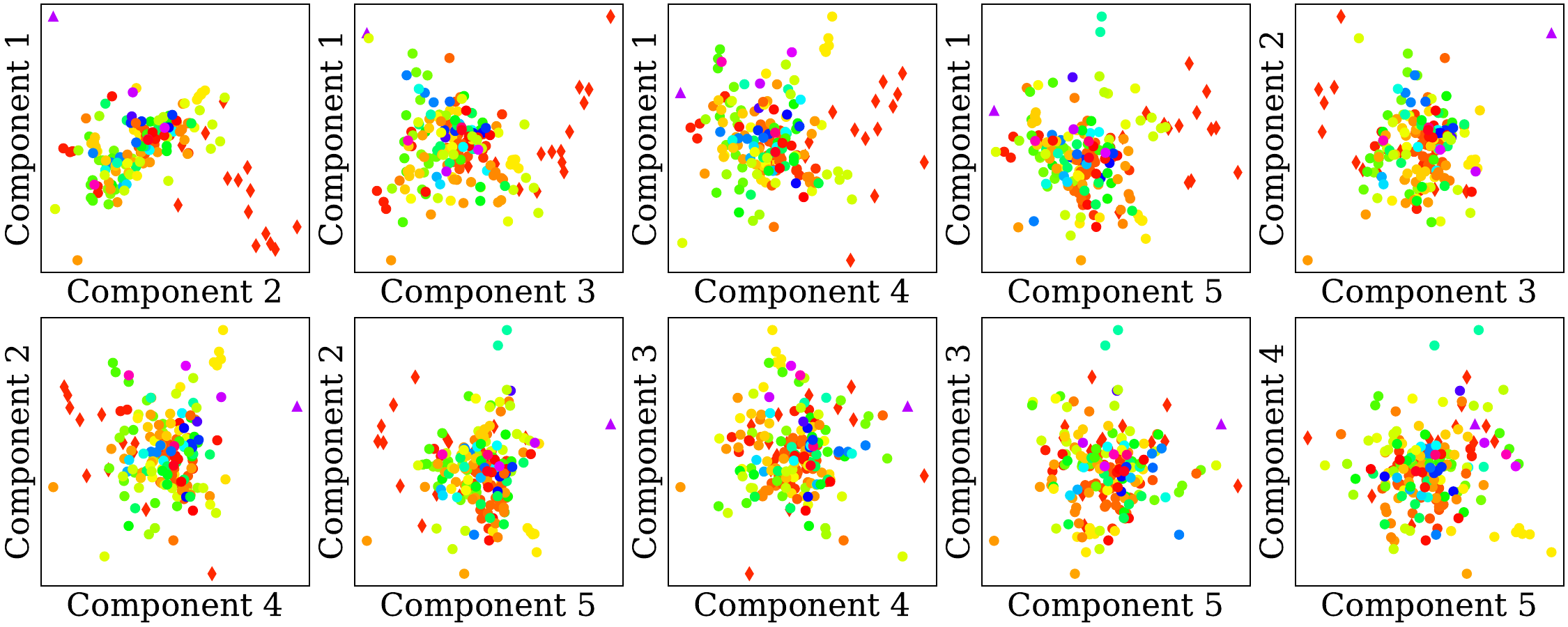}
            \caption{Days 40 - 160}
            \label{fig:35-160_comp_peak}
        \end{subfigure}
    \hfill
        \caption{Representation of the SLSN photospheric spectra in the PCA parameter space in time bins relative to peak. Only the first five components for each bin are plotted for visualisation. Each SLSN event is marked by a unique colour, consistent across all panels. Interesting outliers are marked by different shapes and discussed in Section \ref{sec:clustering}.}
        \label{fig:PCA_components_peak}
    \end{center}
\end{minipage}%
\hfill
\begin{minipage}{.48\textwidth}
\vspace{4mm}
\begin{center}
        \begin{subfigure}{1\columnwidth}
            \centering
            \includegraphics[width=1\columnwidth]{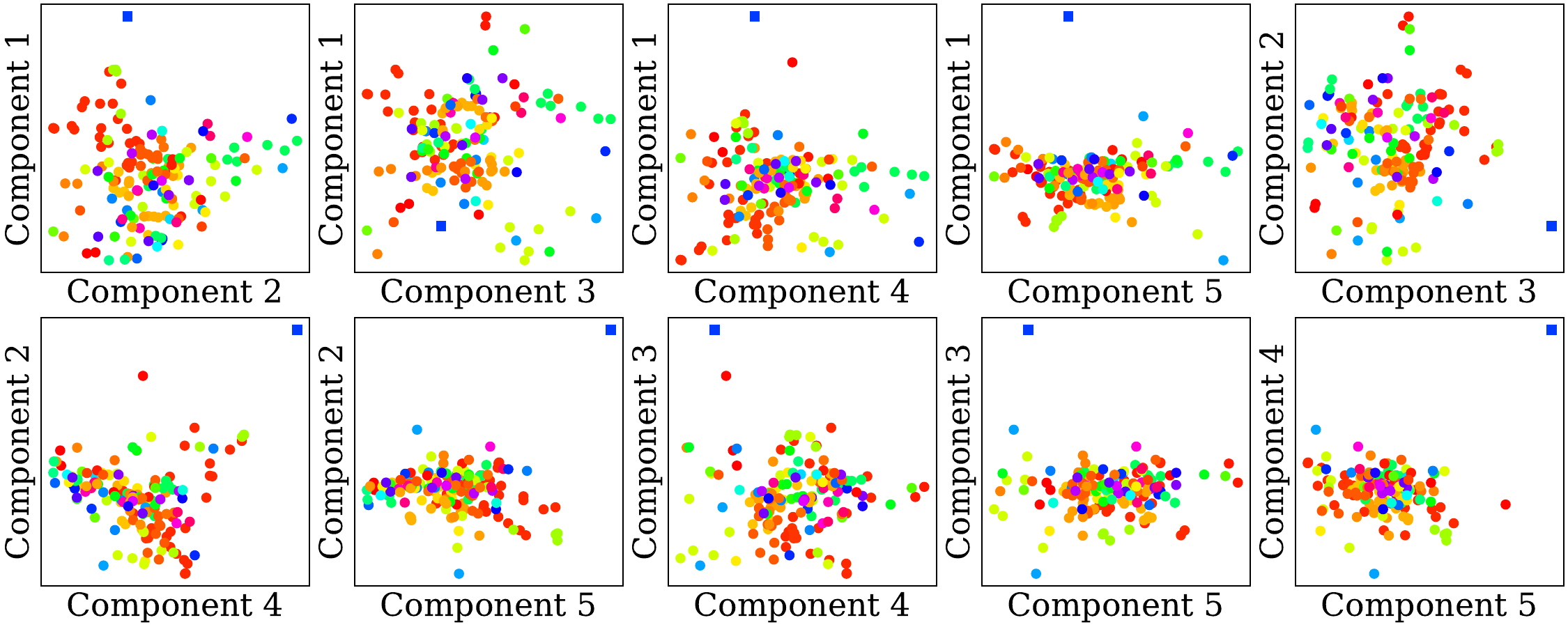}
            \caption{Days 0 - 35}
            \label{fig:-5-35_comp_exp}
        \end{subfigure}
    \par\bigskip
    \hfill
        \begin{subfigure}{1\columnwidth}
            \centering
            \includegraphics[width=1\columnwidth]{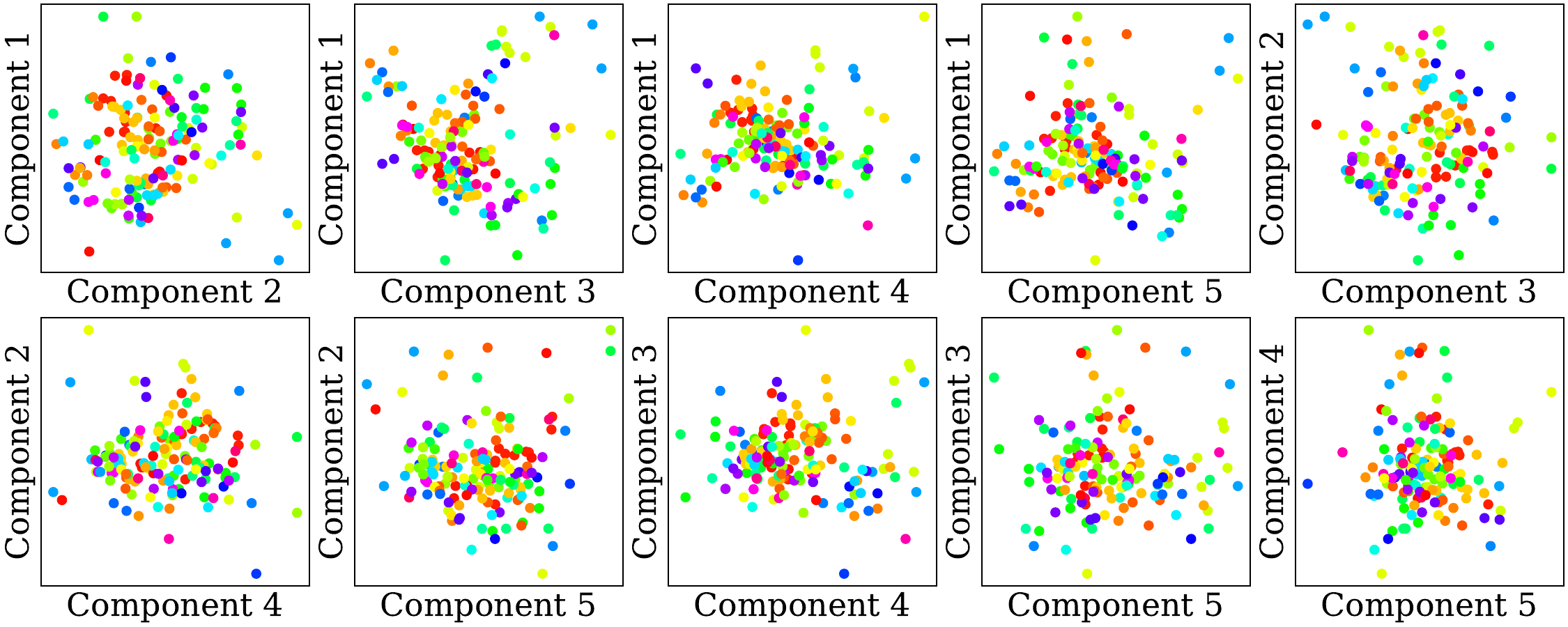}
            \caption{Days 35 - 55}
            \label{fig:35-55_comp_exp}
        \end{subfigure}
    \par\bigskip
    \hfill
        \begin{subfigure}{1\columnwidth}
            \centering
            \includegraphics[width=1\columnwidth]{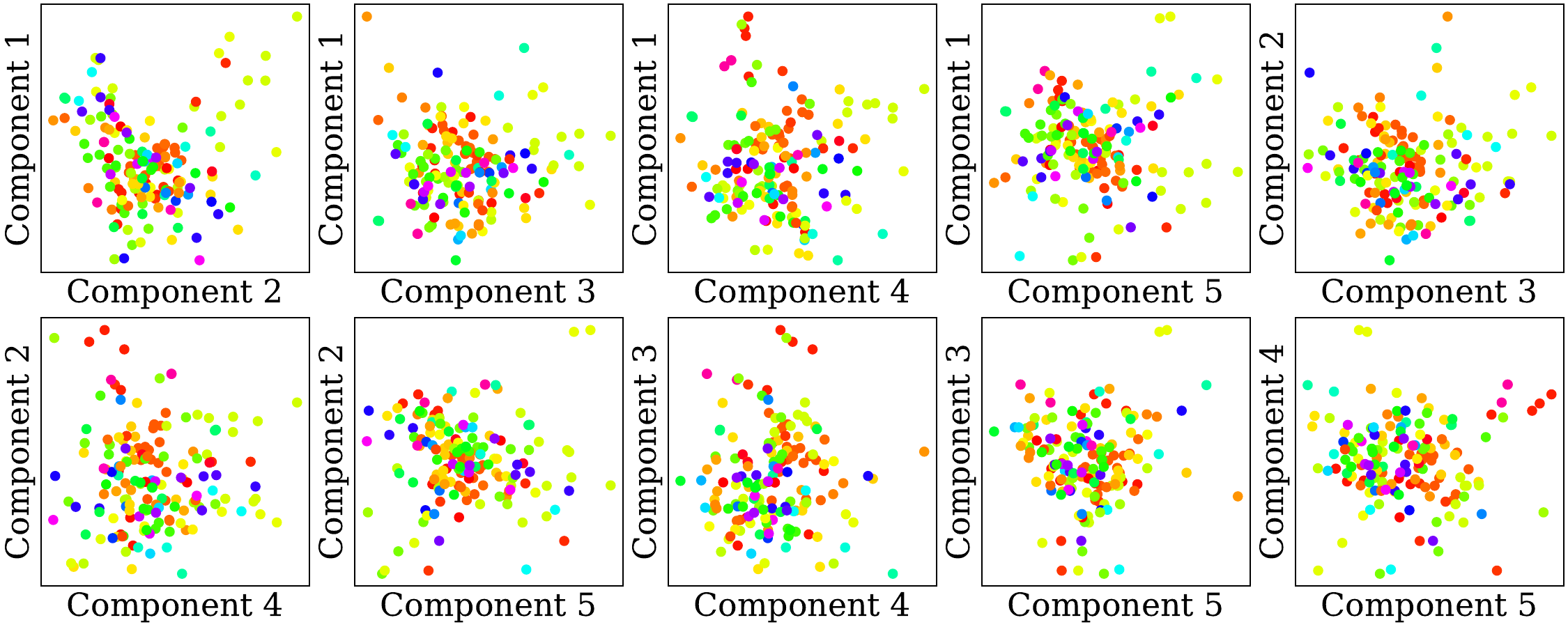}
            \caption{Days 55 - 80}
            \label{fig:55-85_comp_exp}
        \end{subfigure}
    \par\bigskip
    \hfill
        \begin{subfigure}{1\columnwidth}
            \centering
            \includegraphics[width=1\columnwidth]{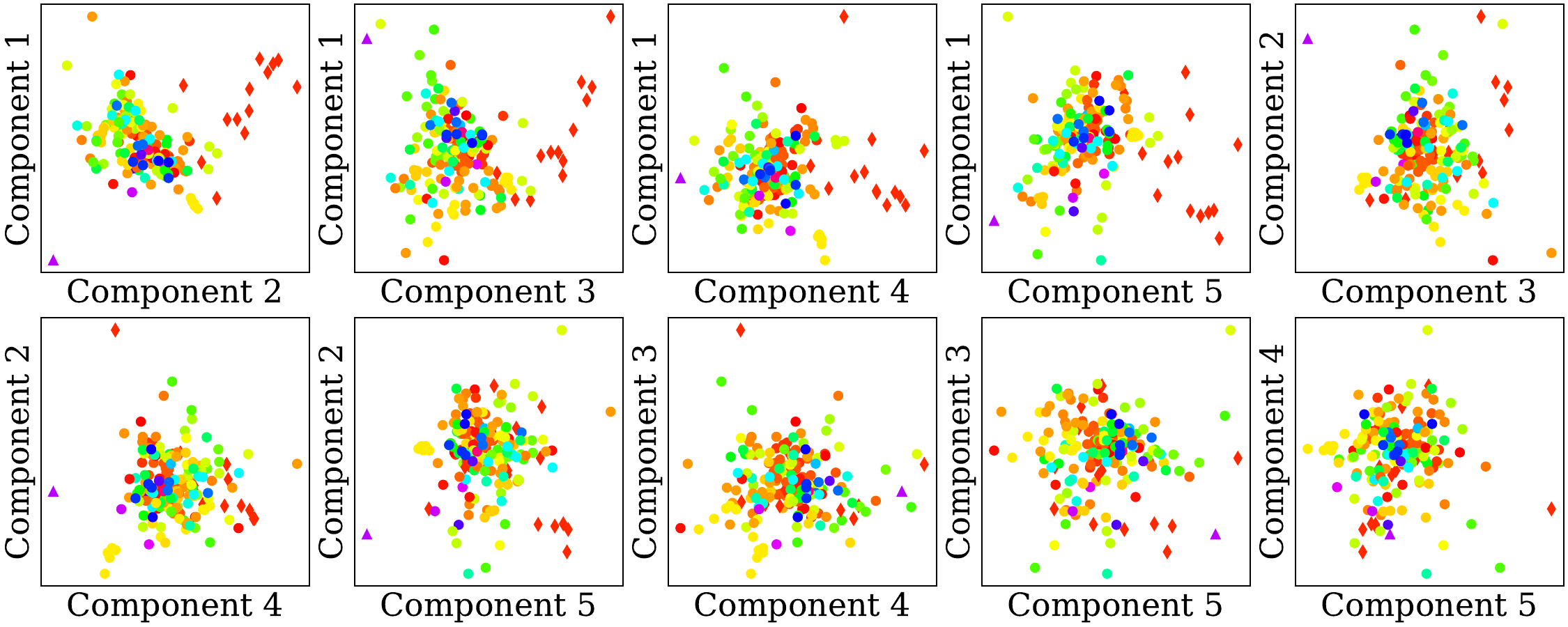}
            \caption{Days 80 - 500}
            \label{fig:80-500_comp_exp}
        \end{subfigure}
     \hfill
        \caption{Representation of the SLSN photospheric spectra in the PCA parameter space in time bins relative to explosions with the phase scaled to match the median e-folding decline time of SLSNe (44 days). Only the first five components for each bin are plotted for visualisation. Each SLSN event is marked by a unique colour, consistent across all panels. Interesting outliers are marked by different shapes and discussed in Section \ref{sec:clustering}.}
        \label{fig:PCA_components_exp}
    \end{center}
\end{minipage}
\end{figure*}

\subsection{Clustering}
\label{sec:clustering}

In order to search for subpopulations, K-means clustering \citep{LLoyd1982} was used on the first two PCA components of each phase bin. This is an unsupervised learning algorithm that separates the data based on the distance of each point from its centroid. In K-means clustering, the number of clusters must be specified in advance. The algorithm then iteratively assigns each data point to the nearest centroid and updates the centroids until the assignments stabilise, resulting in clusters that minimise within-cluster distances. To visualise this, "silhouette plots" were created, which evaluate the quality of clustering results by assessing how well the data points fit within their assigned cluster. The silhouette coefficient, a key metric displayed in these plots, measures how close a point is to the centroid of its own cluster compared to the centroid of the nearest neighbouring cluster. This coefficient ranges from $-1$ to 1, where values close to 1 indicate that the point is well separated from other clusters and strongly belongs to its assigned cluster, while values near 0 suggest the point lies near a cluster boundary. Negative values indicate potential misclassification, where the point is closer to a different cluster than its assigned one. In silhouette plots, each cluster is represented as a bar showing the distribution of silhouette coefficients for all points within the cluster, sorted in descending order. The overall clustering quality is summarised by the mean silhouette coefficient, shown as a vertical line. A mean value above 0.5 typically indicates that the clustering structure is reasonable, while values exceeding 0.7 are considered strong evidence of well-separated and cohesive clusters. Clusters with significant internal variability or many negative silhouette values may indicate overlapping cluster boundaries or misclassified points.

Silhouette plots were created for each phase bin using numbers of clusters between two to five. A representative version of these plots is shown in Figure \ref{fig:sihouette_2} which shows a grouping of the data into two clusters for the spectra from 0 to 35 scaled days post explosion. We can see that although the groups are all of roughly similar size based on the thickness of each plot, and exceed the average silhouette scores, they do not exceed $\sim$0.5 indicating they are not very defined clusters. This is visualised in the right-hand side plots where we can see which points belong to which groups, and see the lack of distinct groups visually. The average cluster score is also $\ll$0.5 indicating no evidence for clusters. For most cluster sizes and time bins, there was no evidence for clusters of any size between two and five. The only exception to this are the silhouette plots created with a cluster size of two for the 80-500 scaled days post explosion data (Figure \ref{fig:sihouette_3}), specifically when comparing component one against components two, three, four, and five. This will be discussed later in this Section. The process was performed for all possible component combinations for the first five PCA components. Overall, this indicates there is no evidence for separate subpopulations within the spectral data when considering the five most important PCA components needed to describe the variance in each time bin.

We also searched for clusters using Bayesian information criterion \cite[BIC;][]{Schwarz1978} shown in Appendix \ref{sec:BIC}. This method assesses the fit of using different numbers of clusters, whilst penalising models with more parameters (i.e. larger numbers of clusters). This is important as using more clusters could result in over-fitting the data. These were then fit using K-means clustering for clusters of sizes one to five, shown in Appendix \ref{sec:BIC}. The plots show there is no evidence of separate clusters as shown by the consistent decrease in BIC values as the number of clusters increases. We would expect to observe a sharp drop at the point corresponding to the optimal number of clusters if this truly represented the number needed to describe the data. This again suggests there is no statistical evidence for multiple populations within the data when accounting for the majority of the variation in the spectra.

Figure \ref{fig:PCA_components_peak} and Figure \ref{fig:PCA_components_exp} show the component spectra from Section \ref{sec:component_specs} represented in PCA parameter space. This plots the PCA coefficients of some of the components against each other to see if there are any correlations between them. Each object is represented by a unique colour consistent across both figures and all panels. This analysis is also explored in Appendix \ref{sec:fast_slow} in the context of ``fast'' and ``slow'' evolving SLSNe. As expected from the cluster search, there do not seem to be any significant trends or groups within the data. However, there do seem to be several individual events that are outliers from the central clumps. Visual assessment of the spectra of all outliers showed that they fall into one of two categories: the spectrum is noisy but still with discernable SN flux, or the spectrum truly deviates from the average spectra calculated in Section \ref{sec:average_spectra}. The spectra of interest are the ones that fall into this latter category and include spectra from SN\,2019pud, PTF12hni, and SN\,2017egm. 

\begin{figure}
	\begin{center}
		\includegraphics[width=1\columnwidth]{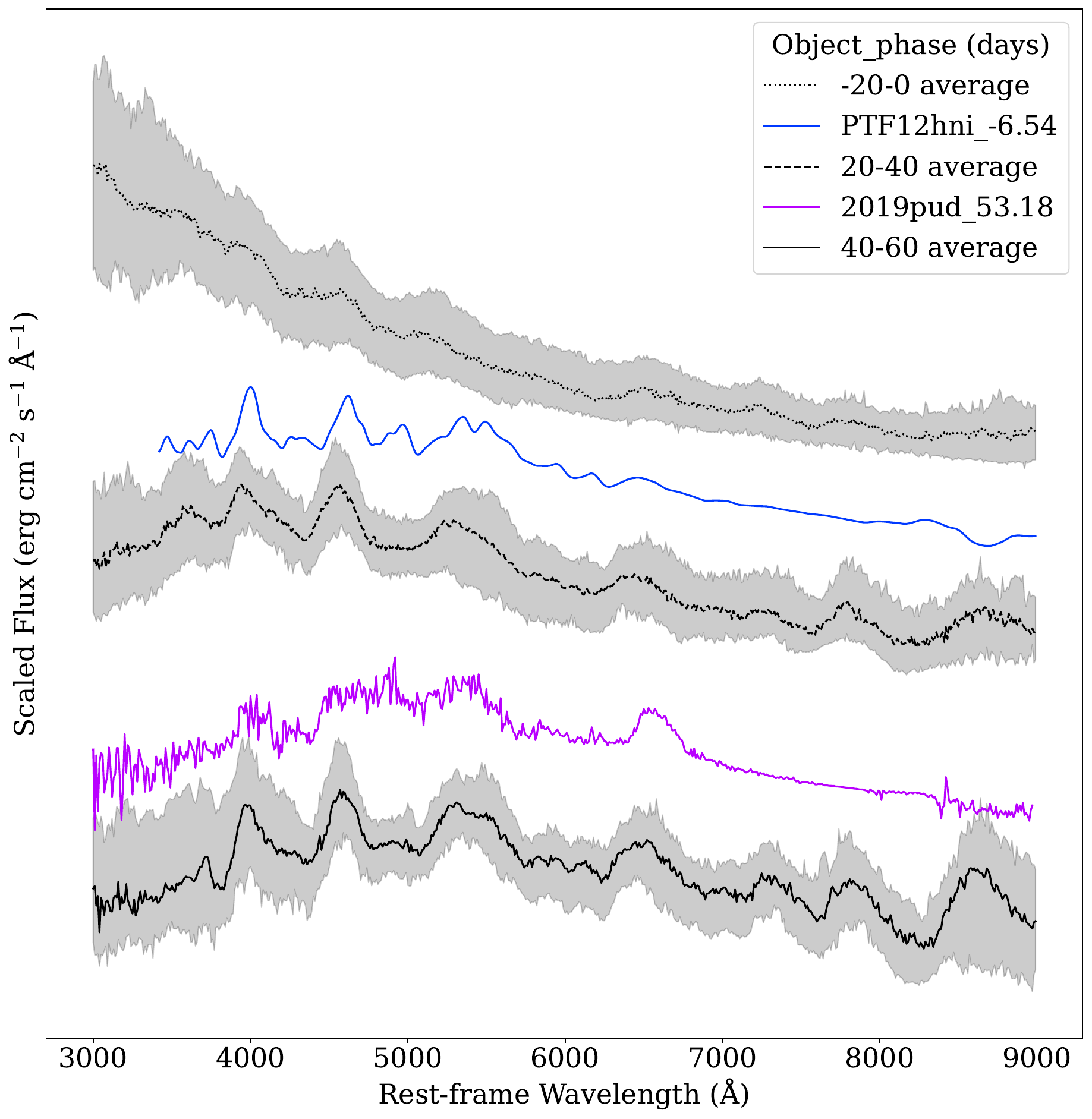}
		\caption{Spectra of PTF12hni and SN\,2019pud that fall outside the central grouping in the PCA parameter space. The colours of the spectra are consistent with the colour of the outlier points in Figure \ref{fig:PCA_components_peak} and Figure \ref{fig:PCA_components_exp}. In black are the relevant average spectra from Figure \ref{fig:av_specs} for comparison. We can see how the spectrum of PTF12hni looks similar to spectra at much later phases. In the case of SN\,2019pud, the spectrum looks different from its respective average with the presence of a strong H$\alpha$ emission line, and lack of features in the red end of the spectrum. The spectrum of PTF12hni is from \citet{Quimby2018}, and the spectrum of SN\,2019pud is from \citet{Kangas2022}.}
            \label{fig:19pud_18bym}
	\end{center}
\end{figure}

In the early phases of Figure \ref{fig:PCA_components_peak} and Figure \ref{fig:PCA_components_exp}, an outlier spectrum marked as a blue square is apparent. This corresponds to a spectrum of PTF12hni, shown in Figure \ref{fig:19pud_18bym} at $-$6.54 days relative to peak, alongside the average spectrum from Section \ref{sec:average_spectra} in the corresponding time bin. In the bluer wavelengths, this spectrum looks more similar to the averages from later time bins such as the average from 20-40 days post peak. However, the weak lines compared to the continuum emission in the redder wavelengths beyond 7000\,\AA\ looks more similar to the average from $-$20-0 days before peak.  The unusual nature of this object is complicated by its discovery after its peak, meaning the time of peak is estimated either from modelling the light curve as is the case for the analysis in \citet{Gomez2024}, or from template matching of the spectra as in \citet{Quimby2018}. Although these methods estimate a similar time of peak, \citet{Quimby2018} also acknowledge plausible matches to SLSN templates up to 100 days post peak. For this reason, PT12hni as an outlier may simply be due to an uncertain peak of the light curve.

An outlier spectrum of SN\,2019pud days is shown in Figure \ref{fig:35-160_comp_peak} and Figure \ref{fig:80-500_comp_exp} as a purple triangle. This spectrum is obtained at 53.18 days post peak is shown in Figure \ref{fig:19pud_18bym}, and compared to the average spectrum from Section \ref{sec:average_spectra} in the corresponding time bins. We can see that SN\,2019pud looks similar to the average at bluer wavelengths except with a slightly broad emission line around 5000\,\AA, that is not visible in the average. However, it also does not display any emission or P-Cygni features redwards of 7000\,\AA. This object appears in a sample paper of H-rich SLSNe \citep{Kangas2022}, where it is also noted for its unusual features even within the class of H-rich SLSNe. The strong emission feature at $\sim$6500\,\AA\ peaks further to the red than the [O I] line in the average spectrum, and was attributed to H$\alpha$. However, this object did not show H emission until $\sim$50 days after peak and was therefore classed as a "SLSNe I.5" by \citet{Kangas2022}. The event also has a steep rise time, and although the colour initially reddens, it becomes bluer about 30 post peak \citep{Kangas2022}. The deviation of this object from the general population at late times demonstrates the effectiveness of PCA in picking out events with unique observables. 

In the final panels from both Figures (\ref{fig:35-160_comp_peak} and \ref{fig:80-500_comp_exp}), we can see a cluster of red diamonds quite distinct from the central grouping. These red points correspond to the spectra of SN\,2017egm. The separation of this object from the rest is particularly pronounced when comparing the first component to the rest. This separation is supported by Figure \ref{fig:sihouette_3} which shows a silhouette plot for the parameter space corresponding to the first two components in the bin at 80-500 scaled days from explosion. This panel had one of the highest silhouette score at $0.47$, close to the $0.5$ threshold generally used to distinguish reasonable clusters. In fact, comparing component one against components three, four, and five resulted in silhouette scores of $0.56$, $0.59$, and $0.60$ respectively. This may initially suggest there are two distinct groups here, however, when we consider that the majority of these points are from a single object it supports the idea that there is a unique object rather than distinct group. We can also see in Figure \ref{fig:sihouette_3} that a few points in the second group, from objects other than SN\,2017egm, have negative silhouette scores.

\begin{figure}
	\begin{center}
		\includegraphics[width=1\columnwidth]{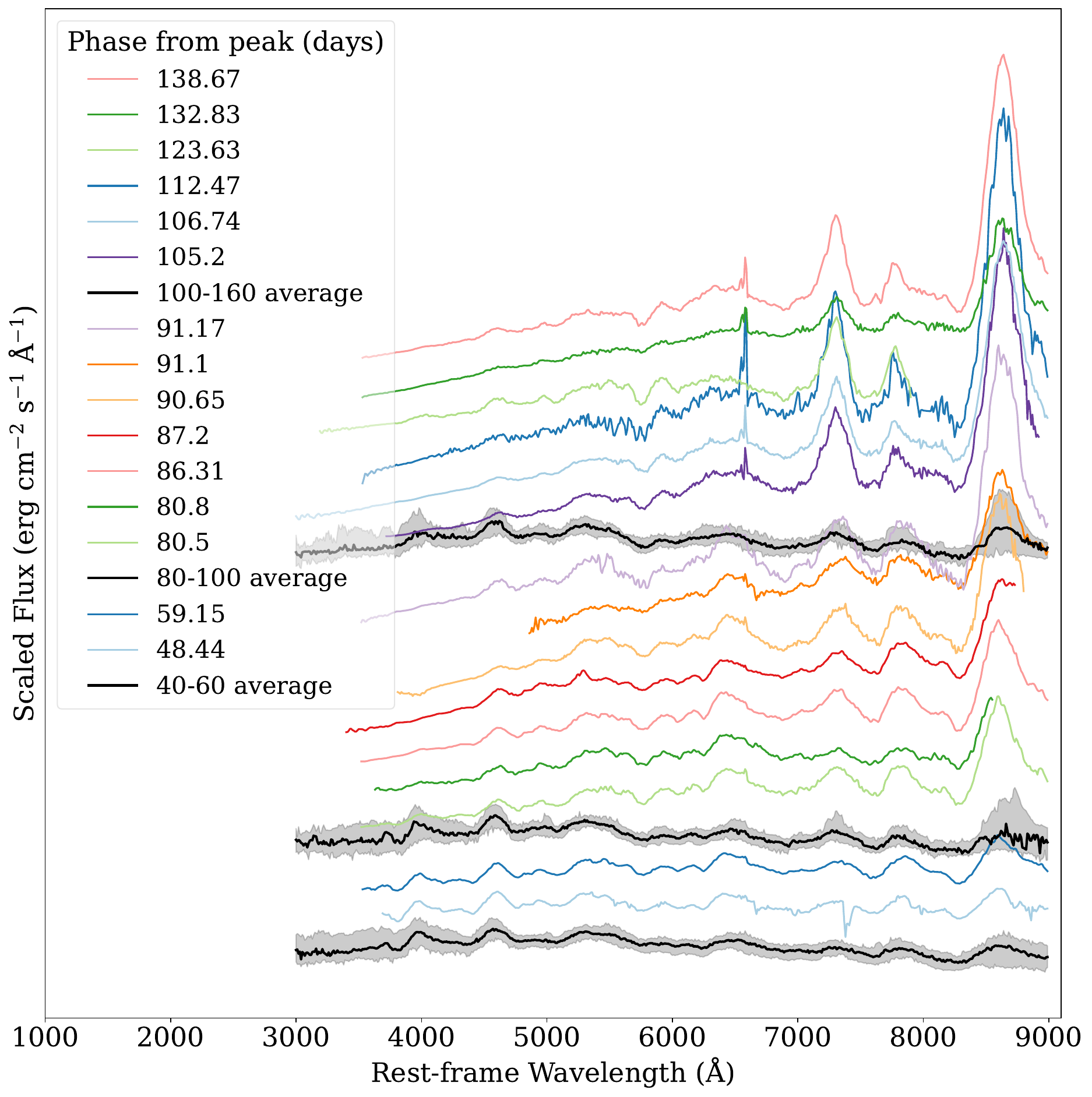}
		\caption{Spectra of SN\,2017egm with phases from peak above 35 days. Inter-spaced are the average spectra from Figure \ref{fig:av_specs} for comparison. All spectra including the averages are scaled to the median flux between 5000$-$6000\,\AA. We can see how the spectra at 40$-$60 days look very similar to the average spectra. The object quickly deviates from the average after this point particularly around the Ca NIR triplet and the Mg I] $\lambda$4571 emission line. Spectra are from \citet{Zhu2023, Lin2023, Nicholl2019c}.}
            \label{fig:2017egm+avs}
	\end{center}
\end{figure}

The spectra of SN\,2017egm in the final time bins are shown in Figure \ref{fig:2017egm+avs}. In this Figure, all spectra including the averages are normalised to the median flux between 5000-6000\,\AA. From this, we can see how the spectra of SN\,2017egm look very similar to the average spectra between 40-60 days, however they deviate increasingly from the averages calculated in Section \ref{sec:average_spectra} at later times. We can see that at later times SN\,2017egm has a cooler underlying blackbody continuum compared to the averages, but the most striking feature is the Ca II NIR triplet, which is significantly stronger than the other emission lines. This is not the case in the average spectra, where this line is much more comparable in strength to other nearby lines. 
The inclusion of SN\,2017egm in the PCA decomposition could be driving the multiple components with strong Ca II NIR features as this object is likely responsible for a lot of the variance in the data set.  This is explored in Appendix \ref{sec:2017egm} where this analysis is rerun with the spectra of SN\,2017egm removed. This event is also unusual in terms of its light curve, displaying undulating features and multiple bumps \citep{Zhu2023, Lin2023}. The time period of the spectra displayed in Figure \ref{fig:2017egm+avs} correspond to a bump in the light curve, followed by a steep decline. This has been attributed to interaction with a shell of CSM, followed by a sharp decline as the shockwave reaches the edge of the material \citep{Zhu2023}. If the late-time line excitation in SN\,2017egm is dominated by interaction, the fact that it is a significant outlier from the average spectrum could imply that a different mechanism dominates the line formation in more typical SLSNe.

\section{Line Velocities}
\label{sec:fe_line}

The dynamics of SN explosions can be investigated through the study of absorption velocities, which probe the structure of the expanding ejecta. Light curve measurements only constrain the ratio of ejecta mass and velocity \citep{Arnett1982}. Instead, an independent velocity measurement is essential to break this degeneracy and estimate the mass. 

Different theoretical models predict different time evolution for line velocities in SLSNe, depending on the source of energy injection. Early 1D magnetar models predicted that observed velocities would initially decline with time as the photosphere receded, followed by a plateau in the velocity evolution \citep[e.g.][]{Kasen2010, Dessart2012}. This behaviour was caused by the internal magnetar wind sweeping up the inner ejecta into a dense shell with a constant velocity. However, multi-dimensional simulations have provided a different perspective. 2D models show that the magnetar energy can induce turbulence in the ejecta, disrupting the formation of a narrow, dense shell \citep{Chen2016, Suzuki2021}. This instead spreads the inner ejecta down to lower velocities. In 3D models \citep{Suzuki2019, Chen2020}, we again see a density profile that declines smoothly as a function of velocity coordinate and therefore would expect to see a smooth decline in the observed line velocity too. 

Within the CSM interaction model, we might expect the velocity of the lines to decrease quickly as the rapid deceleration of ejecta produces the observed luminosity. \citet{Sorokina2016} suggest that the spectra of SLSNe could result from interactions with a fast-moving shell, where the outer edge reaches the observed absorption velocities, while the inner edge remains nearly stationary. However, there are issues with this model such as the broad spectral features of SLSNe pointing towards a significant amount of high-velocity material, which conflicts with the idea of a thin shell. Other theories propose that the interaction layer may not produce the absorption lines, but instead dilute them by enhancing the blue continuum \citep{Branch2000, Sollerman2020}. However, the broad absorption lines in SLSNe seem to show similar equivalent widths even when the continuum luminosity is different, suggesting no such dilution effect in observed events \citep{Nicholl2017d, Yan2018}. Even if broad lines are observed from the SN ejecta, we might also expect to see some lines with lower velocities, from any absorption or emission that takes place in the dense CSM.

Notably, the Fe II $\lambda$5169 absorption line has been identified by \citet{Branch2002} as a valuable tracer of photospheric velocity. Iron is a useful tracer in SN ejecta because its distribution peaks towards the centre, with the most iron-rich region of the line forming region coinciding with the photosphere. From a previous study of 21 SLSNe by \citet{Liu2017}, we know that the velocity of this line decreases over time as the ejecta expands, and inner, slower moving layers of material are revealed. In their analysis, the velocity of the line at 10 days post peak was around 15000\,km\,s$^{-1}$, before gradually decreasing to about 10000\,km\,s$^{-1}$ by 30 days post peak.

\subsection{Velocity Measurements}
\label{sec:velocity_measurement}

For the current analysis, we measured the velocity of the Fe II $\lambda$5169 line. This line is usually blended with Fe II $\lambda$4924 and Fe II $\lambda$5018, making it difficult to measure the absorption minimum of the 5169\,\AA\ line cleanly. Therefore \citet{Liu2017} used a template matching method developed in \citet{Modjaz2016} using template spectra of SNe Ic, and applied to SNe Ic-BL and SLSNe, as this feature is present in all three sub-types. This method applies a convolution with a Gaussian to broaden the line profiles, and a blueshift assumed to reflect the velocity difference between the template and the spectrum of interest.

Here we apply a modified version of the \cite{Modjaz2016} code to our sample of SLSN spectra. This method uses Markov chain Monte Carlo (MCMC) to fit SN Ic template spectra at different phases to the SLSN spectra. It does this by first identifying peaks on either side of the absorption profile, either from the continnum level or the P-Cygni maxima, and using the wavelength range between these peaks for fitting the line. For this analysis, the allowed region used to search for the "blue" peak was constrained to 4300$-$4800\,\AA\, and the "red" peak to 5000$-$5700\,\AA\ across all spectra. Using MCMC, the following parameters were fit: $v$, the blueshift velocity of the line with respect to the template spectrum; $\sigma$, the broadening of the line with respect to the template spectrum; $\Delta w$, a parameter that adjusts the initial wavelength range using the "blue" and "red" peaks; $a$, the amplitude; and $\rm{offset}$, this applies a vertical offset to the template. Here we implement a method with uniform priors for all parameters and ranges between $v$: $-$5000 to 30000\,km\,s$^{-1}$, $\sigma$: 0.1 to 50\,km\,s$^{-1}$, $a$: 0 to 9, $\Delta w$: $-$100 to 100\,\AA, $\rm{offset}$: $-$1 to 1. We also initiated the walkers at randomly sampled values along these priors. This differs from \citet{Modjaz2016}, who used a Gaussian prior for $\Delta w$ centred on 0 and set the 99.7$^{\rm{th}}$ percentile to be 100\,\AA.

\begin{figure}
    \begin{center}
        \includegraphics[width=1\columnwidth]{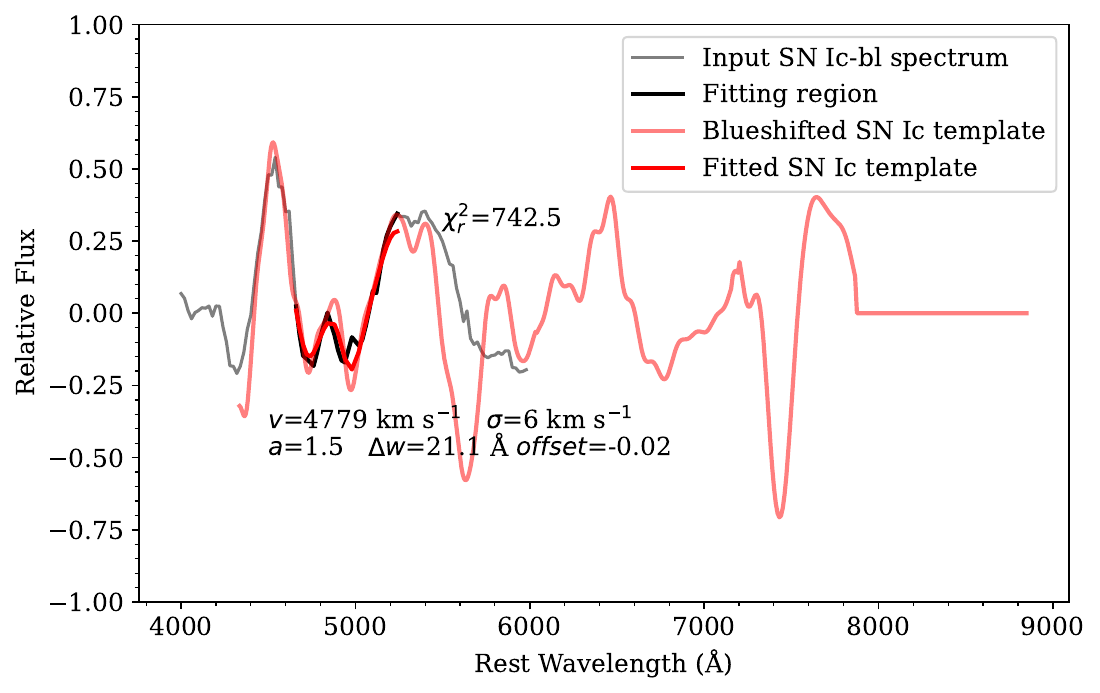}
        \caption{Example fit of the Fe II $\lambda$5169 feature to a spectrum of SN\,2016wi. The blueshifted template spectrum is shown in red matched to the target spectrum in black. The bold lines for both colours show the fitting region chosen by the code with the bold red line also broadened to match the width of the target feature. Median values for the parameter fits are printed.}
            \label{fig:line_fitting}
    \end{center}
\end{figure}

Before employing the MCMC, spectra were first flattened by taking continuum regions on either side of the profile between 4000$-$4500\,\AA\ and 5500$-$6000\,\AA. These regions were used to compute a linear fit for the underlying continuum that was then subtracted from the spectrum. The continuum-subtracted spectrum between 4000-6000\,\AA\ was then used for our analysis.

Only spectra with phases between 10-80 days with respect to peak were fit. This is due to the contamination from Fe III at early times. \citet{Liu2017} disregard all spectra before 15 days due to this contamination, however \citet{Inserra2018b} use phases as low as 10 days. For this reason we include phases from 10 days to be inclusive in our measurements. The phases also had an upper limit imposed of 80 days as the template spectra only cover phases up to 72 days after peak. 

For each velocity measurement, a SN Ic template needs to be selected. The Fe II line in each template also has an inherent velocity, reported in \citet{Modjaz2016}. This was added to the relative blueshift velocity derived by each template fit to calculate a final physical velocity.  The line profiles and relative strengths of the Fe II blend vary with phase from peak, so that different templates provide more appropriate comparisons at different phases. However, as SLSNe evolve more slowly than SNe Ic, the same phase for SNe Ic cannot be assumed for SLSNe. We tried several methods for selecting template spectra at different phases based on the relative decline rates of SLSNe and SNe Ic. In the end we found that this did not make a significant difference to the velocity measurements, and in some cases decreased the fit quality. We therefore report results based on fits that used the template at the same phase from peak as the SLSN spectrum in question.

The difficulty in matching the phases of SLSNe and SNe Ic for such a large sample is a possible caveat in this analysis. Other caveats include that in some cases the Fe II lines could not be fit very well either due to the line being too noisy, or too weak. These spectra were removed manually after visual inspection of the fits. Performing a KS test on the peak apparent magnitudes of the spectra that were retained, and those that were removed gives a p-value of 0.68, indicating they are likely drawn from the same distribution. A KS test was also performed on the redshift distribution of these spectra resulting in a p-value of 0.53. From this we can infer that the the removal of spectra was likely not biased by luminosity or redshift. In total, this resulted in 247 velocities from 111 different objects. 

\begin{figure}
	\begin{center}
        \includegraphics[width=1\columnwidth]{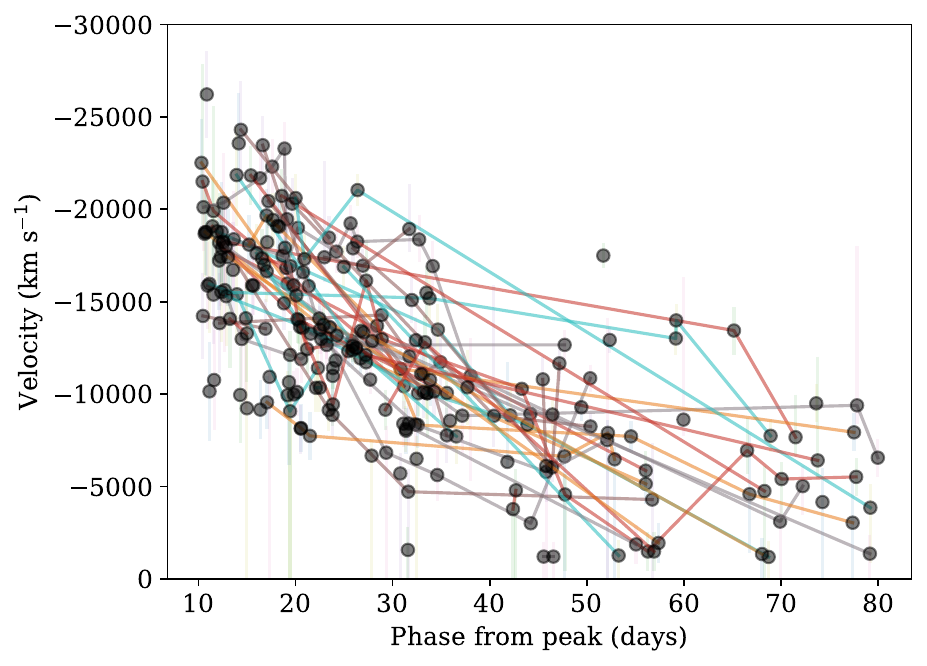}
		\caption{Velocity evolution over time in the rest-frame for the Fe II $\lambda$5169 line. Phase is measured from the peak of the light curve. Displayed are velocities 247 velocity measurements from 111 different objects. Coloured lines connect the velocities for individual events.}
            \label{fig:phase_v}
	\end{center}
\end{figure}

The final velocities of the Fe II $\lambda$5169 line are plotted in Figure \ref{fig:phase_v} against phase from peak. We can see a significant spread in velocities across the entire time frame. Close to 10 days post peak, some events, such as SN\,2019otl, exhibit velocities nearing 27000\,km\,s$^{-1}$, whereas other events such as SN\,2018fcg and PTF10bjp, show velocities closer to 10000\,km\,s$^{-1}$. This wide range of velocities persists even at late times, with a spread of approximately 10000\,km\,s$^{-1}$ still observed at 80 days post peak. We note that the velocity for some events using this method does not match reported velocities in the literature. For example, SN\,2018ibb shows a velocity from $\sim$15000\,km\,s$^{-1}$ to $\sim$6000\,km\,s$^{-1}$ over the time frame. Comparing this to the velocity measured in \citet{Schulze2024}, which remains consistent around $\sim$8500\,km\,s$^{-1}$ over this time frame. This may be a result of the different methods used to measure the velocity, and also the broad range of priors required to the fit the line profiles in this sample. Figure \ref{fig:phase_v} also displays a gradual decrease in measured velocity over time. This is due to the photosphere receding inwards, allowing us to see slower moving material closer to the SN core. 

Comparing our results to the sample of 21 SLSNe from \citet{Liu2017}, we can see that their sample exhibit comparable velocities to our sample at similar epochs. They also applied this method to measure the velocity of the Fe line of SNe Ic and SNe Ic-BL. To investigate the evolution of each population, they employed moving weighted averages to compare velocities across the three SN classes. Their analysis showed that SLSNe and SNe Ic-BL follow a similar pattern of velocity evolution, with both displaying significantly higher velocities than SNe Ic at early times. By approximately 40 days after peak brightness, however, the velocities of all three classes converged to similar values.

Further insights into this relationship were provided by \citet{Finneran2024}, who analysed 61 SNe Ic-BL, and 13 SNe Ic associated with gamma-ray bursts (GRB-SNe). Their findings showed that SNe Ic-BL have velocities comparable to GRB-SNe, though the latter consistently occupy the upper end of the velocity distribution. The velocities of SLSNe in this sample overlap with those measured in \citet{Finneran2024}, but they generally fall below the velocities observed for GRB-SNe.

Our findings and previous work show that SLSNe have similar velocities to some SNe Ic-BL, but at the lower end of the distribution. This may provide an important clue to the energy sources that accelerate and heat the ejecta. The discovery of a SLSN associated with an ultra-long GRB \citep{Greiner2015} suggests that SLSNe and GRBs may represent different manifestations of the same type of stellar explosion. 

This is supported by their spectral similarities in the nebular phase \citep{Nicholl2016a} and their preference for low-metallicity host environments \citep{Lunnan2014}, with SLSNe typically found in galaxies with less than 0.4 solar metellicity and long GRBs in environments with less than 0.7 solar metallicity \citep{Leloudas2015b, Schulze2018}. The distinction could lie in the timescale of energy injection: SLSNe are thought to involve prolonged energy injection, which drives late-time heating, whereas GRBs are characterised by a rapid release of energy that accelerates the ejecta \citep{Metzger2015, Metzger2018, Margalit2018}. In the magnetar central engine model, this variation in timescale could arise from differences in magnetar field strength.

\subsection{Where are the lines and the continuum formed?}
\label{sec:v_comparison}

As mentioned in Section \ref{sec:velocity_measurement}, the velocity of the Fe II $\lambda$5169 line is often used as a proxy for the photospheric velocity for rapidly expanding ejecta. This is often an assumption made in the expanding photosphere method for SNe II \citep{Kirshner1974}. However, depending on the assumed powering mechanism of SLSNe, this is not necessarily at the same location where the bulk of the continuum luminosity is produced. For example, in SNe with strong CSM interaction, most of the continuum emission will be generated in the shocked CSM exterior to the fast-expanding ejecta. In order to gain an understanding of where the line and continuum form in relation to one another, we can compare the implied photospheric radius derived from fitting a blackbody model to the SED, to the photospheric radius derived from the Fe II $\lambda$5169 line velocity. 

\begin{figure}
	\begin{center}
		\includegraphics[width=1\columnwidth]{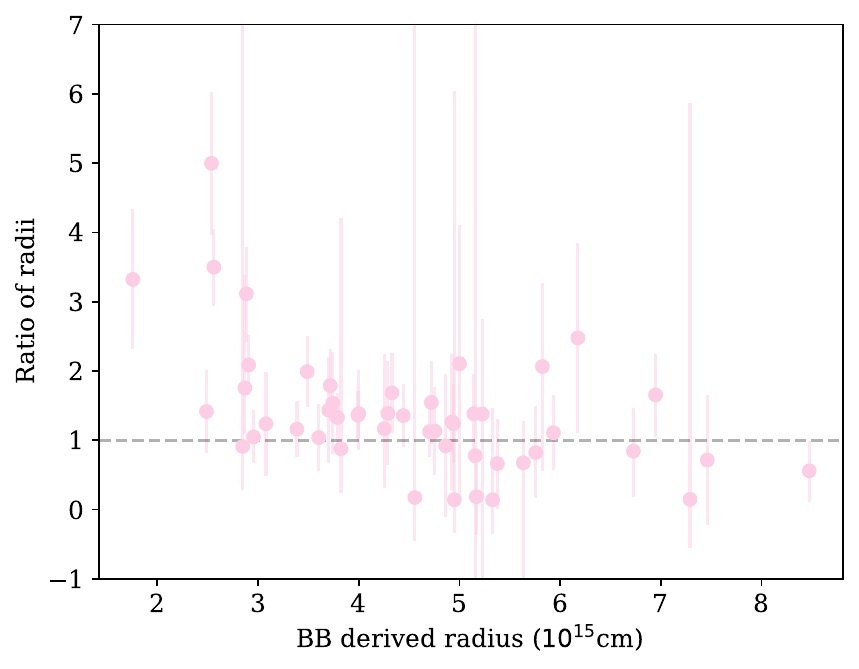}
		\caption{The blackbody radius of the SNe at peak plotted against the ratio of the radius derived from the iron line velocity, to the blackbody radius. The blackbody derived values are taken from \citet{Gomez2024}. The dashed horizontal line indicates the value at which both radii would be equal.}
            \label{fig:r_rphot}
	\end{center}
\end{figure}

The Fe II radius was derived by taking the earliest velocity measurement for each object and multiplying it by the time from explosion to when the spectrum was obtained in rest frame days. This provides an estimated upper limit for the distance the material could have travelled in that time. This was then compared to the blackbody radius derived from fitting the light curve SED in \citet{Gomez2024} using \texttt{extrabol} \citep{Thornton2024} at the same phase. Figure \ref{fig:r_rphot} shows the ratio of the Fe II radius to the blackbody radius, plotted against the blackbody radius. The mean of the ratio is $1.39\pm0.90$ indicating the radii measured using both methods are consistent with one another. From this we can infer that the velocity measured by the Fe II $\lambda$5169 line is indeed a good tracer for the photospheric velocity and therefore the Fe line forms in a region close to the rapidly expanding photosphere. This suggests that the bulk of the light is coming from a region close to the fast-moving ejecta. This picture is consistent with centrally-powered models, where the electron-scattering photosphere that produces the continuum lies within the SN ejecta and the Fe II lines form just outside, but still within the ejecta. It could also be consistent with interaction models, if the CSM is quite confined, and the optically thick CSM is not much larger than the radius reached by the fast ejecta at maximum light. However, this situation may require an aspherical CSM and a viewing angle that allows us to see the fast ejecta at the same time as the dense CSM.

\begin{figure}
	\begin{center}
		\includegraphics[width=1\columnwidth]{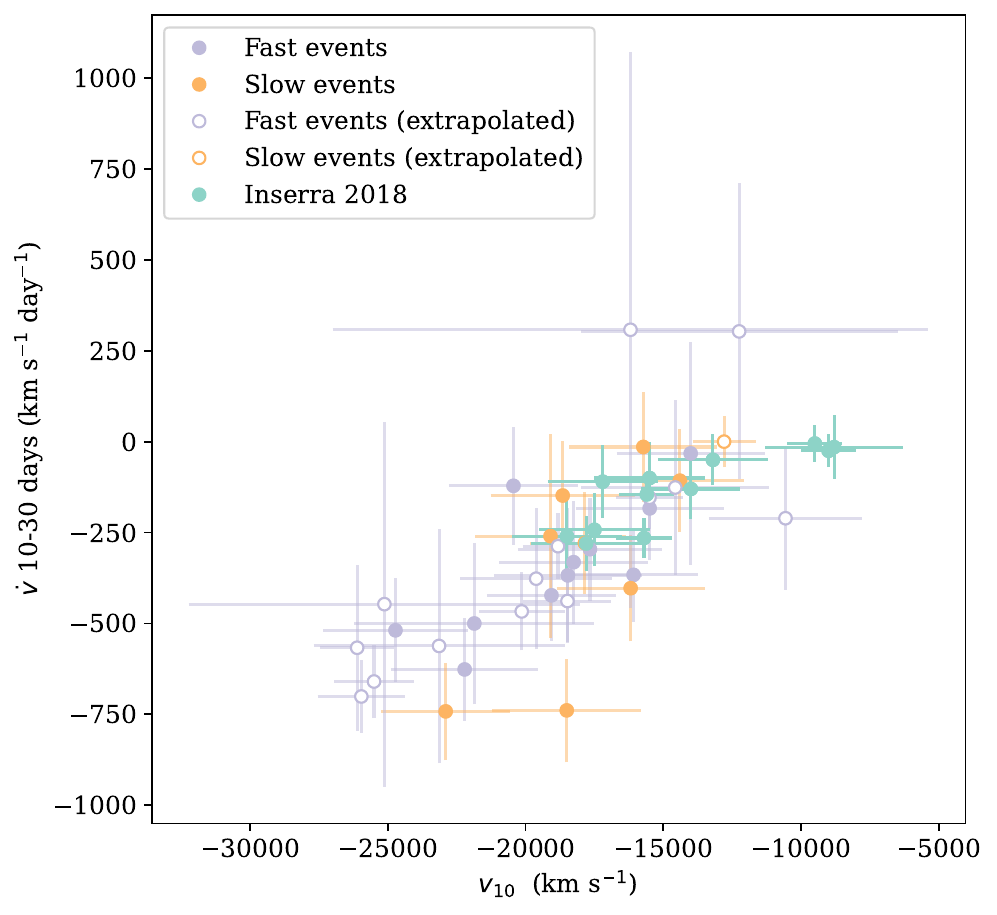}
		\caption{Velocity at 10 days versus the change in velocity from 10 to 30 days for 34 events. Events with "fast" light curve evolutions (defined as $\tau_{e} \leq $ 30 days) are shown in purple. Unfilled circles indicate extrapolated points for those events without a spectrum between 10-15 days post peak, and therefore the velocity at 10 days has been extrapolated from the line of best fit. Plotted in blue are the measurements from \citet{Inserra2018b}.}
            \label{fig:v_vdot}
	\end{center}
\end{figure}

\begin{figure}
	\begin{center}
		\includegraphics[width=1\columnwidth]{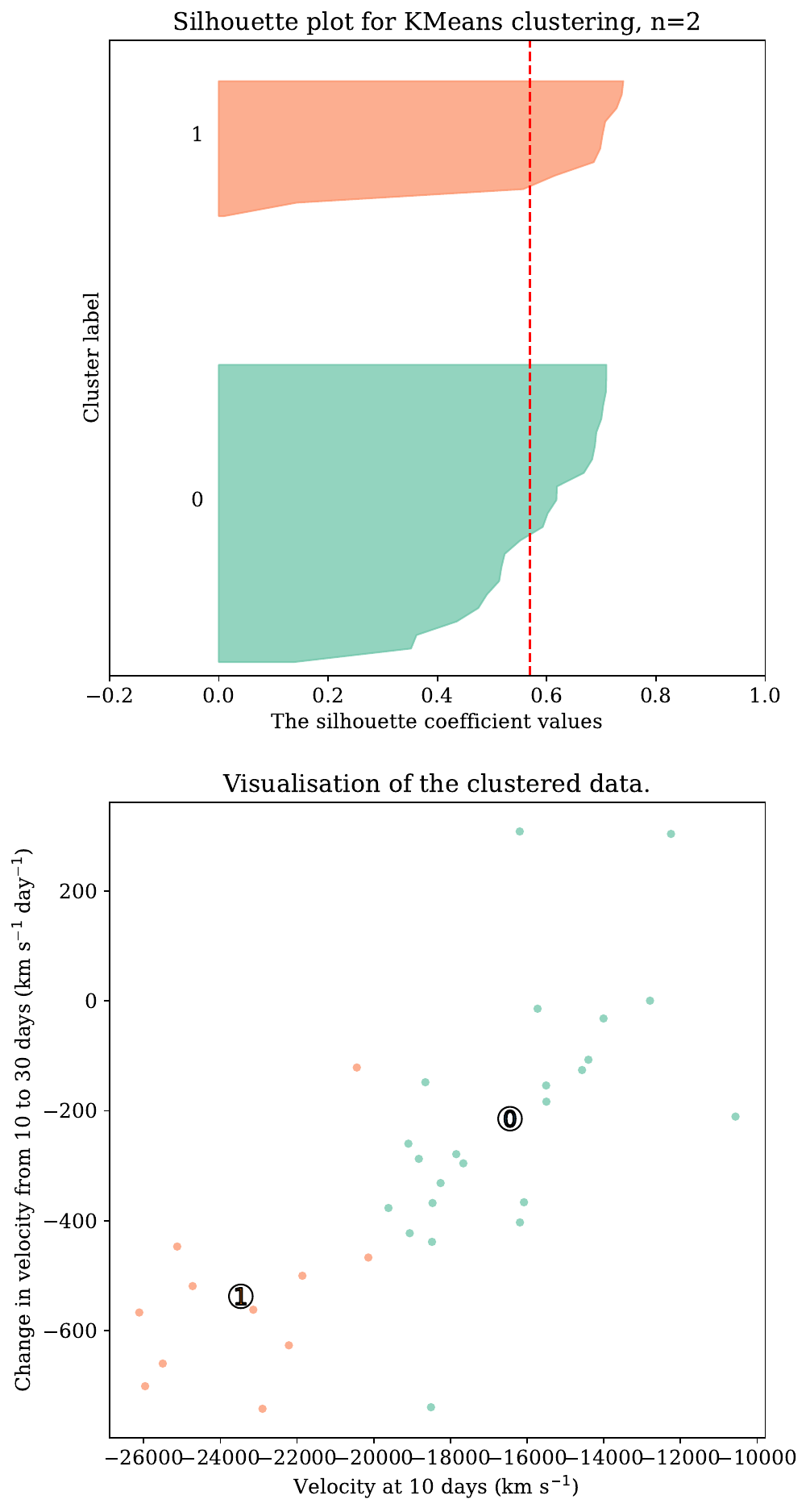}
		\caption{Silhouette plot using a cluster size of two applied to the data from Figure \ref{fig:v_vdot}. These plots show the quality of clustering by displaying the silhouette coefficient for each point, which measures how well the point fits within its assigned cluster compared to the nearest neighbouring cluster. Values range from -1 to 1, with higher values indicating well-separated clusters, while negative values suggest possible misclassification. The mean silhouette score, shown as a red vertical line, summarises clustering performance. With a score of 0.57, this suggests a preference for two distinct clusters. However, these clusters do not coincide with the separate groupings of "fast" and "slow" events.}
            \label{fig:v_vdot_clustering}
	\end{center}
\end{figure}

\subsection{Change in Photospheric Velocity}
\label{sec:v_vdot}

We investigated the photospheric evolution in this sample by looking at the velocity from the Fe II $\lambda$5169 line at 10 days ($v_{10}$), versus the rate of change of velocity from 10 to 30 days ($\dot{v}$) following \citet{Inserra2018b}. Objects were first checked to identify those with two or more spectra taken between 10 and 50 days post peak, with a minimum time span of five days between the earliest and latest observations within this window. For these objects, the velocity evolution over time was linearly fit with the reciprocal of the errors used as weights. If there was a spectrum between 10-15 days post peak, the spectrum closest to 10 days was used for $v_{10}$. Otherwise the linear fit was used to extrapolate $v_{10}$ and these points are marked as unfilled circles in Figure \ref{fig:v_vdot}. The velocity at 30 days ($v_{30}$), was extrapolated from the line of best fit. If only two spectra were available, the errors on $v_{10}$ and $v_{30}$ were obtained by using the errors on the first and last spectrum. If more than two spectra were available, the errors were derived from the covariance matrix of the fit. There are large error bars for a few points with extrapolated values for $v_{10}$. This can be explained by the need to extend the linear fits of the velocity evolution in some cases by over 10 days. 


Comparing our sample to that in \citet{Inserra2018b}, we find
both samples follow the same shape as shown in Figure \ref{fig:v_vdot}. However, our sample extends further into higher $v$ and $\dot{v}$ values. This systematic offset could be due to differences in the way the velocities were calculated. In \citet{Inserra2018b}, the velocity is measured by fitting the absorption minima of the profile but this is affected by blending from Fe II $\lambda$4924 and Fe II $\lambda$5018 at early times. \citet{Inserra2018b} state that the strong linear relationship of their version of this plot indicates that the faster the photospheric velocity, the faster the velocity decreases. For this dataset we find a Spearman's rank correlation coefficient of 0.78, with a p-value $\ll 0.05$, indicating a strong positive correlation between $\dot{v}$ and $v_{10}$ that is statistically significant. We can therefore conclude a linear relationship between these parameters and so also find that for faster initial velocities, the decrease in velocity is also faster. 

Physically, we can interpret this result by considering the SN ejecta to be in homologous expansion as shown:

\begin{equation}
    v(r) = v_{0} \frac{r}{R}
    \label{eq:v}
\end{equation}

In Equation \ref{eq:v}, $v(r)$ represents the velocity at a given radius $r$. $R$ is the radius of the outer most ejecta, and $v_{0}$ is the scale velocity. The velocity at the photosphere follows $v_{\rm{phot}} = v(r_{\rm{phot}})$. We know the location of the photosphere relates to the optical depth ($\tau$) through the relation:

\begin{equation}
    \tau = \int^{R}_{r_{\rm{phot}}} \kappa\,\rho(r,t) dr = \frac{2}{3}
    \label{eq:tau}
\end{equation}

Here $\kappa$ is the opacity, and $\rho$ is the density. This density is a function of the radius $r$, which itself evolves with time (and is proportional to $v_0$). Therefore:

\begin{equation}
    \rho(t) \propto \frac{1}{R^3} \propto \frac{1}{v_{0}^3}
    \label{eq:rho}
\end{equation}

For events with a larger scale velocity, the density decreases faster and from Equation \ref{eq:tau} we see deeper into the ejecta to slower moving material at any given time. This results in a larger change in the measured photospheric velocity within a given time interval. Using $v_{10}$ as a proxy for $v_{0}$, we do indeed see that events with a larger scale velocity show larger velocity changes within 30 days. This suggests that the observed correlation is consistent with all SLSNe forming these lines close to the photosphere of the homologously expanding ejecta of comparable density, with a scale velocity that varies between events.








We also split our sample into events with ``fast'' and ``slow'' light curve evolution, coloured by purple and orange points respectively. The split of events into fast and slow categories was assigned based on having an e-folding decline time in the light curve of above or below 30 days. This is motivated by the debate in the literature over whether they represent distinct populations or a continuum of properties. Some argue that these systems arise from the same explosion mechanism with differences in evolution explained by different ejecta masses \citep{Nicholl2015a} and heating timescales \citep{Nicholl2017a}. Others argue that the diversity of the population points to distinct sub-groups \citep{Quimby2018, Konyves-Toth2022} that could be powered by different engines \citep{Inserra2018b}.

\citet{Inserra2018b} claimed that when splitting the data based on light curve timescale, the slow events also showed low $v$ and slow $\dot{v}$, while the faster events showed both high $v$ and $\dot{v}$. Although this divide looks clear in \citet{Inserra2018b}, in our sample the two groups overlap significantly with no clear distinction between fast and slow events. To support this visual assessment,  KMeans clustering was applied to this data to see if the data could be explained by two groups. We can see a silhouette plot to represent this in Figure \ref{fig:v_vdot_clustering}. The top panel shows the silhouette scores for each point, with an average silhouette score of 0.57 for the whole sample. This indicates a preference for two clusters however it is not a strong clustering. From the bottom panel of this Figure, we can see that the grouping of points do not have a distinct cut off and are blended into one another. This is supported by calculating the median decline time in each group, producing a comparable 38 days for group 0, and 37 days for group 1. This indicates that both clusters in this case contain a mix of both fast and slow events. We also performed BIC analysis on this data for 2-5 clusters and find no strong evidence for any of these numbers of clusters.

\section{Conclusions}
\label{sec:conclusions}

In this paper we have presented the largest sample of photospheric spectra for Type I SLSNe. This has been collated from ePESSTO+, the Finding Luminous and Exotic Extragalactic Transients (FLEET) search, and all published spectra up to December 2022. The main conclusions from this analysis are:

\begin{itemize}[leftmargin=0.5cm]
    \item[\ding{73}] Creating average spectra shows that SLSNe initially have high temperatures of around 10000$-$11000\,K before peak, with blue continua and relatively weak lines. This is followed by a rapid spectral change in a period of cooling up to 40 days after peak, where the lines transition to P-Cygni profiles, and strengths increase relative to the continuum. 
    This indicates spectral changes are driven by the temperature evolution.\\
    
    \item[\ding{73}] In the late photospheric phase from 40 days post peak, temperatures plateau to 5000$-$6000\,K. These spectra show much more pronounced line features, though still with a strong underlying blackbody continuum. This  temperature range matches the results from light curve models such as \texttt{MOSFiT} and \textsc{Redback} which introduce a final temperature parameter that generally converges to a similar value. \\

    \item[\ding{73}] The average spectra suggest there may be a small amount of He retained by many SLSNe at the time of their explosion. This is seen by the consistent troughs for prominent He lines at $\sim$7500\,km\,s$^{-1}$ for spectra 20$-$40 days post peak, and at $\sim$6000\,km\,s$^{-1}$ at 60$-$80 days post peak. \\
    
    \item[\ding{73}] There is no evidence for distinct spectral subclasses within the SLSN population when decomposing using KMeans clustering and testing with the Bayesian Information Criterion, however, there are a few outliers. Notably SN\,2017egm deviates from the general population due to its very strong Ca NIR triplet emission. This coincides with when the event undergoes bumps in its light curve due to interaction with shells of CSM. This suggests that significant CSM interaction can influence both the light curve and spectral evolution, and line emission of these events. \\
    
    \item[\ding{73}] We find that the variance between spectra within a given time bin was slightly reduced when binning with respect to the time of explosion, rather than peak, and after normalising the phase using the decline timescale of each light curve. This is supported by the PCA decomposition which required fewer components to explain the same level of variation when grouping the spectra based on scaled phase from explosion rather than phase from peak. However, we find that the overall reduction in variance is surprisingly small. We suggest that this is because most SLSNe have a similar temperature at maximum light, and it is the temperature evolution, rather than time since explosion, that controls the spectrum.\\
    
    \item[\ding{73}] We find good agreement between the velocity based on blackbody fits \citep{Gomez2024}, and the Fe II $\lambda$5169 line velocity at peak. From this we can infer the Fe line forming region is close to the photosphere where the bulk of the thermal emission is produced. This is naturally consistent with a centrally powered model, where the continuum-forming photosphere lies within the SN ejecta, with Fe II lines just outside. Interaction models could also be viable if the CSM is compact and the Fe II lines form in the fast ejecta close to the CSM but that has not yet collided with it. However, this may require an aspherical CSM and a favourable viewing angle to observe both the fast ejecta and dense CSM simultaneously. \\
    
    \item[\ding{73}] We find events with faster photospheric velocities, also have a faster velocity gradient, consistent with previous findings in the literature. We suggest that this correlation arises naturally from a physical interpretation in which larger scale velocities within a homologously expanding ejecta lead us to see deeper into the ejecta after any given time interval, and therefore lead to larger changes in line velocities. KMeans clustering indicates a slight preference for two clusters with a silhouette score of 0.57, however each cluster group is composed of a mixture of events with ``fast'' and ``slow'' light curve evolutions, complicating any picture where SLSNe separate cleanly by both timescale and velocity.

\end{itemize}

The analysis in this paper suggests that most SLSNe can be explained as a single population of events, with the bulk of the spectral lines forming above a photosphere within homologously expanding ejecta. We suggest that this is more naturally consistent with an internal powering mechanism, but the identification of outliers such as SN\,2017egm shows that interaction, when present, can significantly alter this evolution.

Next-generation surveys will dramatically expand our sample of SLSNe. These facilities will provide an unprecedented opportunity for spectroscopic follow-up, allowing us to characterise the diversity of SLSNe, track their evolution, and test explosion models. The spectral dataset presented here will serve as an important benchmark for interpreting these future observations, and in particular in identifying outliers that follow-up efforts should focus on, maximising the scientific return from these upcoming surveys.


\section*{Acknowledgements}
We would like to thank Maryam Modjaz for helpful inputs and discussions.

A. Aamer and MN are supported by the European Research Council (ERC) under the European Union’s Horizon 2020 research and innovation programme (grant agreement No. 948381).

SG is supported by an STScI Postdoctoral Fellowship.

The Berger Time-Domain Group is supported in part by NSF and NASA funds.

A. Aryan and T-WC acknowledge the Yushan Young Fellow Program by the Ministry of Education, Taiwan, for the financial support (MOE-111-YSFMS-0008-001-P1).

This work was funded by ANID, Millennium Science Initiative, ICN12\_009.

LG and CPG acknowledge financial support from Centro Superior de Investigaciones Cient\'ificas (CSIC) under the PIE project 20215AT016, 
the Spanish Ministerio de Ciencia e Innovaci\'on (MCIN) and the Agencia Estatal de Investigaci\'on (AEI) 10.13039/501100011033 under the PID2023-151307NB-I00 SNNEXT project, 
the program Unidad de Excelencia Mar\'ia de Maeztu CEX2020-001058-M, 
and the Departament de Recerca i Universitats de la Generalitat de Catalunya through the 2021-SGR-01270 grant.
LG also acknowledges financial support from AGAUR, ILINK23001, and COOPB2304.
CPG acknowledges financial support from the Secretary of Universities and Research (Government of Catalonia) and by the Horizon 2020 Research and Innovation Programme of the European Union under the Marie Sk\l{}odowska-Curie and the Beatriu de Pin\'os 2021 BP 00168 programme.

HK was funded by the Research Council of Finland projects 324504, 328898, and 353019.

GL was supported by a research grant (VIL60862) from VILLUM FONDEN.

TEMB is funded by Horizon Europe ERC grant no. 101125877.

MR acknowledge support from National Agency for Research and Development (ANID) grants ANID-PFCHA/Doctorado Nacional/2020-21202606

AS acknowledges support from the Knut and Alice Wallenberg foundation through the “Gravity Meets Light” project.

S. Schulze is partially supported by LBNL Subcontract 7707915.

This paper includes data gathered with the 6.5 meter Magellan Telescopes located at Las Campanas Observatory, Chile. Observations reported here were obtained at the MMT Observatory, a joint facility of the Smithsonian Institution and the University of Arizona.

Based on observations collected at the European Organisation for Astronomical Research in the Southern Hemisphere, Chile, as part of PESSTO (the Public ESO Spectroscopic Survey for Transient Objects - PI: Smartt), ePESSTO, and ePESSTO+ (the advanced Public ESO Spectroscopic Survey for Transient Objects Survey – PI: Inserra), under ESO program IDs 188.D-3003, 191.D-0935, 1103.D-0328, 106.216C and 108.220C.


This work made use of the data products generated by the Modjaz group (formerly NYU SN group), and 
released under DOI:10.5281/zenodo.58766, 
available at \url{https://github.com/nyusngroup/SESNtemple/}.


\section*{Data Availability}

This data will be publicly available on Github\footnote{https://github.com/gmzsebastian/SLSNe}. Previously unpublished raw spectra will also be publicly uploaded to the Weizmann Interactive Supernova Data Repository (WISeREP\footnote{https://www.wiserep.org})



\bibliographystyle{mnras}
\bibliography{Photospheric_spectra} 

\section*{Affiliations}
\noindent
{\it
$^{1}$Astrophysics Research Centre, School of Mathematics and Physics, Queen’s University Belfast, Belfast BT7 1NN, UK\\
$^{2}$Center for Astrophysics \textbar{} Harvard \& Smithsonian, 60 Garden Street, Cambridge, MA 02138-1516, USA\\
$^{3}$The NSF AI Institute for Artificial Intelligence and Fundamental Interactions, Cambridge, MA 02138, USA\\
$^{4}$European Southern Observatory, Alonso de C\'ordova 3107, Casilla 19, Santiago, Chile\\
$^{5}$Millennium Institute of Astrophysics MAS, Nuncio Monsenor Sotero Sanz 100, Off. 104, Providencia, Santiago, Chile\\
$^{6}$Graduate Institute of Astronomy, National Central University, 300 Jhongda Road, 32001 Jhongli, Taiwan\\
$^{7}$Institute for Astronomy, University of Hawai'i, 2680 Woodlawn Drive, Honolulu HI 96822, USA\\
$^{8}$Department of Physics, Lancaster University, Lancaster LA1 4YB, UK\\
$^{9}$Institute of Space Sciences (ICE-CSIC), Campus UAB, Carrer de Can Magrans, s/n, E-08193 Barcelona, Spain\\
$^{10}$Institut d'Estudis Espacials de Catalunya (IEEC), 08860 Castelldefels (Barcelona), Spain\\
$^{11}$The Oskar Klein Centre, Department of Astronomy, Stockholm University, Albanova University Center, 106 91 Stockholm, Sweden\\
$^{12}$Astronomical Observatory, University of Warsaw, Al. Ujazdowskie 4,
00-478 Warszawa, Poland\\
$^{13}$Department of Astronomy \& Astrophysics, University of California, San Diego, 9500 Gilman Drive, MC 0424, La Jolla, CA 92093-0424, USA\\
$^{14}$Cardiff Hub for Astrophysics Research and Technology, School of Physics \& Astronomy, Cardiff University, Queens Buildings, The Parade, Cardiff, CF24 3AA, UK\\
$^{15}$Department of Physics, Royal Holloway, University of London, Egham TW20 0EX, UK\\
$^{16}$Department of Physics, University of Warwick, Gibbet Hill Road, Coventry CV4 7AL, UK\\
$^{17}$ Tuorla Observatory, Department of Physics and Astronomy, FI-20014 University of Turku, Finland\\
$^{18}$Finnish Centre for Astronomy with ESO (FINCA), FI-20014 University of Turku, Finland\\
$^{19}$DTU Space, National Space Institute, Technical University of Denmark, Elektrovej 327, 2800 Kgs. Lyngby, Denmark\\
$^{20}$Astrophysics Research Institute, Liverpool John Moores University, 146 Brownlow Hill, Liverpool L3 5RF, UK\\
$^{21}$Max-Planck-Institut für Astrophysik, Karl-Schwarzschild Straße 1, 85748 Garching, Germany\\
$^{22}$School of Physics, Trinity College Dublin, The University of Dublin, Dublin 2, Ireland\\
$^{23}$Instituto de Ciencias Exactas y Naturales (ICEN), Universidad Arturo Prat, Chile\\
$^{24}$Instituto de Astrofisica, Departamento de Fisica, Facultad de Ciencias Exactas, Universidad Andres Bello, Fernandez Concha 700, Las Condes, Santiago RM, Chile\\
$^{25}$Center for Interdisciplinary Exploration and Research in Astrophysics (CIERA) and Department of Physics and Astronomy, Northwestern University, Evanston, IL 60208, USA\\
$^{26}$Astrophysics sub-Department, Department of Physics, University of Oxford, Keble Road, Oxford OX1 3RH, UK\\
}



\appendix


\clearpage
\onecolumn
\section{Objects and data in Sample}
\label{tab:all_spec}
\input{Figures/spectra_data_latex.txt}

\clearpage
\twocolumn

\section{Bayesian Information Criterion}
\label{sec:BIC}

\begin{figure}
	\begin{center}
		\includegraphics[width=1\columnwidth]{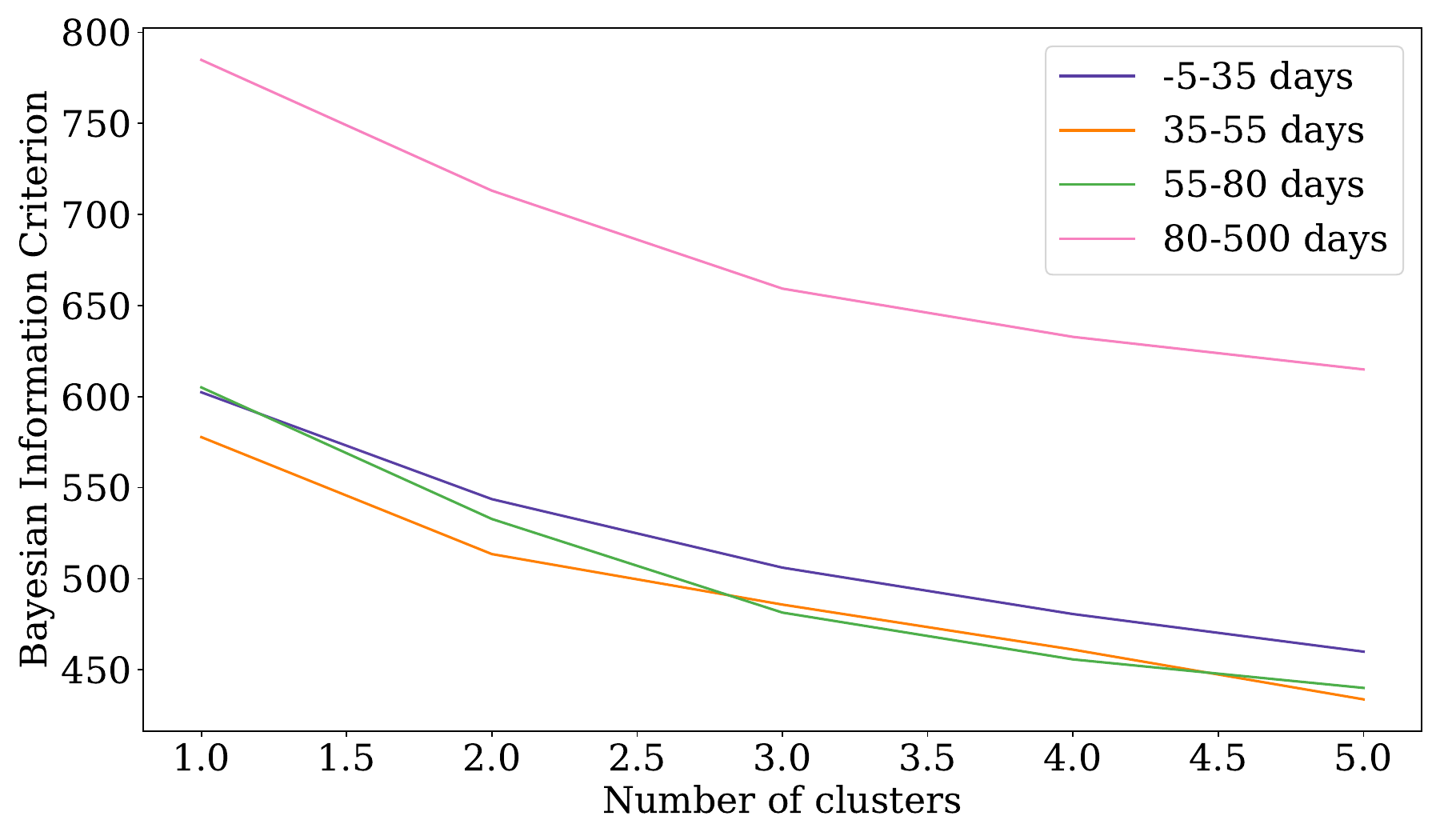}
		\caption{Bayesian information criterion (BIC) evaluated for varying numbers of clusters using KMeans clustering. Spectra have been split into varying time bins with phases relative to the peak of the light curve.}
            \label{fig:BIC_peak}
	\end{center}
\end{figure}

\begin{figure}
	\begin{center}
		\includegraphics[width=1\columnwidth]{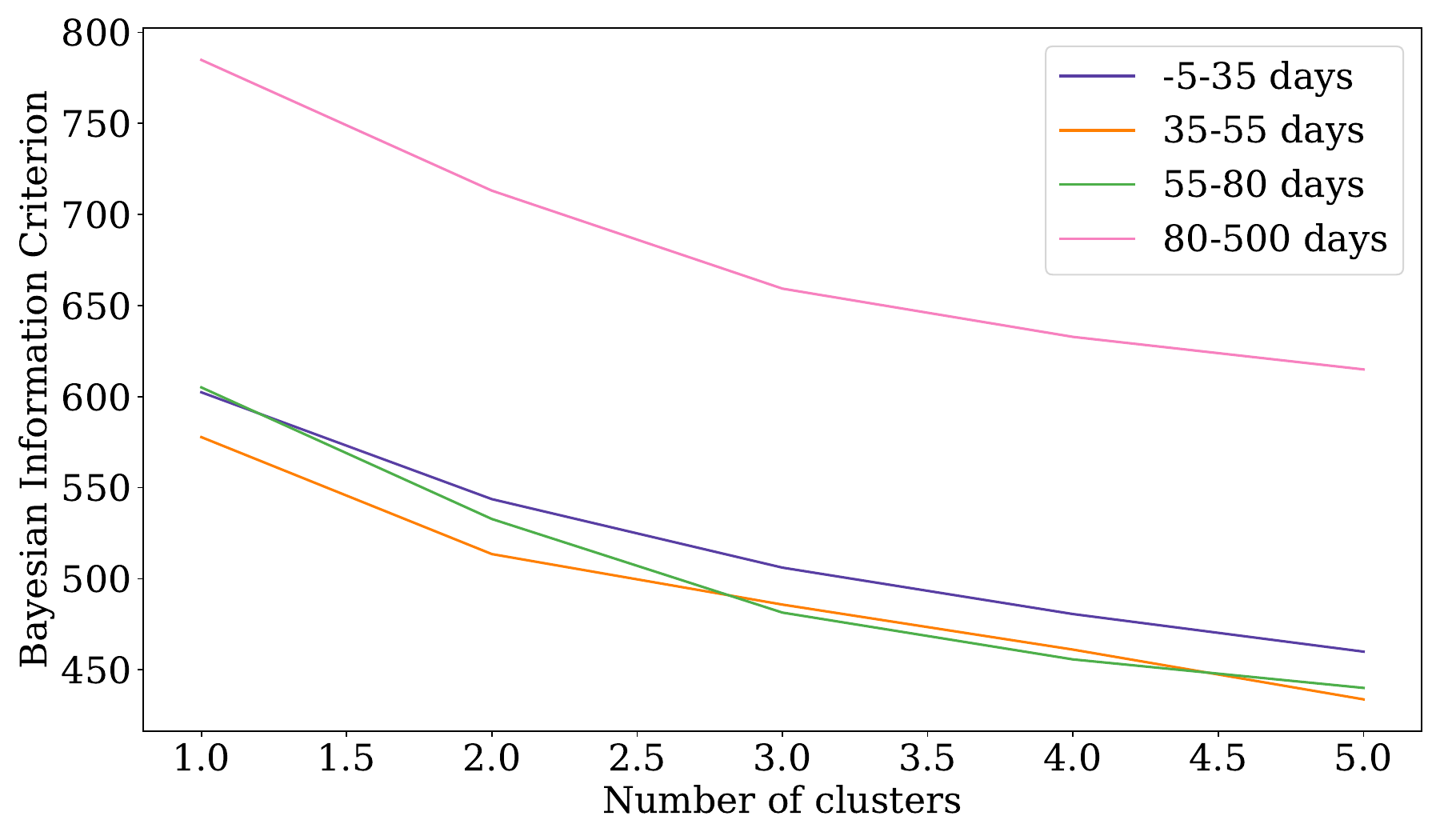}
		\caption{Bayesian information criterion (BIC) evaluated for varying numbers of clusters using KMeans clustering. Spectra have been split into varying time bins with phases relative to explosion and scaled by the median e-folding decline time (44 days).}
            \label{fig:BIC_exp}
	\end{center}
\end{figure}

The Bayesian Information Criterion (BIC) is a statistical metric used for model selection \citep{Schwarz1978}, based on Bayesian probability principles. It penalises models with more parameters to prevent overfitting \citep{Kass1995}, favouring simpler models when possible. A lower BIC score indicates a better model. In this analysis, K-means clustering was applied to the first two components of the PCA parameter space, and BIC values were computed for cluster sizes ranging from one to five across all time bins and both binning schemes. The results, shown in Figure \ref{fig:BIC_peak} and Figure \ref{fig:BIC_exp}, indicate a smooth decrease in BIC values as the number of clusters increases, suggesting no strong evidence for distinct clusters.

\section{The Effect of ``Fast'' and ``Slow'' events on Clustering}
\label{sec:fast_slow}

\begin{figure*}
\centering
\begin{minipage}{0.48\textwidth}
\begin{center}
        \begin{subfigure}{1\columnwidth}
            \centering
            \includegraphics[width=1\columnwidth]{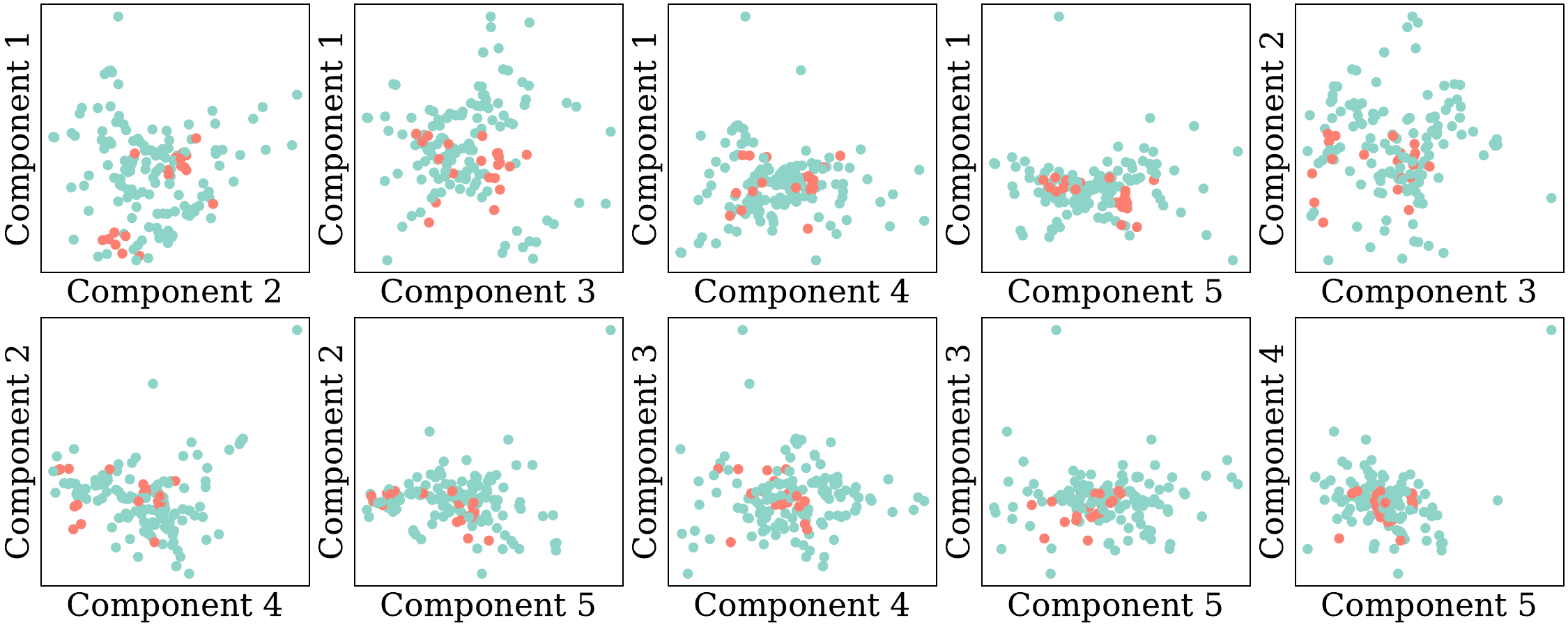}
            \caption{Days -80 - -5}
            \label{fig:-80--5_fs}
        \end{subfigure}
    \par\bigskip
    \hfill
        \begin{subfigure}{1\columnwidth}
            \centering
            \includegraphics[width=1\columnwidth]{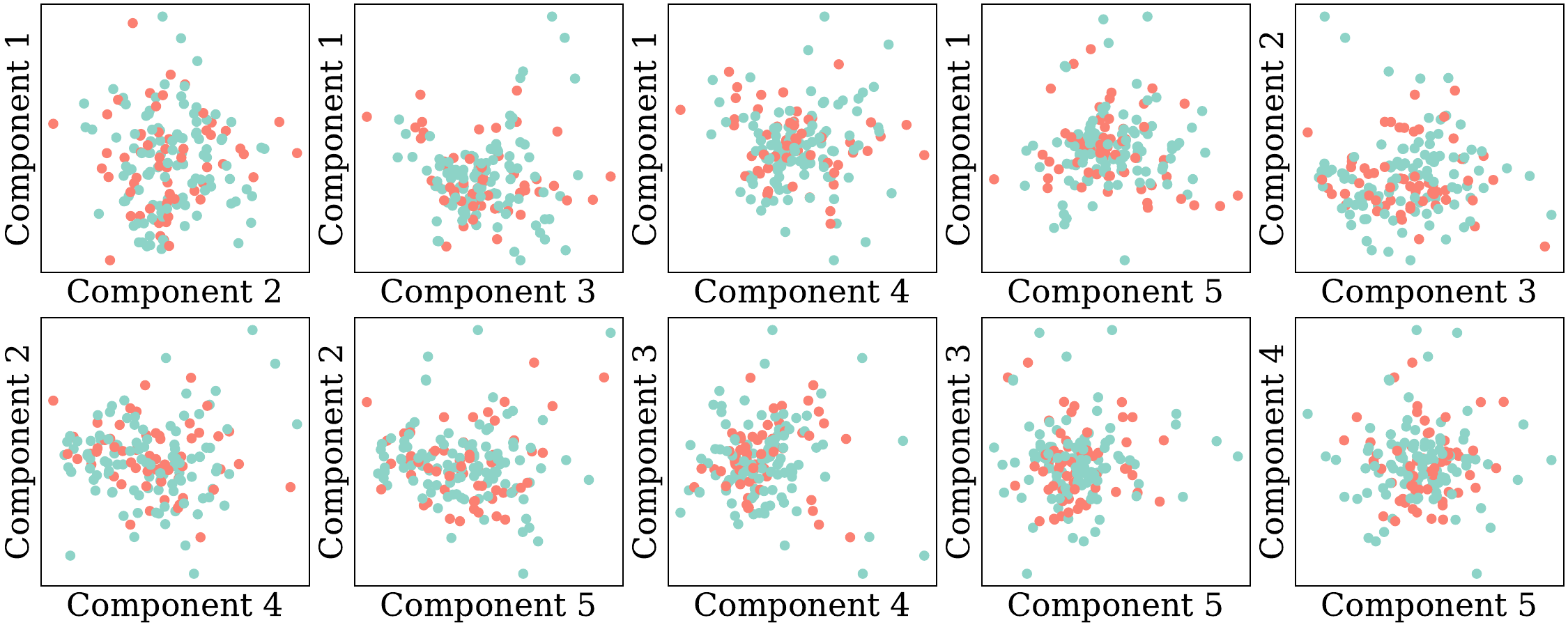}
            \caption{Days -5 - 15}
            \label{fig:-5-15_fs}
        \end{subfigure}
    \par\bigskip
    \hfill
        \begin{subfigure}{1\columnwidth}
            \centering
            \includegraphics[width=1\columnwidth]{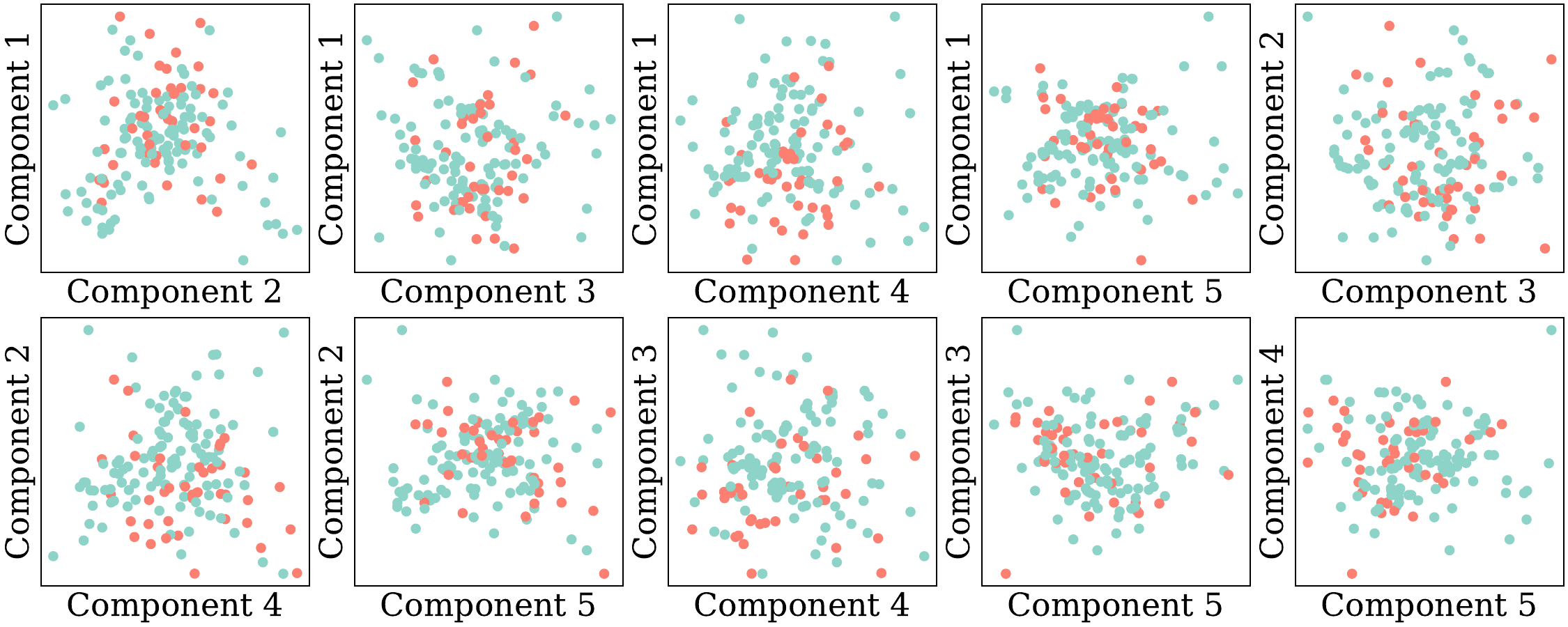}
            \caption{Days 15 - 40}
            \label{fig:15-35_fs}
        \end{subfigure}
    \par\bigskip
    \hfill
        \begin{subfigure}{1\columnwidth}
            \centering
            \includegraphics[width=1\columnwidth]{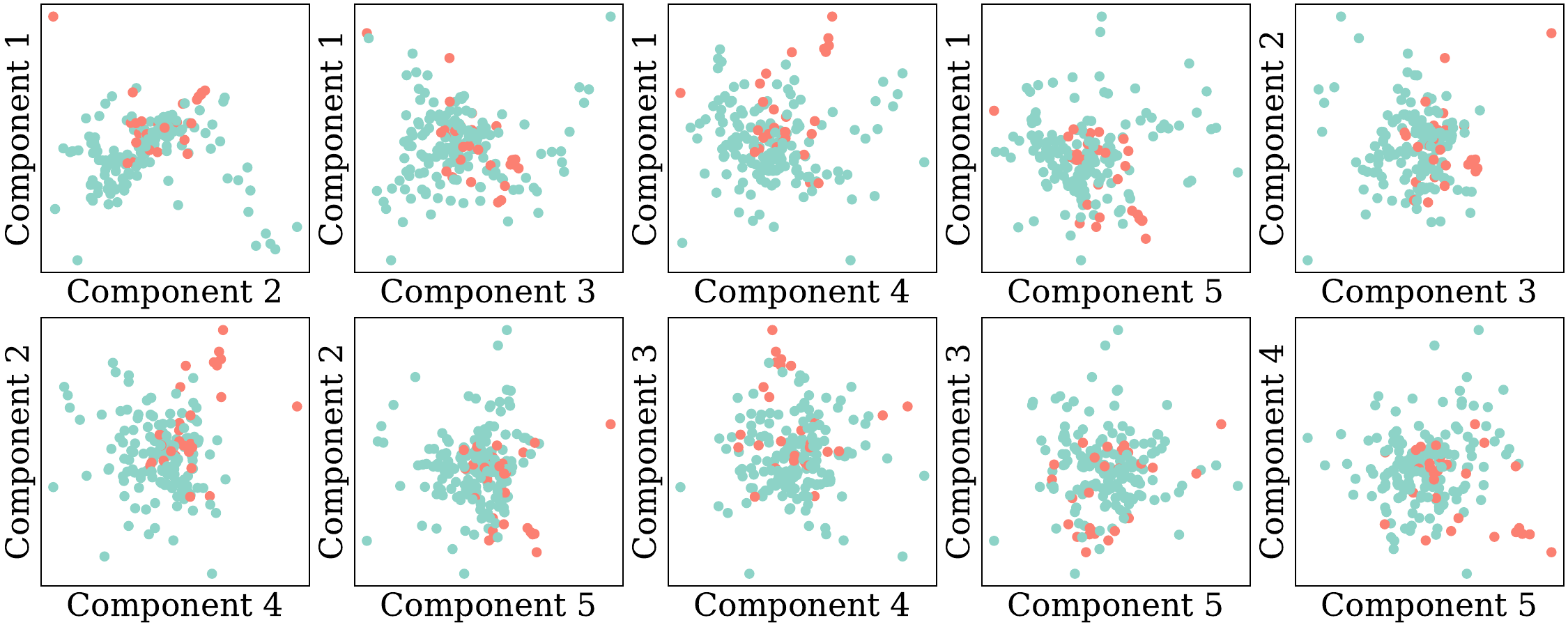}
            \caption{Days 40 - 160}
            \label{fig:35-160_fs}
        \end{subfigure}
    \hfill
        \caption{Representation of the SLSN photospheric spectra in the PCA parameter space in time bins relative to peak. Only the first five components for each bin are plotted for visualisation. Events are split into fast and slow categories based on having an e-folding decline time in the light curve of above or below 30 days, and coloured red and blue respectively. }
        \label{fig:PCA_components_peak_fs}
    \end{center}
\end{minipage}%
\hfill
\begin{minipage}{.48\textwidth}
\vspace{4mm}
\begin{center}
        \begin{subfigure}{1\columnwidth}
            \centering
            \includegraphics[width=1\columnwidth]{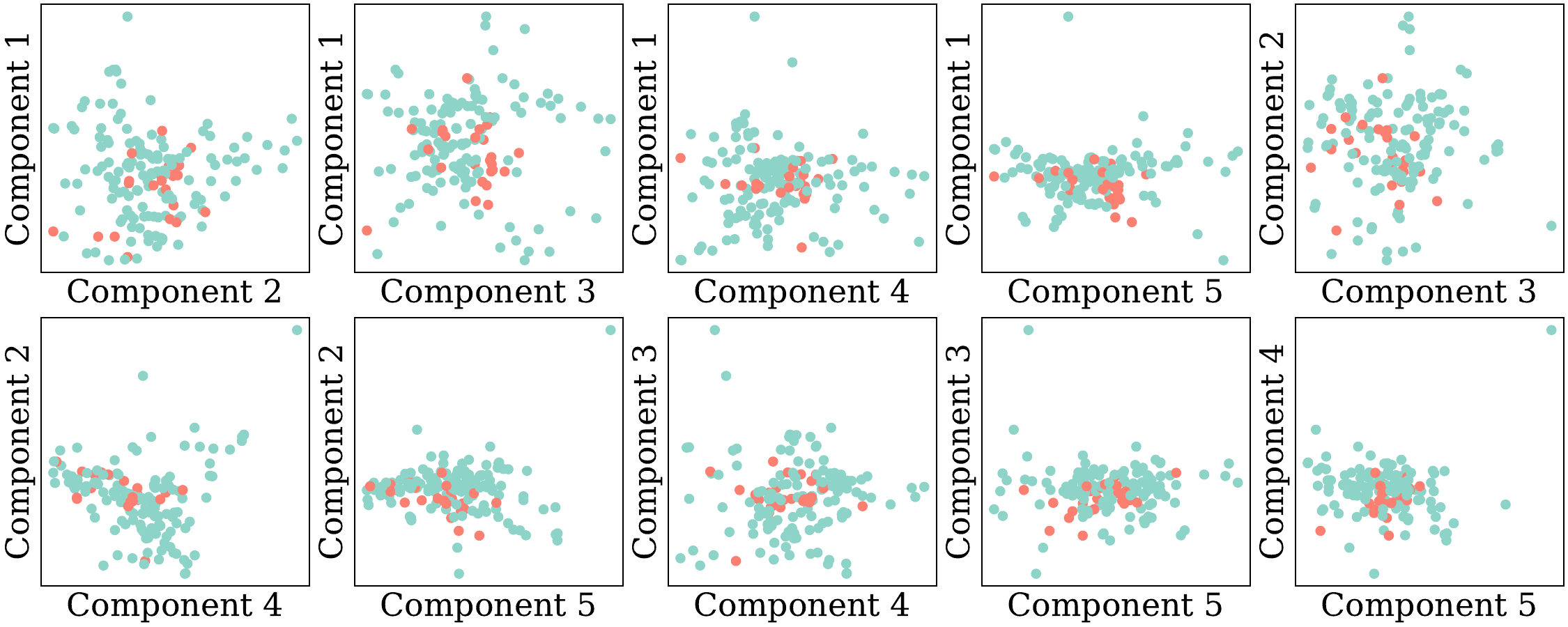}
            \caption{Days 0 - 35}
            \label{fig:-5-35_fs}
        \end{subfigure}
    \par\bigskip
    \hfill
        \begin{subfigure}{1\columnwidth}
            \centering
            \includegraphics[width=1\columnwidth]{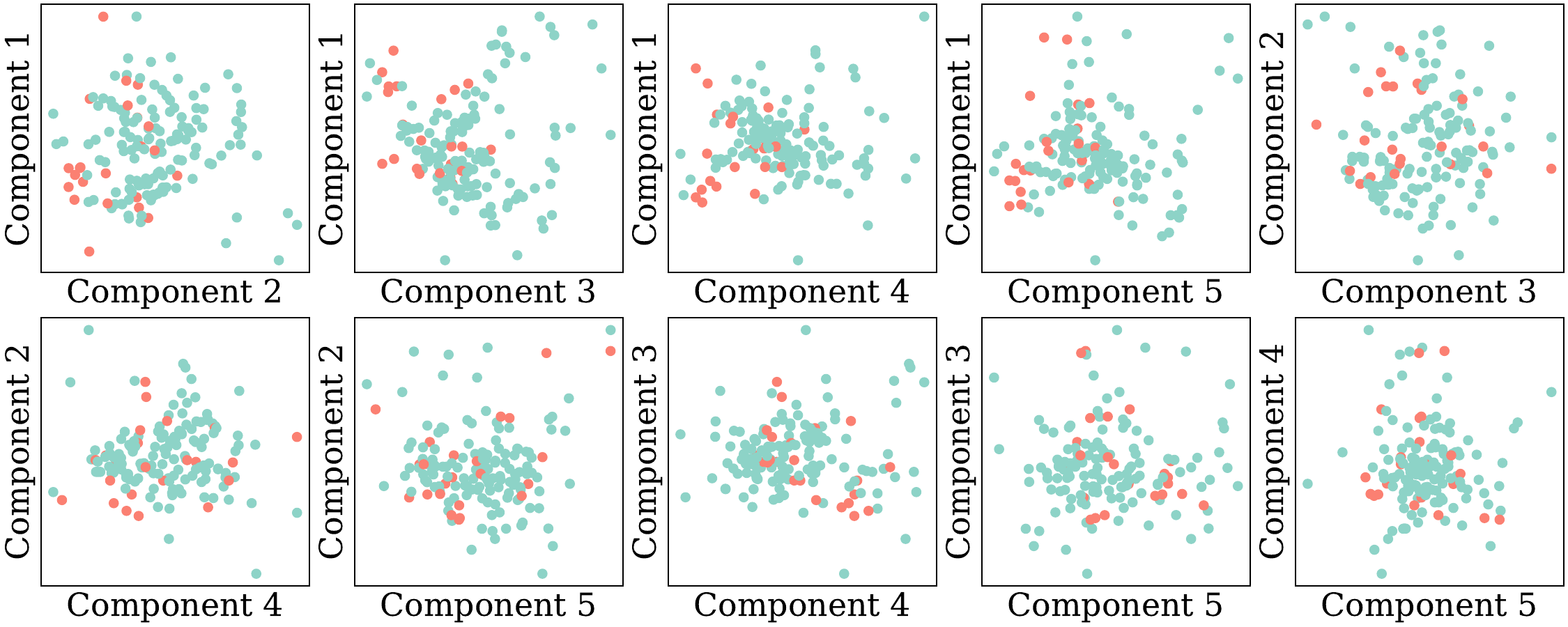}
            \caption{Days 35 - 55}
            \label{fig:35-55_fs}
        \end{subfigure}
    \par\bigskip
    \hfill
        \begin{subfigure}{1\columnwidth}
            \centering
            \includegraphics[width=1\columnwidth]{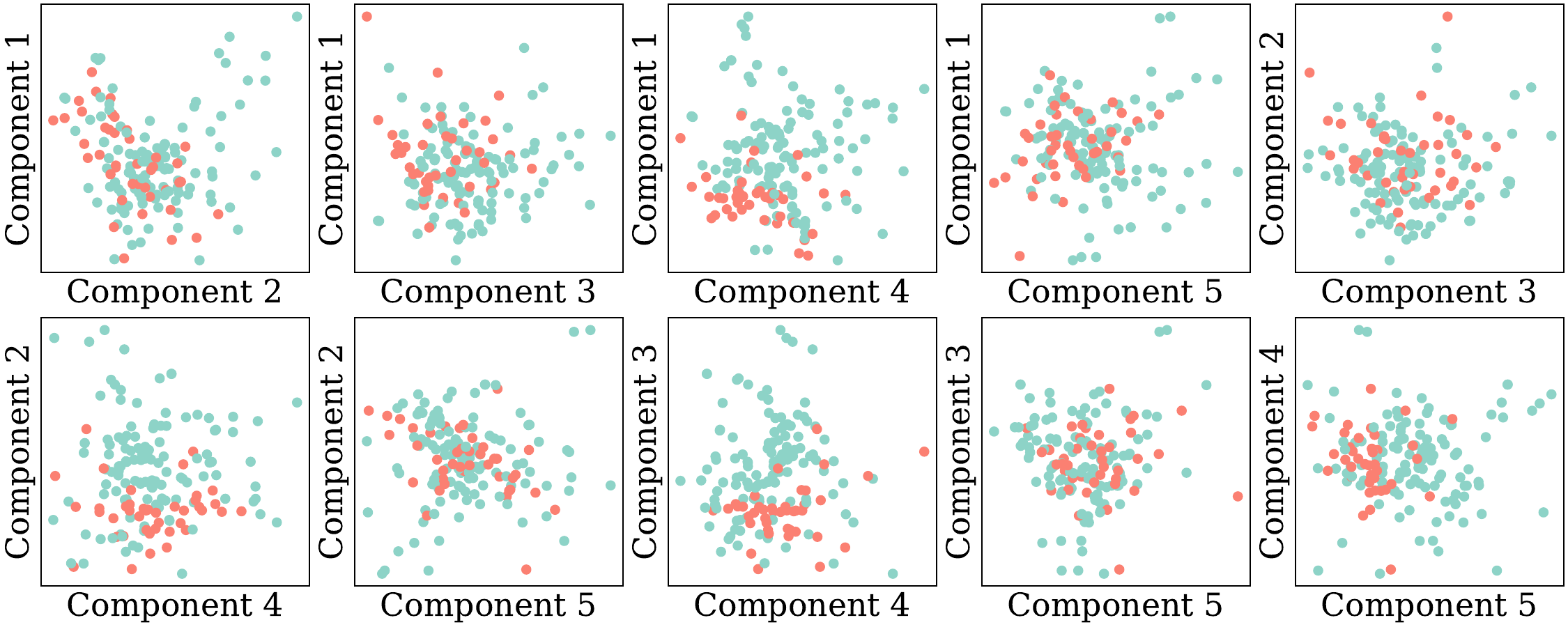}
            \caption{Days 55 - 80}
            \label{fig:55-85_fs}
        \end{subfigure}
    \par\bigskip
    \hfill
        \begin{subfigure}{1\columnwidth}
            \centering
            \includegraphics[width=1\columnwidth]{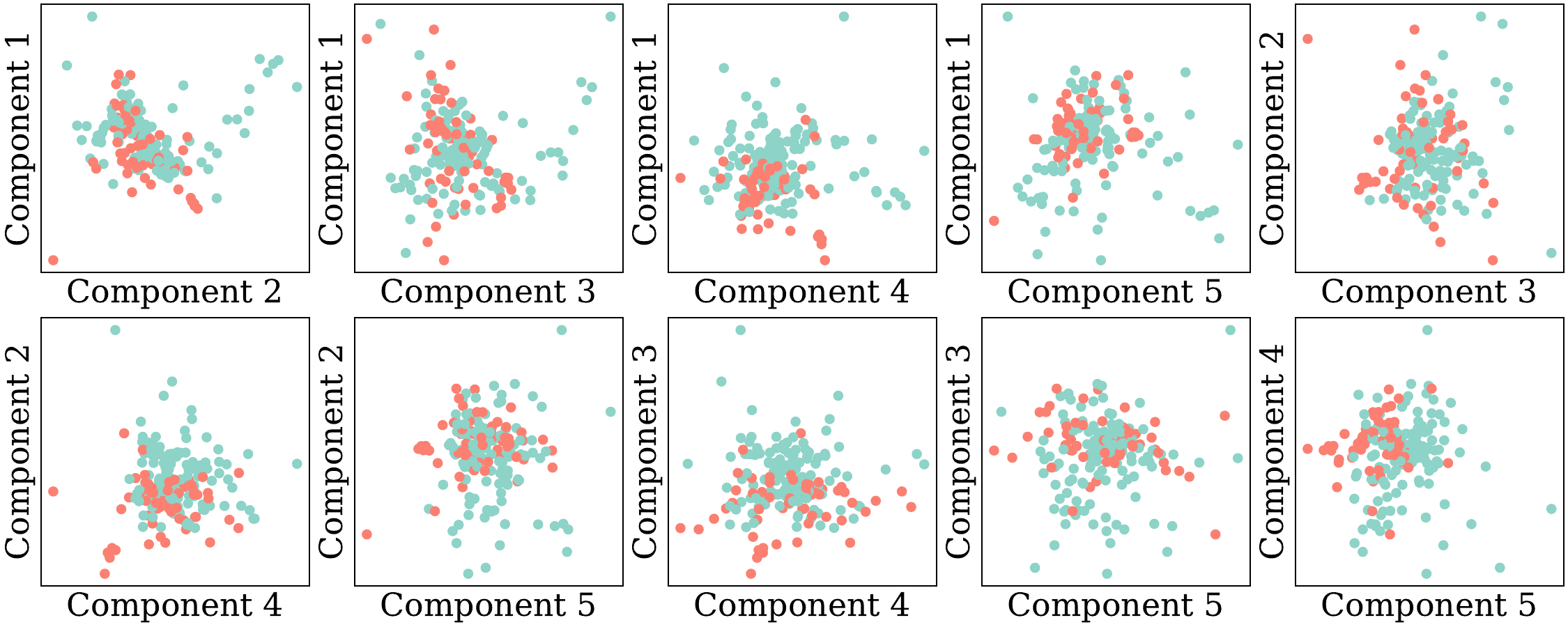}
            \caption{Days 80 - 500}
            \label{fig:80-500_fs}
        \end{subfigure}
     \hfill
        \caption{Representation of the SLSN photospheric spectra in the PCA parameter space in time bins relative to explosions with the phase scaled to match the median e-folding decline time of SLSNe (44 days). Only the first five components for each bin are plotted for visualisation. Events are split into fast and slow categories based on having an e-folding decline time in the light curve of above or below 30 days, and coloured red and blue respectively. }
        \label{fig:PCA_components_exp_fs}
    \end{center}
\end{minipage}
\end{figure*}

In Section \ref{sec:fe_line}, we discussed previous studies that identified spectral differences between fast and slow evolving SLSNe. To investigate whether similar trends appear in our sample, we repeat the analysis from Section \ref{sec:pca}, this time separating the two populations. The fast events are defined as those with an e-folding decline time of less than 30 days from \citet{Gomez2024}, and slow events with a time greater or equal to 30 days.

Figure \ref{fig:PCA_components_peak_fs} and Figure \ref{fig:PCA_components_exp_fs} show the component spectra from this decomposition represented in parameter space. This plots the coefficients of first few the PCA components against one another. Fast events are coloured in red, and slow events in blue. Here we see no correlation between the fast and slow events indicating there is no significant spectroscopic variation between these groups of SLSNe.

\section{The Effect of SN\,2017egm on Clustering}
\label{sec:2017egm}

\begin{figure*}
\centering
\begin{minipage}{.48\textwidth}
\begin{center}
        \begin{subfigure}{1\columnwidth}
            \centering
            \includegraphics[width=1\columnwidth]{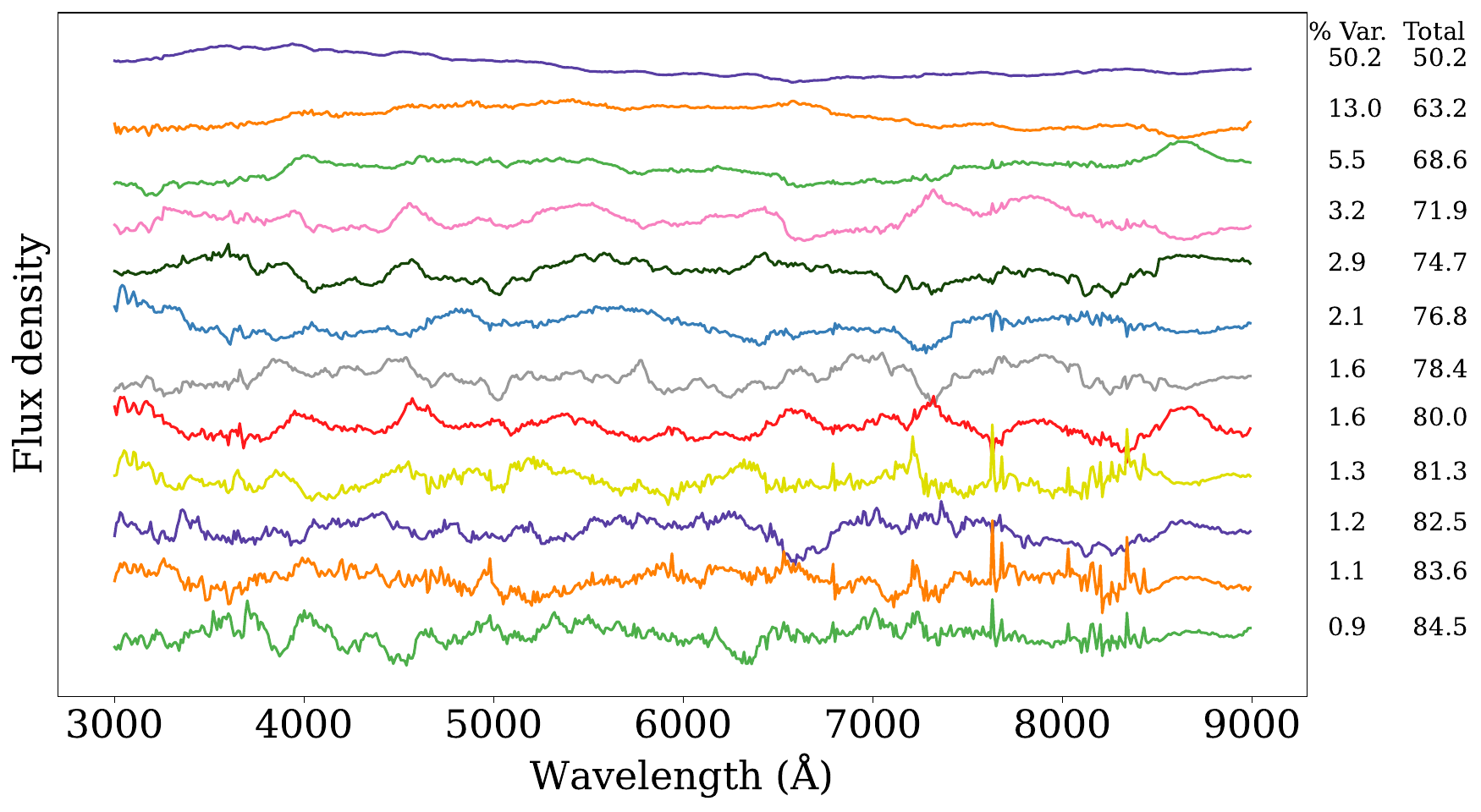}
            \caption{PCA component spectra with the spectra of SN\,2017egm removed, in the time bin 40 - 160 days relative to peak.}
            \label{fig:40-160_17egm}
        \end{subfigure}
    \par\bigskip
    \hfill
        \begin{subfigure}{1\columnwidth}
            \centering
            \includegraphics[width=1\columnwidth]{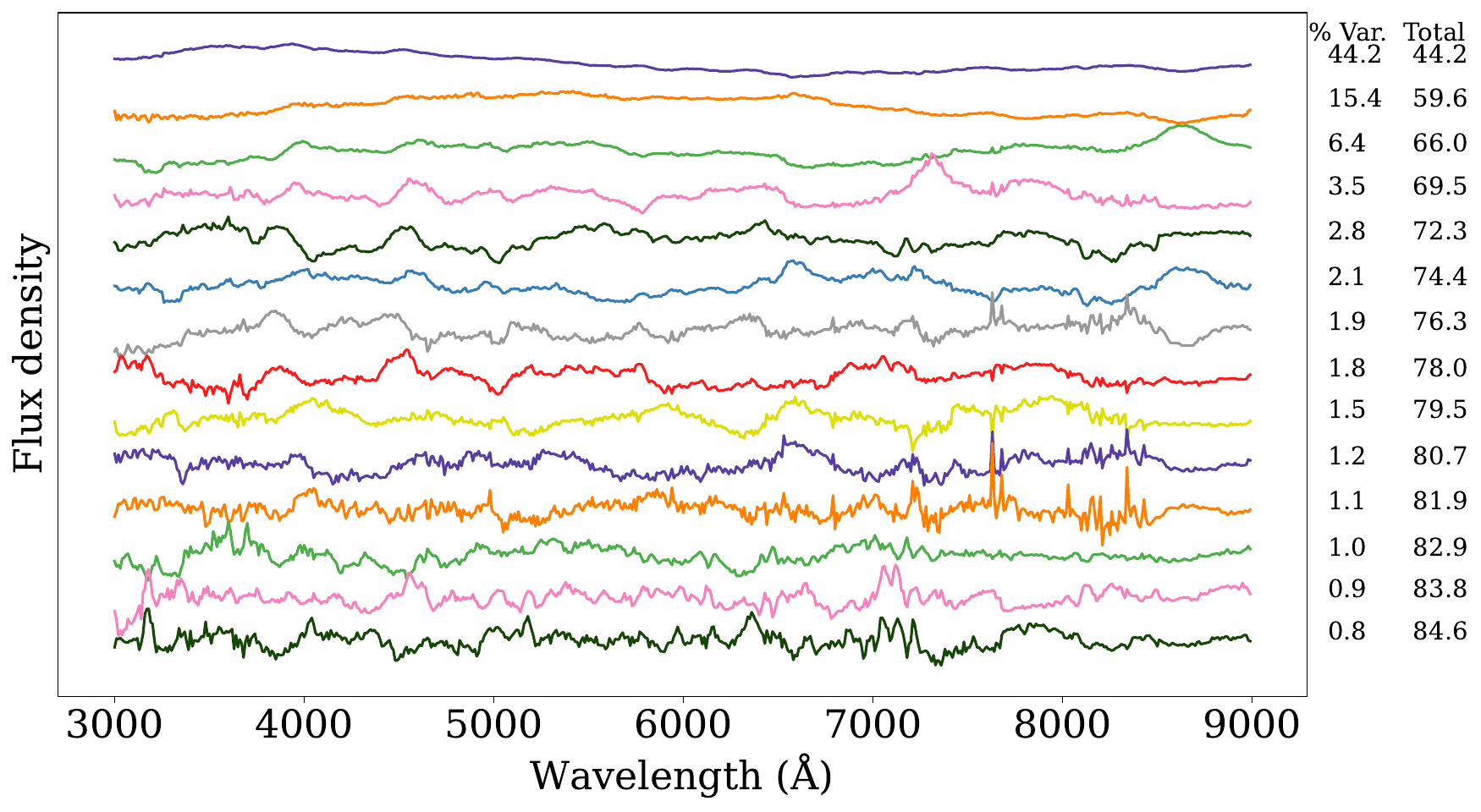}
            \caption{PCA component spectra with the spectra of SN\,2017egm removed, in the time bin 80 - 500 days relative to explosion with the phase scaled to match the median e-folding decline time of SLSNe (44 days).}
            \label{fig:80-500_17egm}
        \end{subfigure}
    \hfill
        \caption{PCA decomposition of SLSN spectra with the spectra of SN\,2017egm removed. Explaining variations up to 85\% of the sample. More components are needed to explain the same level of variation with SN\,2017egm removed supporting the idea that there is more diversity within the population at late times.}
        \label{fig:PCA_components_17egm}
    \end{center}
\end{minipage}%
\hfill
\begin{minipage}{.48\textwidth}
\begin{center}
    \begin{subfigure}{1\columnwidth}
            \centering
            \includegraphics[width=1\columnwidth]{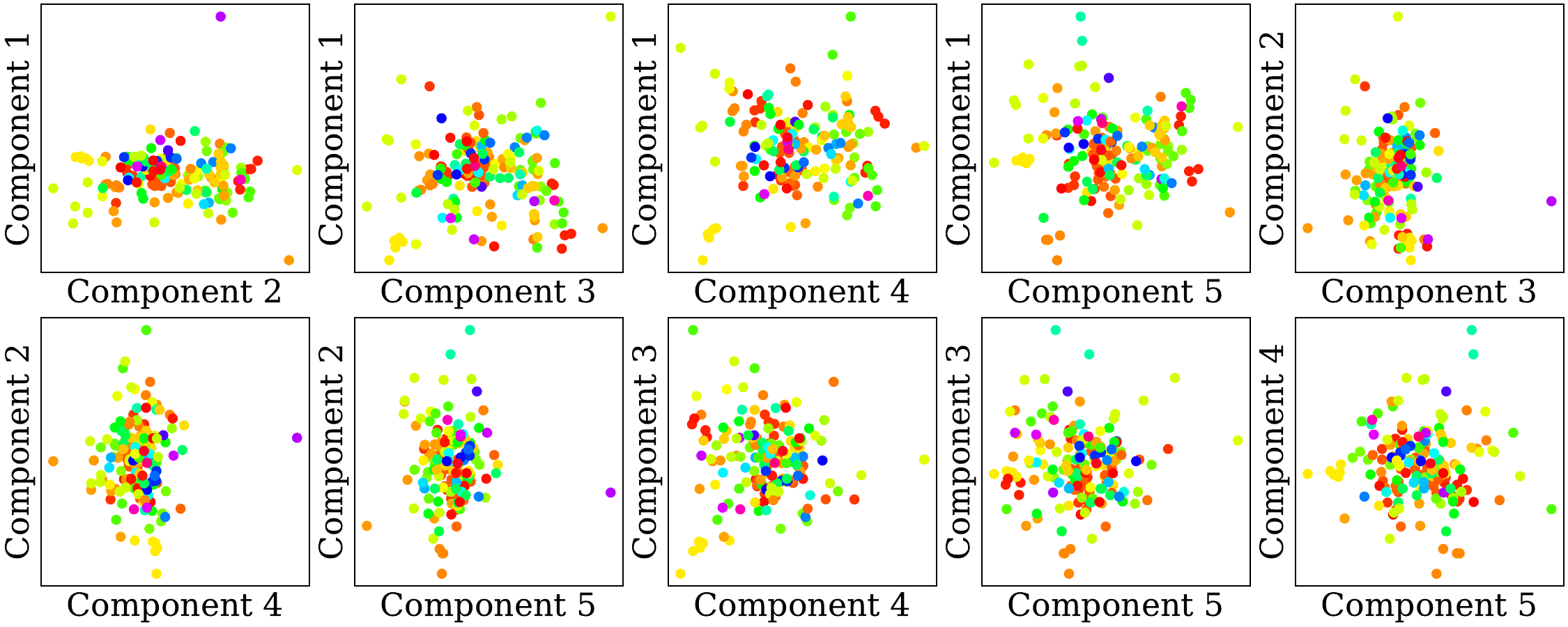}
            \caption{PCA parameter space with the spectra of SN\,2017egm removed, in the time bin 40 - 160 days relative to peak.}
            \label{fig:40-160_17egm_params}
        \end{subfigure}
        
    \par\bigskip
    \hfill
        \begin{subfigure}{1\columnwidth}
            \centering
            \includegraphics[width=1\columnwidth]{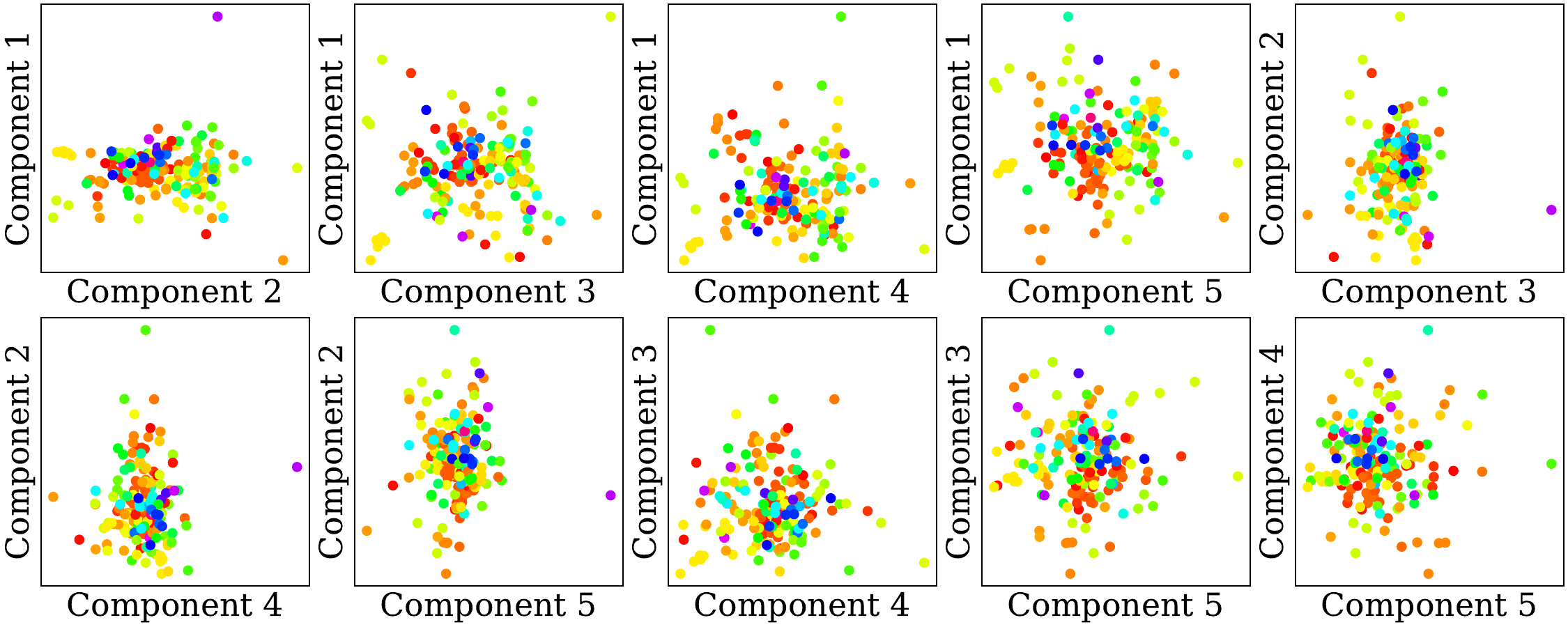}
            \caption{PCA parameter space with the spectra of SN\,2017egm removed. Explaining variations up to 85\% of the sample. More components are needed to explain the same level of variation with SN\,2017egm removed supporting the idea that there is more diversity within the population at late times.}
            \label{fig:80-500_17egm_params}
        \end{subfigure}
    \hfill
        \caption{Representation of the SLSNe photospheric spectra in the PCA parameter space without SN\,2017egm. Only the first five components for each bin are plotted for ease of visualisation. Each SLSN event is marked by a unique colour, consistent across all panels as well as Figure \ref{fig:PCA_components_peak} and Figure \ref{fig:PCA_components_exp}.}
        \label{fig:PCA_params_17egm}
    \end{center}
\end{minipage}
\end{figure*}

\begin{center}
\begin{table}
    \begin{tabular}{|P{2.5cm}|P{2cm}|P{2.8cm}|} 
    \hline\hline
    Time bin (days) & Components & Components\\ 
     & (all) & (without SN\,2017egm) \\
    \hline
    40 \hspace{0.1em} -- \hspace{0.1em} 160 & 8 & 12 \\ 
    80 \hspace{0.1em} -- \hspace{0.1em} 500 (scaled) & 8 & 14 \\ 
    \hline\hline
    \end{tabular}

    \caption{The bins used for the PCA analysis using phases from peak and scaled phases from explosion, as well as the corresponding number of components needed to explain 85\% of the variation with and without the spectra of SN\,2017egm included.}
    \label{tab:pca_17egm}
\end{table}
\end{center}

In Section \ref{sec:pca}, we performed PCA on different time bins in order to investigate spectral features and diversity of the population. We saw that at late times, the spectra of SN\,2017egm deviated from the central clustering which was apparent when comparing the spectra to the average spectra generated. This resulted in the K-Means clustering favouring two separate clusters for the final time bin when grouping by scaled phase from explosion. 

Here we present the same analysis with the spectra from SN\,2017egm removed from the final time bins in Section \ref{sec:pca}. This resulted in 170 spectra for the grouped by phase from peak bin, and 173 spectra in the grouped by scaled phases from explosion bin. Compared to their counterparts in \ref{sec:component_specs}, we can see many more components are needed to explain the same level of variation as shown in Table \ref{tab:pca_17egm}. However, the latter components in both binning schemes contribute $\lesssim 1\%$ to the variation level. These smaller components also have few discernible features, appearing to be dominated by noise. This may signify the search for variation is limited by the noise in our data sample at this point. The largest component in both panels of Figure \ref{fig:PCA_components_17egm} is a general blackbody shape and represents $\gtrsim 40\%$ of the variation signifying the largest variation is due to the underlying blackbody. The increase in number of components needed with the spectra of SN\,2017egm removed, reinforces the idea that the spectra of SN\,2017egm were driving the diversity in the original sample.




\bsp	
\label{lastpage}
\end{document}

%% file: Figures/spectra_data_latex.txt
\begin{longtable}{llp{5cm}p{5cm}p{3cm}} \\
\caption{Phases of spectra used in the analysis of this paper with both phases from peak, and phases from explosion presented. Superscript $^{a}$ denotes extra levels of processing where telluric removal was rerun on smoothed and rebinned spectra. Superscript $^{b}$ denotes where host galaxy line removal was not performed, and superscript $^{c}$ indicates spectra that had their edges cut to remove noise. Superscript $^{d}$ indicates where spectra had chip gaps or noisy middles which were masked, and superscript $^{e}$ denotes spectra that were not used in the analysis of this paper.} \\
\hline \hline \\
Object & Redshift & Phase from Peak & Phase from Explosion & Source \\
\hline \\
\endhead
1991D&0.04179&29.17$^{}$, \enspace43.04$^{}$&45.71$^{}$, \enspace59.58$^{}$ & \cite{Matheson2001} \\
1999as&0.127&0.98$^{}$, \enspace10.75$^{}$, \enspace10.75$^{}$, \enspace31.15$^{}$, \enspace36.48$^{}$&31.18$^{}$, \enspace40.94$^{}$, \enspace40.94$^{}$, \enspace61.35$^{}$, \enspace66.67$^{}$ & \\
2002gh&0.3653&35.59$^{}$, \enspace58.29$^{}$, \enspace61.96$^{}$, \enspace-0.3$^{}$, \enspace7.02$^{}$, \enspace9.95$^{}$, \enspace16.55$^{}$, \enspace19.48$^{}$, \enspace20.21$^{}$&101.15$^{}$, \enspace123.86$^{}$, \enspace127.52$^{}$, \enspace65.26$^{}$, \enspace72.58$^{}$, \enspace75.51$^{}$, \enspace82.11$^{}$, \enspace85.04$^{}$, \enspace85.77$^{}$ & \\
2005ap&0.2832&-4.13$^{}$, \enspace2.88$^{}$, \enspace26.26$^{}$&10.06$^{}$, \enspace17.07$^{}$, \enspace40.45$^{}$ & \cite{Quimby2007, Gal-Yam2007}\\
2006oz&0.376&-17.65$^{ace}$&23.7$^{ace}$ & \cite{Leloudas2012}\\
2007bi&0.1279&51.13$^{ac}$, \enspace52.9$^{}$, \enspace99.01$^{}$, \enspace130.92$^{}$, \enspace45.46$^{}$, \enspace45.81$^{}$&104.68$^{ac}$, \enspace106.45$^{}$, \enspace152.56$^{}$, \enspace184.48$^{}$, \enspace99.01$^{}$, \enspace99.36$^{}$ & \cite{Shivvers2019, Gal-Yam2009, Young2010}\\
2009cb&0.1867&15.66$^{}$&33.28$^{}$ & \cite{Quimby2018}\\
2009jh&0.3499&77.5$^{}$, \enspace117.51$^{}$, \enspace-8.43$^{}$, \enspace-8.43$^{}$, \enspace13.05$^{}$, \enspace30.83$^{}$&135.16$^{}$, \enspace175.16$^{}$, \enspace49.23$^{}$, \enspace49.23$^{}$, \enspace70.71$^{}$, \enspace88.49$^{}$ & \cite{Quimby2011, Quimby2018}\\
2010gx&0.2297&-7.34$^{}$, \enspace-4.9$^{}$, \enspace-4.08$^{ad}$, \enspace4.05$^{}$, \enspace9.74$^{}$, \enspace10.56$^{}$, \enspace21.13$^{acd}$, \enspace29.26$^{}$, \enspace33.33$^{}$, \enspace56.91$^{ac}$, \enspace66.67$^{}$&16.08$^{}$, \enspace18.52$^{}$, \enspace19.33$^{ad}$, \enspace27.46$^{}$, \enspace33.15$^{}$, \enspace33.97$^{}$, \enspace44.54$^{acd}$, \enspace52.67$^{}$, \enspace56.74$^{}$, \enspace80.32$^{ac}$, \enspace90.08$^{}$ & \cite{Quimby2011, Quimby2018, Pastorello2010}\\
2010hy&0.1901&16.78$^{}$, \enspace38.66$^{}$, \enspace40.12$^{}$, \enspace60.47$^{}$&56.04$^{}$, \enspace77.92$^{}$, \enspace79.38$^{}$, \enspace99.73$^{}$ & \cite{Quimby2018, Shivvers2019}\\
2010kd&0.101&83.0$^{}$, \enspace126.58$^{}$, \enspace-29.42$^{}$, \enspace-25.81$^{}$, \enspace-24.91$^{}$, \enspace-23.97$^{}$, \enspace9.53$^{}$, \enspace26.75$^{}$&134.88$^{}$, \enspace178.47$^{}$, \enspace22.46$^{}$, \enspace26.07$^{}$, \enspace26.97$^{}$, \enspace27.91$^{}$, \enspace61.41$^{}$, \enspace78.63$^{}$ & \cite{Kumar2020}\\
2010md&0.0987&67.34$^{ac}$, \enspace74.62$^{}$, \enspace79.18$^{}$, \enspace80.09$^{}$, \enspace100.11$^{}$, \enspace126.5$^{}$, \enspace-22.76$^{}$, \enspace-3.65$^{}$, \enspace20.01$^{}$, \enspace20.92$^{}$, \enspace26.39$^{}$, \enspace30.94$^{}$, \enspace31.85$^{}$, \enspace51.87$^{}$&107.96$^{ac}$, \enspace115.25$^{}$, \enspace119.8$^{}$, \enspace120.71$^{}$, \enspace140.73$^{}$, \enspace167.12$^{}$, \enspace17.86$^{}$, \enspace36.97$^{}$, \enspace60.64$^{}$, \enspace61.55$^{}$, \enspace67.01$^{}$, \enspace71.56$^{}$, \enspace72.47$^{}$, \enspace92.49$^{}$ & \cite{Inserra2013b, Quimby2018, Shivvers2019}\\
2011ke&0.1428&10.96$^{}$, \enspace12.71$^{}$, \enspace14.46$^{}$, \enspace20.59$^{}$, \enspace23.21$^{}$, \enspace27.59$^{}$, \enspace29.34$^{}$, \enspace29.34$^{}$, \enspace34.59$^{}$, \enspace36.34$^{}$, \enspace44.22$^{}$, \enspace50.34$^{ac}$, \enspace56.47$^{}$&36.88$^{}$, \enspace38.63$^{}$, \enspace40.38$^{}$, \enspace46.51$^{}$, \enspace49.13$^{}$, \enspace53.51$^{}$, \enspace55.26$^{}$, \enspace55.26$^{}$, \enspace60.51$^{}$, \enspace62.26$^{}$, \enspace70.13$^{}$, \enspace76.26$^{ac}$, \enspace82.39$^{}$ & \cite{Inserra2013b, Quimby2018} \\
2011kf&0.245&17.33$^{}$, \enspace43.83$^{}$&42.44$^{}$, \enspace68.95$^{}$ & \cite{Inserra2013b}\\
2011kg&0.1924&-11.51$^{}$, \enspace157.05$^{}$, \enspace-8.16$^{}$, \enspace-0.61$^{}$, \enspace6.1$^{}$, \enspace6.1$^{}$, \enspace6.94$^{}$, \enspace13.64$^{a}$, \enspace18.68$^{}$, \enspace43.0$^{}$, \enspace44.67$^{ac}$&18.47$^{}$, \enspace187.03$^{}$, \enspace21.82$^{}$, \enspace29.37$^{}$, \enspace36.08$^{}$, \enspace36.08$^{}$, \enspace36.92$^{}$, \enspace43.63$^{a}$, \enspace48.66$^{}$, \enspace72.98$^{}$, \enspace74.66$^{ac}$ & \cite{Inserra2013b, Quimby2018} \\
2011kl&0.677&-3.97$^{}$&12.06$^{}$ & \\
2012aa&0.083&94.63$^{}$, \enspace-2.37$^{}$, \enspace17.04$^{}$, \enspace36.55$^{}$&132.07$^{}$, \enspace35.07$^{}$, \enspace54.48$^{}$, \enspace74.0$^{}$ & \cite{Shivvers2019}\\
2012il&0.175&12.4$^{}$, \enspace40.49$^{}$, \enspace52.4$^{a}$&32.78$^{}$, \enspace60.87$^{}$, \enspace72.78$^{a}$ & \cite{Inserra2013b}\\
2013dg&0.265&5.02$^{ac}$, \enspace7.39$^{a}$, \enspace8.43$^{}$, \enspace16.11$^{a}$, \enspace23.21$^{a}$, \enspace35.87$^{e}$, \enspace47.71$^{}$&31.92$^{ac}$, \enspace34.29$^{a}$, \enspace35.33$^{}$, \enspace43.01$^{ae}$, \enspace50.11$^{a}$, \enspace62.77$^{e}$, \enspace74.61$^{}$ & \cite{Nicholl2014, Shivvers2019}\\
2013hy&0.663&33.44$^{}$&60.18$^{}$ & \\
2015bn&0.1136&21.56$^{}$, \enspace22.66$^{}$, \enspace22.68$^{}$, \enspace32.34$^{}$, \enspace32.34$^{}$, \enspace33.23$^{}$, \enspace36.98$^{}$, \enspace45.81$^{}$, \enspace52.09$^{}$, \enspace59.28$^{}$, \enspace73.64$^{}$, \enspace85.32$^{}$, \enspace91.6$^{}$, \enspace100.58$^{}$, \enspace107.77$^{}$, \enspace-26.03$^{}$, \enspace-24.91$^{}$, \enspace-18.6$^{ae}$, \enspace-17.7$^{}$, \enspace-17.68$^{a}$, \enspace-15.26$^{a}$, \enspace-6.08$^{ae}$, \enspace-5.29$^{}$, \enspace-5.27$^{}$, \enspace0.01$^{}$, \enspace1.14$^{}$, \enspace1.16$^{ac}$, \enspace2.8$^{ae}$, \enspace5.4$^{}$, \enspace9.22$^{ac}$, \enspace9.24$^{}$&108.27$^{}$, \enspace109.37$^{}$, \enspace109.39$^{}$, \enspace119.05$^{}$, \enspace119.05$^{}$, \enspace119.94$^{}$, \enspace123.69$^{}$, \enspace132.52$^{}$, \enspace138.8$^{}$, \enspace145.99$^{}$, \enspace160.35$^{}$, \enspace172.03$^{}$, \enspace178.31$^{}$, \enspace187.29$^{}$, \enspace194.48$^{}$, \enspace60.68$^{}$, \enspace61.8$^{}$, \enspace68.11$^{ae}$, \enspace69.01$^{}$, \enspace69.03$^{a}$, \enspace71.45$^{a}$, \enspace80.63$^{ae}$, \enspace81.42$^{}$, \enspace81.44$^{}$, \enspace86.72$^{}$, \enspace87.85$^{}$, \enspace87.87$^{ac}$, \enspace89.51$^{ae}$, \enspace92.11$^{}$, \enspace95.93$^{ac}$, \enspace95.95$^{}$ & \cite{Nicholl2016b, Shivvers2019}\\
2016ard&0.2025&79.13$^{}$, \enspace119.88$^{}$, \enspace-4.03$^{}$, \enspace-3.7$^{ac}$, \enspace6.78$^{}$, \enspace24.24$^{}$, \enspace31.73$^{}$&109.63$^{}$, \enspace150.38$^{}$, \enspace26.47$^{}$, \enspace26.8$^{ac}$, \enspace37.28$^{}$, \enspace54.74$^{}$, \enspace62.23$^{}$ & \cite{Blanchard2018b, Chornock2018}\\
2016eay&0.1013&-19.68$^{}$, \enspace150.4$^{}$, \enspace159.27$^{}$, \enspace-16.88$^{}$, \enspace-14.17$^{}$, \enspace-10.53$^{ac}$, \enspace-10.53$^{ac}$, \enspace-8.72$^{}$, \enspace-6.9$^{}$, \enspace-5.08$^{}$, \enspace-4.18$^{}$, \enspace2.18$^{}$, \enspace2.18$^{}$, \enspace12.14$^{a}$, \enspace16.71$^{}$, \enspace20.34$^{}$, \enspace21.2$^{a}$, \enspace25.79$^{}$, \enspace31.17$^{}$, \enspace42.95$^{}$&18.92$^{}$, \enspace189.0$^{}$, \enspace197.87$^{}$, \enspace21.72$^{}$, \enspace24.43$^{}$, \enspace28.07$^{ac}$, \enspace28.07$^{ac}$, \enspace29.88$^{}$, \enspace31.7$^{}$, \enspace33.51$^{}$, \enspace34.42$^{}$, \enspace40.78$^{}$, \enspace40.78$^{}$, \enspace50.74$^{a}$, \enspace55.31$^{}$, \enspace58.94$^{}$, \enspace59.8$^{a}$, \enspace64.39$^{}$, \enspace69.77$^{}$, \enspace81.55$^{}$ & \cite{Kangas2017, Yan2017b, Nicholl2017d}\\
2016inl&0.3057&15.02$^{}$&74.39$^{}$ & \cite{Blanchard2021a}\\
2016wi&0.224&84.64$^{ac}$, \enspace105.07$^{ac}$, \enspace-10.34$^{}$, \enspace7.63$^{}$, \enspace12.2$^{}$, \enspace28.87$^{}$, \enspace33.16$^{ac}$, \enspace34.92$^{}$, \enspace56.11$^{}$&126.92$^{ac}$, \enspace147.36$^{ac}$, \enspace31.94$^{}$, \enspace49.92$^{}$, \enspace54.49$^{}$, \enspace71.16$^{}$, \enspace75.44$^{ac}$, \enspace77.21$^{}$, \enspace98.4$^{}$ & \cite{Yan2017a}\\
2017dwh&0.13&6.88$^{}$&29.63$^{}$ & \cite{Blanchard2018a}\\
2017egm&0.03072&-61.16$^{}$, \enspace-57.25$^{}$, \enspace-56.92$^{}$, \enspace-56.91$^{}$, \enspace-56.3$^{}$, \enspace48.44$^{}$, \enspace-46.23$^{}$, \enspace59.15$^{}$, \enspace-45.3$^{}$, \enspace80.5$^{}$, \enspace80.8$^{}$, \enspace86.31$^{}$, \enspace87.2$^{}$, \enspace90.65$^{}$, \enspace91.1$^{ad}$, \enspace91.17$^{}$, \enspace-41.4$^{}$, \enspace105.2$^{}$, \enspace106.74$^{}$, \enspace-40.18$^{}$, \enspace-40.16$^{ac}$, \enspace112.47$^{}$, \enspace-39.8$^{ac}$, \enspace-39.43$^{}$, \enspace123.63$^{}$, \enspace-38.39$^{e}$, \enspace132.83$^{}$, \enspace138.67$^{}$, \enspace-36.52$^{}$, \enspace-35.32$^{}$, \enspace-32.66$^{}$, \enspace-30.69$^{a}$, \enspace-25.87$^{}$, \enspace-9.43$^{}$, \enspace-51.35$^{}$, \enspace-49.47$^{}$, \enspace-49.08$^{}$&-3.2$^{}$, \enspace0.71$^{}$, \enspace1.04$^{}$, \enspace1.05$^{}$, \enspace1.66$^{}$, \enspace106.4$^{}$, \enspace11.73$^{}$, \enspace117.11$^{}$, \enspace12.66$^{}$, \enspace138.46$^{}$, \enspace138.76$^{}$, \enspace144.27$^{}$, \enspace145.16$^{}$, \enspace148.6$^{}$, \enspace149.06$^{ad}$, \enspace149.13$^{}$, \enspace16.55$^{}$, \enspace163.16$^{}$, \enspace164.7$^{}$, \enspace17.78$^{}$, \enspace17.8$^{ac}$, \enspace170.43$^{}$, \enspace18.16$^{ac}$, \enspace18.53$^{}$, \enspace181.59$^{}$, \enspace19.57$^{e}$, \enspace190.79$^{}$, \enspace196.63$^{}$, \enspace21.44$^{}$, \enspace22.63$^{}$, \enspace25.3$^{}$, \enspace27.27$^{a}$, \enspace32.09$^{}$, \enspace48.53$^{}$, \enspace6.61$^{}$, \enspace8.49$^{}$, \enspace8.88$^{}$ & \cite{Xiang2017, Dong2017, Bose2018, Nicholl2017c, Zhu2023, Lin2023}\\
2017gci&0.087&72.85$^{}$, \enspace103.2$^{}$, \enspace131.03$^{}$, \enspace133.62$^{}$, \enspace155.67$^{}$, \enspace-7.02$^{}$, \enspace-5.17$^{}$, \enspace-3.35$^{}$, \enspace-2.43$^{}$, \enspace3.1$^{e}$, \enspace23.65$^{}$, \enspace29.17$^{e}$, \enspace32.46$^{}$, \enspace37.43$^{e}$, \enspace42.05$^{e}$, \enspace50.87$^{b}$&100.09$^{}$, \enspace130.43$^{}$, \enspace158.26$^{}$, \enspace160.85$^{}$, \enspace182.9$^{}$, \enspace20.21$^{}$, \enspace22.06$^{}$, \enspace23.88$^{}$, \enspace24.8$^{}$, \enspace30.33$^{e}$, \enspace50.88$^{}$, \enspace56.4$^{e}$, \enspace59.69$^{}$, \enspace64.67$^{e}$, \enspace69.28$^{e}$, \enspace78.1$^{b}$ & \cite{Fiore2021}\\
2017hbx&0.1652&9.43$^{b}$&32.05$^{b}$ & \cite{Neill2017} \\
2018avk&0.132&-21.54$^{a}$, \enspace-21.54$^{a}$, \enspace-13.46$^{}$, \enspace-12.7$^{}$&43.93$^{a}$, \enspace43.93$^{a}$, \enspace52.0$^{}$, \enspace52.77$^{}$ & \cite{Nicholl2018c, Lunnan2020}\\
2018beh&0.06&-9.4$^{e}$, \enspace-8.49$^{}$, \enspace-5.49$^{b}$, \enspace-2.87$^{}$, \enspace7.5$^{a}$&28.45$^{e}$, \enspace29.36$^{}$, \enspace32.36$^{b}$, \enspace34.98$^{}$, \enspace45.35$^{a}$ & \cite{Mcbrien2018}\\
2018bgv&0.0795&-1.81$^{ac}$, \enspace0.07$^{}$, \enspace2.85$^{}$, \enspace11.14$^{ac}$, \enspace13.97$^{}$, \enspace13.97$^{}$, \enspace19.44$^{ac}$, \enspace22.31$^{}$, \enspace53.8$^{}$&13.79$^{ac}$, \enspace15.67$^{}$, \enspace18.45$^{}$, \enspace26.74$^{ac}$, \enspace29.57$^{}$, \enspace29.57$^{}$, \enspace35.04$^{ac}$, \enspace37.91$^{}$, \enspace69.4$^{}$ & \cite{Dong2018b, Lunnan2020}\\
2018bsz&0.0267&-12.73$^{}$, \enspace88.39$^{}$, \enspace102.95$^{}$, \enspace102.98$^{}$, \enspace-10.72$^{ac}$, \enspace-10.71$^{}$, \enspace115.61$^{a}$, \enspace-2.24$^{}$, \enspace5.55$^{}$, \enspace12.37$^{a}$, \enspace19.19$^{a}$, \enspace27.95$^{ad}$, \enspace33.8$^{e}$, \enspace59.16$^{}$, \enspace59.19$^{}$, \enspace60.16$^{}$, \enspace60.2$^{a}$, \enspace68.91$^{}$, \enspace68.95$^{a}$&11.1$^{}$, \enspace112.21$^{}$, \enspace126.78$^{}$, \enspace126.81$^{}$, \enspace13.1$^{ac}$, \enspace13.12$^{}$, \enspace139.44$^{a}$, \enspace21.58$^{}$, \enspace29.38$^{}$, \enspace36.19$^{a}$, \enspace43.01$^{a}$, \enspace51.78$^{ad}$, \enspace57.62$^{e}$, \enspace82.98$^{}$, \enspace83.01$^{}$, \enspace83.99$^{}$, \enspace84.03$^{a}$, \enspace92.74$^{}$, \enspace92.77$^{a}$ & \cite{Hiramatsu2018, Clark2018, Anderson2018, Chen2021, Pursiainen2022}\\
2018bym&0.274&73.76$^{}$, \enspace115.99$^{ac}$, \enspace116.86$^{e}$, \enspace-5.75$^{}$, \enspace3.67$^{a}$, \enspace3.67$^{a}$, \enspace7.59$^{}$, \enspace27.21$^{}$&115.84$^{}$, \enspace158.07$^{ac}$, \enspace158.94$^{e}$, \enspace36.33$^{}$, \enspace45.75$^{a}$, \enspace45.75$^{a}$, \enspace49.67$^{}$, \enspace69.29$^{}$ & \cite{Lunnan2020, Hosseinzadeh2022}\\
2018cxa&0.19&19.83$^{ac}$, \enspace20.15$^{ac}$, \enspace71.41$^{e}$&43.26$^{ac}$, \enspace43.58$^{ac}$, \enspace94.84$^{e}$ & \cite{Gomez2021i}\\
2018don&0.0734&36.03$^{}$, \enspace-21.73$^{}$, \enspace-8.69$^{}$, \enspace0.62$^{}$, \enspace0.62$^{}$, \enspace1.56$^{ac}$, \enspace15.53$^{}$&108.99$^{}$, \enspace51.23$^{}$, \enspace64.27$^{}$, \enspace73.59$^{}$, \enspace73.59$^{}$, \enspace74.52$^{ac}$, \enspace88.49$^{}$ & \cite{Lunnan2018b, Fremling2019c}\\
2018fcg&0.1011&-6.42$^{}$, \enspace11.15$^{a}$, \enspace60.02$^{a}$&14.87$^{}$, \enspace32.44$^{a}$, \enspace81.31$^{a}$ & \\
2018fd&0.263&56.7$^{e}$&167.23$^{e}$ & \cite{Gomez2021i}\\
2018ffs&0.141&-8.85$^{e}$, \enspace49.67$^{a}$&32.64$^{e}$, \enspace91.16$^{a}$ & \cite{Fremling2018c}\\
2018gbw&0.3454&-2.74$^{}$&29.43$^{}$ & \\
2018gft&0.232&-44.96$^{}$&31.26$^{}$ & \\
2018gkz&0.2405&12.87$^{b}$, \enspace26.2$^{}$&81.51$^{b}$, \enspace94.84$^{}$ & \cite{Fremling2018b}\\
2018hpq&0.124&14.98$^{}$&65.47$^{}$ & \\
2018hti&0.0612&47.34$^{}$, \enspace49.38$^{}$, \enspace49.4$^{}$, \enspace61.46$^{ac}$, \enspace64.87$^{}$, \enspace68.95$^{}$, \enspace69.89$^{}$, \enspace72.22$^{}$, \enspace81.87$^{}$, \enspace84.42$^{}$, \enspace101.39$^{}$, \enspace-37.12$^{}$, \enspace-34.53$^{}$, \enspace-32.89$^{}$, \enspace-30.63$^{}$, \enspace-30.59$^{}$, \enspace-27.72$^{a}$, \enspace-26.36$^{}$, \enspace-21.79$^{}$, \enspace-18.51$^{}$, \enspace-17.14$^{}$, \enspace-16.88$^{}$, \enspace-15.71$^{}$, \enspace-11.25$^{}$, \enspace-10.51$^{}$, \enspace-8.89$^{}$, \enspace-8.87$^{a}$, \enspace-1.39$^{ac}$, \enspace-1.38$^{a}$, \enspace4.54$^{}$, \enspace5.69$^{}$, \enspace9.34$^{}$, \enspace10.53$^{}$, \enspace12.59$^{}$, \enspace16.36$^{}$, \enspace17.6$^{}$, \enspace18.87$^{}$, \enspace19.06$^{}$, \enspace20.25$^{}$, \enspace20.26$^{}$, \enspace24.2$^{}$, \enspace25.67$^{}$, \enspace26.36$^{}$, \enspace32.73$^{}$, \enspace34.13$^{}$, \enspace38.03$^{}$, \enspace44.16$^{}$&102.51$^{}$, \enspace104.55$^{}$, \enspace104.56$^{}$, \enspace116.62$^{ac}$, \enspace120.03$^{}$, \enspace124.11$^{}$, \enspace125.05$^{}$, \enspace127.38$^{}$, \enspace137.03$^{}$, \enspace139.58$^{}$, \enspace156.56$^{}$, \enspace18.04$^{}$, \enspace20.64$^{}$, \enspace22.27$^{}$, \enspace24.54$^{}$, \enspace24.58$^{}$, \enspace27.45$^{a}$, \enspace28.8$^{}$, \enspace33.37$^{}$, \enspace36.66$^{}$, \enspace38.02$^{}$, \enspace38.28$^{}$, \enspace39.45$^{}$, \enspace43.91$^{}$, \enspace44.66$^{}$, \enspace46.27$^{}$, \enspace46.29$^{a}$, \enspace53.77$^{ac}$, \enspace53.79$^{a}$, \enspace59.7$^{}$, \enspace60.85$^{}$, \enspace64.5$^{}$, \enspace65.69$^{}$, \enspace67.75$^{}$, \enspace71.52$^{}$, \enspace72.77$^{}$, \enspace74.04$^{}$, \enspace74.23$^{}$, \enspace75.41$^{}$, \enspace75.43$^{}$, \enspace79.36$^{}$, \enspace80.83$^{}$, \enspace81.52$^{}$, \enspace87.89$^{}$, \enspace89.29$^{}$, \enspace93.2$^{}$, \enspace99.33$^{}$ & \cite{Burke2018, Lin2020, Fiore2022}\\
2018ibb&0.166&-5.51$^{}$, \enspace-5.21$^{}$, \enspace1.57$^{b}$, \enspace4.12$^{b}$, \enspace5.01$^{}$, \enspace6.72$^{}$, \enspace6.74$^{}$, \enspace17.9$^{e}$, \enspace20.47$^{}$, \enspace23.85$^{}$, \enspace27.28$^{}$, \enspace27.3$^{}$, \enspace28.88$^{ae}$, \enspace28.88$^{ae}$, \enspace43.3$^{ac}$, \enspace43.3$^{ac}$, \enspace43.53$^{e}$, \enspace47.75$^{}$, \enspace47.77$^{}$, \enspace56.28$^{a}$, \enspace66.49$^{}$, \enspace70.04$^{}$, \enspace70.06$^{}$, \enspace77.72$^{a}$, \enspace85.53$^{}$&107.75$^{}$, \enspace108.05$^{}$, \enspace114.83$^{b}$, \enspace117.38$^{b}$, \enspace118.27$^{}$, \enspace119.98$^{}$, \enspace120.0$^{}$, \enspace131.16$^{e}$, \enspace133.73$^{}$, \enspace137.11$^{}$, \enspace140.54$^{}$, \enspace140.56$^{}$, \enspace142.14$^{ae}$, \enspace142.14$^{ae}$, \enspace156.56$^{ac}$, \enspace156.56$^{ac}$, \enspace156.79$^{e}$, \enspace161.01$^{}$, \enspace161.03$^{}$, \enspace169.54$^{a}$, \enspace179.75$^{}$, \enspace183.3$^{}$, \enspace183.32$^{}$, \enspace190.98$^{a}$, \enspace198.79$^{}$ & \cite{Fremling2018a, Schulze2024}\\
2018kyt&0.108&9.49$^{a}$, \enspace49.4$^{}$&46.95$^{a}$, \enspace86.86$^{}$ & \cite{Fremling2019h}\\
2018lfd&0.2686&-4.33$^{}$&36.55$^{}$ & \\
2018lfe&0.35&82.63$^{a}$, \enspace32.93$^{}$, \enspace33.86$^{ac}$, \enspace39.84$^{ac}$, \enspace57.37$^{ad}$, \enspace64.19$^{e}$&111.89$^{a}$, \enspace62.19$^{}$, \enspace63.12$^{ac}$, \enspace69.1$^{ac}$, \enspace86.63$^{ad}$, \enspace93.45$^{e}$ & \cite{Gomez2019b}\\
2018lzv&0.434&-19.74$^{}$, \enspace-19.74$^{}$&31.21$^{}$, \enspace31.21$^{}$ & \cite{Perley2022d}\\
2018lzw&0.3198&31.36$^{}$, \enspace31.36$^{}$&79.88$^{}$, \enspace79.88$^{}$ & \cite{Yan2022b}\\
2018lzx&0.4373&20.09$^{}$&108.55$^{}$ & \\
2019J&0.1346&18.13$^{}$, \enspace18.31$^{}$&66.49$^{}$, \enspace66.67$^{}$ & \cite{Fremling2019f}\\
2019aamp&0.404&-3.6$^{b}$, \enspace10.65$^{}$&25.38$^{b}$, \enspace39.63$^{}$ & \cite{Yan2022a}\\
2019aamq&0.386&2.55$^{}$&75.19$^{}$ & \\
2019aamr&0.42&5.32$^{}$&29.68$^{}$ & \\
2019aams&0.636&4.63$^{}$&31.56$^{}$ & \\
2019aamt&0.2138&16.31$^{}$&51.19$^{}$ & \\
2019aamu&0.259&16.07$^{}$&82.49$^{}$ & \\
2019aamv&0.3996&-6.65$^{}$&62.35$^{}$ & \\
2019aamw&0.22&140.85$^{}$&198.48$^{}$ & \\
2019aamx&0.41&22.45$^{}$&67.94$^{}$ & \\
2019bgu&0.148&18.82$^{}$&43.78$^{}$ & \\
2019cca&0.4103&31.38$^{}$&66.5$^{}$ & \\
2019cdt&0.153&12.71$^{}$, \enspace12.86$^{}$&37.88$^{}$, \enspace38.03$^{}$ & \cite{Fremling2019g}\\
2019cwu&0.32&-18.56$^{}$, \enspace23.86$^{a}$&27.42$^{}$, \enspace69.84$^{a}$ & \cite{Smith2019a}\\
2019dgr&0.3815&17.09$^{}$&46.35$^{}$ & \\
2019dlr&0.26&60.71$^{c}$&101.71$^{c}$ & \\
2019dwa&0.082&3.49$^{b}$&37.8$^{b}$ & \cite{Fremling2019e}\\
2019eot&0.3057&-11.84$^{}$, \enspace-11.52$^{}$&39.84$^{}$, \enspace40.16$^{}$ & \cite{Fremling2019d}\\
2019gam&0.1235&-7.66$^{}$&61.98$^{}$ & \\
2019gfm&0.18167&-8.74$^{}$, \enspace-6.05$^{}$, \enspace13.26$^{}$, \enspace14.99$^{}$, \enspace44.59$^{e}$, \enspace45.41$^{e}$, \enspace56.4$^{e}$&12.12$^{}$, \enspace14.81$^{}$, \enspace34.12$^{}$, \enspace35.85$^{}$, \enspace65.45$^{e}$, \enspace66.27$^{e}$, \enspace77.26$^{e}$ & \cite{Frohmaier2019}\\
2019gqi&0.3642&25.92$^{}$, \enspace31.98$^{e}$, \enspace45.7$^{a}$, \enspace46.43$^{}$, \enspace47.17$^{}$&56.01$^{}$, \enspace62.08$^{e}$, \enspace75.79$^{a}$, \enspace76.52$^{}$, \enspace77.26$^{}$ & \\
2019hge&0.0866&87.94$^{}$, \enspace87.97$^{}$, \enspace97.15$^{}$, \enspace97.18$^{}$, \enspace-19.37$^{}$, \enspace-18.83$^{}$, \enspace29.03$^{}$, \enspace33.74$^{}$, \enspace33.77$^{}$, \enspace33.8$^{a}$, \enspace33.82$^{}$, \enspace33.86$^{a}$, \enspace33.88$^{a}$&146.0$^{}$, \enspace146.03$^{}$, \enspace155.21$^{}$, \enspace155.24$^{}$, \enspace38.69$^{}$, \enspace39.23$^{}$, \enspace87.09$^{}$, \enspace91.8$^{}$, \enspace91.83$^{}$, \enspace91.86$^{a}$, \enspace91.88$^{}$, \enspace91.92$^{a}$, \enspace91.94$^{a}$ & \cite{Dahiwale2019a, Prentice2019b, Prentice2021}\\
2019hno&0.26&3.15$^{}$&32.66$^{}$ & \\
2019ieh&0.032&19.33$^{b}$, \enspace-2.89$^{}$&32.21$^{b}$, \enspace9.99$^{}$ & \cite{Dahiwale2019b, Zheng2019}\\
2019itq&0.481&7.92$^{a}$, \enspace10.72$^{a}$, \enspace13.87$^{a}$, \enspace13.95$^{a}$, \enspace28.92$^{a}$&39.76$^{a}$, \enspace42.56$^{a}$, \enspace45.71$^{a}$, \enspace45.79$^{a}$, \enspace60.77$^{a}$ & \cite{Gomez2021h}\\
2019kcy&0.399&-18.78$^{}$, \enspace-3.04$^{a}$, \enspace-3.02$^{a}$, \enspace1.27$^{a}$, \enspace16.94$^{}$, \enspace16.95$^{}$&27.78$^{}$, \enspace43.53$^{a}$, \enspace43.55$^{a}$, \enspace47.84$^{a}$, \enspace63.51$^{}$, \enspace63.52$^{}$ & \cite{Perley2019c}\\
2019kwq&0.49&74.25$^{}$&111.52$^{}$ & \\
2019kws&0.1977&20.54$^{}$&56.72$^{}$ & \\
2019kwt&0.3562&22.89$^{}$&86.67$^{}$ & \\
2019kwu&0.6&27.27$^{}$&52.26$^{}$ & \\
2019lsq&0.1295&-4.46$^{}$, \enspace-4.13$^{}$&32.78$^{}$, \enspace33.12$^{}$ & \cite{Fremling2019a}\\
2019neq&0.1059&-10.39$^{}$, \enspace12.32$^{}$&19.6$^{}$, \enspace42.31$^{}$ & \cite{Perley2019b}\\
2019nhs&0.189&-8.72$^{}$, \enspace-8.4$^{ac}$, \enspace37.68$^{}$, \enspace49.43$^{}$&24.79$^{}$, \enspace25.1$^{ac}$, \enspace71.19$^{}$, \enspace82.94$^{}$ & \cite{Perley2019a}\\
2019obk&0.1656&14.31$^{}$&54.02$^{}$ & \\
2019otl&0.514&80.89$^{a}$, \enspace88.26$^{a}$, \enspace10.87$^{}$&137.58$^{a}$, \enspace144.96$^{a}$, \enspace67.57$^{}$ & \cite{Gomez2021h}\\
2019pud&0.1136&-7.81$^{b}$, \enspace-4.18$^{ac}$, \enspace-4.18$^{ac}$, \enspace53.18$^{}$&17.0$^{b}$, \enspace20.62$^{ac}$, \enspace20.62$^{ac}$, \enspace77.98$^{}$ & \cite{Fremling2019b, Kangas2022}\\
2019pvs&0.167&60.64$^{ac}$, \enspace40.95$^{ac}$&114.94$^{ac}$, \enspace95.24$^{ac}$ & \cite{Gomez2021h}\\
2019qgk&0.3468&41.82$^{}$&73.92$^{}$ & \\
2019sgg&0.5726&-8.43$^{}$&37.61$^{}$ & \\
2019sgh&0.344&15.41$^{a}$, \enspace32.61$^{}$, \enspace36.99$^{}$&36.16$^{a}$, \enspace53.36$^{}$, \enspace57.74$^{}$ & \cite{Gomez2021h}\\
2019szu&0.212&-12.4$^{}$, \enspace-0.87$^{a}$, \enspace-0.06$^{a}$, \enspace-62.8$^{e}$, \enspace-47.86$^{}$, \enspace-44.65$^{}$, \enspace-42.06$^{a}$, \enspace-33.88$^{}$, \enspace-23.95$^{a}$&105.89$^{}$, \enspace117.42$^{a}$, \enspace118.22$^{a}$, \enspace55.49$^{e}$, \enspace70.42$^{}$, \enspace73.64$^{}$, \enspace76.22$^{a}$, \enspace84.41$^{}$, \enspace94.34$^{a}$ & \cite{Nicholl2019a, Aamer2024}\\
2019ujb&0.2008&68.03$^{}$, \enspace-4.3$^{a}$, \enspace26.99$^{}$&106.01$^{}$, \enspace33.68$^{a}$, \enspace64.97$^{}$ & \cite{Gomez2021h}\\
2019unb&0.0635&53.19$^{}$, \enspace67.4$^{}$, \enspace68.28$^{}$, \enspace97.37$^{}$, \enspace-34.14$^{}$, \enspace-11.51$^{a}$, \enspace-9.63$^{}$, \enspace7.93$^{}$, \enspace18.59$^{}$, \enspace21.47$^{}$, \enspace25.06$^{ae}$, \enspace35.46$^{e}$, \enspace37.34$^{ac}$, \enspace41.11$^{}$&111.55$^{}$, \enspace125.76$^{}$, \enspace126.64$^{}$, \enspace155.73$^{}$, \enspace24.23$^{}$, \enspace46.86$^{a}$, \enspace48.73$^{}$, \enspace66.29$^{}$, \enspace76.96$^{}$, \enspace79.84$^{}$, \enspace83.43$^{ae}$, \enspace93.82$^{e}$, \enspace95.71$^{ac}$, \enspace99.47$^{}$ & \cite{Prentice2019a, Prentice2021, Dahiwale2020b}\\
2019une&0.1666&72.2$^{e}$&113.57$^{e}$ & \\
2019vvc&0.3314&25.9$^{}$&67.65$^{}$ & \\
2019xaq&0.2&25.43$^{a}$, \enspace25.87$^{a}$&48.47$^{a}$, \enspace48.91$^{a}$ & \cite{Gomez2021h}\\
2019xdy&0.2206&30.79$^{}$&78.31$^{}$ & \\
2019zbv&0.3785&-3.69$^{}$, \enspace6.18$^{}$&38.94$^{}$, \enspace48.81$^{}$ & \cite{Gomez2021h}\\
2019zeu&0.39&31.57$^{}$&58.57$^{}$ & \cite{Gomez2021h}\\
2020abjc&0.219&33.48$^{e}$, \enspace-34.36$^{}$&140.62$^{e}$, \enspace72.78$^{}$ & \cite{Blanchard2020a}\\
2020abjx&0.39&20.09$^{}$&115.32$^{}$ & \\
2020adkm&0.226&32.4$^{a}$, \enspace-13.91$^{}$, \enspace-7.5$^{}$, \enspace3.07$^{}$&106.18$^{a}$, \enspace59.87$^{}$, \enspace66.27$^{}$, \enspace76.85$^{}$ & \cite{Blanchard2021b}\\
2020afag&0.3815&14.16$^{}$&54.98$^{}$ & \\
2020afah&0.3754&-23.8$^{}$&36.67$^{}$ & \\
2020ank&0.2485&-2.6$^{}$, \enspace1.54$^{}$, \enspace7.94$^{}$, \enspace27.9$^{}$&24.63$^{}$, \enspace28.78$^{}$, \enspace35.17$^{}$, \enspace55.13$^{}$ & \cite{Poidevin2020, Dahiwale2020c}\\
2020aup&0.31&14.89$^{}$&35.6$^{}$ & \\
2020auv&0.28&3.49$^{}$&26.58$^{}$ & \\
2020dlb&0.398&0.67$^{}$&46.49$^{}$ & \\
2020exj&0.1216&42.39$^{}$, \enspace42.69$^{ac}$&63.28$^{}$, \enspace63.58$^{ac}$ & \cite{Dahiwale2020a}\\
2020fvm&0.2428&23.84$^{}$&127.81$^{}$ & \\
2020fyq&0.1765&46.35$^{}$&116.57$^{}$ & \\
2020htd&0.3515&10.83$^{}$&78.08$^{}$ & \\
2020iyj&0.369&2.75$^{}$&47.08$^{}$ & \\
2020jii&0.396&2.55$^{}$, \enspace36.22$^{a}$&38.29$^{}$, \enspace71.96$^{a}$ & \cite{Gomez2020c}\\
2020kox&0.456&-6.37$^{}$&41.16$^{}$ & \\
2020myh&0.283&73.61$^{e}$, \enspace115.0$^{ac}$&101.69$^{e}$, \enspace143.09$^{ac}$ & \cite{Gomez2020c}\\
2020onb&0.153&-27.91$^{}$, \enspace24.13$^{}$&18.74$^{}$, \enspace70.78$^{}$ & \cite{Gomez2020c}\\
2020qef&0.1831&7.63$^{a}$, \enspace30.12$^{}$&57.08$^{a}$, \enspace79.57$^{}$ & \cite{Terreran2020c}\\
2020qlb&0.1585&53.27$^{}$, \enspace-56.88$^{b}$, \enspace110.5$^{}$, \enspace-26.33$^{}$, \enspace3.56$^{b}$, \enspace24.97$^{ac}$&127.83$^{}$, \enspace17.67$^{b}$, \enspace185.06$^{}$, \enspace48.23$^{}$, \enspace78.11$^{b}$, \enspace99.53$^{ac}$ & \cite{Perez-Fournon2020, West2023}\\
2020rmv&0.2621&56.06$^{}$, \enspace-8.18$^{}$, \enspace12.42$^{}$&114.14$^{}$, \enspace49.9$^{}$, \enspace70.5$^{}$ & \cite{Terreran2020b}\\
2020tcw&0.064&2.83$^{b}$&39.94$^{b}$ & \cite{Perley2020b}\\
2020uew&0.22367&77.36$^{ae}$, \enspace95.27$^{a}$, \enspace119.76$^{}$, \enspace9.54$^{}$, \enspace15.22$^{}$, \enspace17.02$^{}$, \enspace28.4$^{a}$, \enspace32.59$^{}$, \enspace47.99$^{a}$, \enspace54.54$^{a}$, \enspace66.74$^{}$&103.23$^{ae}$, \enspace121.14$^{a}$, \enspace145.63$^{}$, \enspace35.41$^{}$, \enspace41.09$^{}$, \enspace42.89$^{}$, \enspace54.27$^{a}$, \enspace58.46$^{}$, \enspace73.86$^{a}$, \enspace80.41$^{a}$, \enspace92.61$^{}$ & \cite{Ihanec2020, Jaeger2020}\\
2020vpg&0.257&10.42$^{}$&54.91$^{}$ & \cite{Terreran2020a}\\
2020wnt&0.032&33.52$^{}$, \enspace41.08$^{}$, \enspace47.69$^{}$, \enspace51.19$^{ac}$, \enspace52.16$^{ac}$, \enspace52.85$^{}$, \enspace75.16$^{}$, \enspace80.97$^{ac}$, \enspace101.26$^{}$, \enspace-39.83$^{}$, \enspace-38.64$^{}$, \enspace-38.08$^{}$, \enspace-38.08$^{}$, \enspace-31.38$^{}$, \enspace-29.23$^{}$, \enspace-23.75$^{}$, \enspace-20.54$^{}$, \enspace-11.82$^{}$, \enspace-8.96$^{}$, \enspace-7.22$^{}$, \enspace-6.01$^{}$, \enspace-3.41$^{}$, \enspace8.14$^{}$, \enspace8.86$^{}$, \enspace17.07$^{}$, \enspace20.57$^{}$, \enspace21.49$^{}$, \enspace24.67$^{}$, \enspace28.32$^{}$&105.11$^{}$, \enspace112.66$^{}$, \enspace119.28$^{}$, \enspace122.77$^{ac}$, \enspace123.75$^{ac}$, \enspace124.44$^{}$, \enspace146.74$^{}$, \enspace152.56$^{ac}$, \enspace172.85$^{}$, \enspace31.76$^{}$, \enspace32.95$^{}$, \enspace33.51$^{}$, \enspace33.51$^{}$, \enspace40.21$^{}$, \enspace42.35$^{}$, \enspace47.84$^{}$, \enspace51.05$^{}$, \enspace59.77$^{}$, \enspace62.63$^{}$, \enspace64.37$^{}$, \enspace65.58$^{}$, \enspace68.18$^{}$, \enspace79.73$^{}$, \enspace80.45$^{}$, \enspace88.66$^{}$, \enspace92.16$^{}$, \enspace93.08$^{}$, \enspace96.26$^{}$, \enspace99.91$^{}$ & \cite{Tinyanont2020, Gutierrez2022}\\
2020xga&0.44&-4.3$^{}$, \enspace2.79$^{}$, \enspace3.39$^{}$, \enspace17.22$^{}$, \enspace33.21$^{a}$&32.85$^{}$, \enspace39.94$^{}$, \enspace40.54$^{}$, \enspace54.37$^{}$, \enspace70.36$^{a}$ & \cite{Gkini2025}\\
2020xgd&0.454&-1.47$^{}$, \enspace-0.02$^{}$, \enspace8.27$^{a}$, \enspace9.53$^{a}$, \enspace10.32$^{ac}$, \enspace19.18$^{}$&28.94$^{}$, \enspace30.4$^{}$, \enspace38.68$^{a}$, \enspace39.94$^{a}$, \enspace40.74$^{ac}$, \enspace49.6$^{}$ & \cite{Gomez2020a, Weil2020}\\
2020xkv&0.241&-13.52$^{}$&64.41$^{}$ & \\
2020zbf&0.35&-2.66$^{}$, \enspace1.78$^{}$, \enspace18.04$^{e}$, \enspace29.14$^{}$, \enspace38.03$^{a}$, \enspace49.82$^{a}$&23.15$^{}$, \enspace27.58$^{}$, \enspace43.84$^{e}$, \enspace54.94$^{}$, \enspace63.84$^{a}$, \enspace75.63$^{a}$ & \cite{Gkini2024}\\
2020znr&0.1&44.18$^{a}$, \enspace0.69$^{ac}$, \enspace18.91$^{ac}$, \enspace19.79$^{}$, \enspace20.78$^{ac}$, \enspace22.52$^{ac}$, \enspace23.54$^{ac}$&117.77$^{a}$, \enspace74.28$^{ac}$, \enspace92.5$^{ac}$, \enspace93.38$^{}$, \enspace94.37$^{ac}$, \enspace96.11$^{ac}$, \enspace97.13$^{ac}$ & \\
2020zzb&0.1659&5.76$^{ac}$&78.43$^{ac}$ & \\
2021een&0.16&104.07$^{a}$, \enspace31.1$^{}$, \enspace55.06$^{e}$&132.93$^{a}$, \enspace59.96$^{}$, \enspace83.92$^{e}$ & \cite{Dahiwale2021}\\
2021ejo&0.44&39.09$^{a}$&91.86$^{a}$ & \cite{Gomez2021e}\\
2021ek&0.193&-10.47$^{}$, \enspace-6.29$^{}$, \enspace-6.27$^{}$, \enspace7.96$^{}$&13.33$^{}$, \enspace17.5$^{}$, \enspace17.53$^{}$, \enspace31.75$^{}$ & \cite{Srivastav2021}\\
2021fpl&0.121&59.94$^{a}$, \enspace115.07$^{ac}$, \enspace-5.24$^{}$, \enspace9.07$^{}$, \enspace9.09$^{a}$&123.64$^{a}$, \enspace178.77$^{ac}$, \enspace58.46$^{}$, \enspace72.77$^{}$, \enspace72.79$^{a}$ & \cite{Deckers2021, Poidevin2023}\\
2021gtr&0.303&84.56$^{}$&184.8$^{}$ & \cite{Gomez2021g}\\
2021hpc&0.24&15.9$^{a}$&56.56$^{a}$ & \cite{Gomez2021f}\\
2021hpx&0.213&-11.01$^{}$, \enspace12.13$^{}$&36.33$^{}$, \enspace59.46$^{}$ & \cite{Gonzalez2021}\\
2021kty&0.159&-6.45$^{a}$&31.67$^{a}$ & \cite{Yao2021c}\\
2021lwz&0.065&-4.72$^{}$&5.22$^{}$ & \cite{Perley2021}\\
2021mkr&0.28&-5.68$^{a}$, \enspace6.27$^{ad}$&49.11$^{a}$, \enspace61.05$^{ad}$ & \cite{Chu2021b, Poidevin2021}\\
2021nxq&0.15&-26.1$^{}$&49.12$^{}$ & \cite{Weil2021b}\\
2021rwz&0.19&33.39$^{}$&67.36$^{}$ & \cite{Weil2021a}\\
2021txk&0.46&10.43$^{a}$, \enspace11.56$^{}$, \enspace33.58$^{ace}$&40.82$^{a}$, \enspace41.95$^{}$, \enspace63.97$^{ace}$ & \cite{Gomez2021d}\\
2021uvy&0.095&45.95$^{}$, \enspace56.94$^{}$, \enspace65.1$^{}$, \enspace65.14$^{}$, \enspace71.48$^{}$, \enspace91.62$^{ae}$, \enspace125.35$^{}$, \enspace-12.49$^{ac}$, \enspace-12.33$^{ac}$, \enspace1.37$^{}$, \enspace6.85$^{}$, \enspace11.45$^{ac}$, \enspace12.59$^{be}$, \enspace17.73$^{}$, \enspace27.88$^{}$, \enspace35.02$^{ac}$&101.06$^{}$, \enspace112.04$^{}$, \enspace120.21$^{}$, \enspace120.24$^{}$, \enspace126.58$^{}$, \enspace146.72$^{ae}$, \enspace180.46$^{}$, \enspace42.61$^{ac}$, \enspace42.77$^{ac}$, \enspace56.48$^{}$, \enspace61.95$^{}$, \enspace66.56$^{ac}$, \enspace67.7$^{be}$, \enspace72.83$^{}$, \enspace82.99$^{}$, \enspace90.12$^{ac}$ & \cite{Chu2021a, Ridley2021}\\
2021vuw&0.2&61.53$^{a}$, \enspace141.44$^{}$, \enspace22.11$^{a}$, \enspace22.51$^{a}$, \enspace46.59$^{a}$&107.62$^{a}$, \enspace187.53$^{}$, \enspace68.2$^{a}$, \enspace68.6$^{a}$, \enspace92.68$^{a}$ & \cite{Gomez2021a}\\
2021xfu&0.32&110.27$^{a}$, \enspace-19.66$^{}$, \enspace4.68$^{}$&172.39$^{ae}$, \enspace42.47$^{}$, \enspace66.81$^{}$ & \cite{Gomez2021c}\\
2021ybf&0.13&11.47$^{a}$, \enspace15.0$^{}$&61.44$^{a}$, \enspace64.97$^{}$ & \cite{Bruch2021}\\
2021ynn&0.22&84.19$^{ac}$, \enspace45.55$^{ac}$, \enspace46.55$^{ac}$&110.18$^{ac}$, \enspace71.54$^{ac}$, \enspace72.55$^{ac}$ & \cite{Gomez2023b}\\
2021yrp&0.3&3.47$^{a}$&38.5$^{a}$ & \cite{Gomez2021b}\\
2021zcl&0.117&56.06$^{}$, \enspace63.34$^{a}$, \enspace89.22$^{a}$, \enspace-0.25$^{a}$, \enspace1.6$^{}$, \enspace7.84$^{a}$, \enspace10.71$^{a}$, \enspace26.69$^{a}$, \enspace27.57$^{a}$, \enspace33.8$^{}$, \enspace41.81$^{a}$&104.26$^{}$, \enspace111.54$^{a}$, \enspace137.42$^{a}$, \enspace47.95$^{a}$, \enspace49.8$^{}$, \enspace56.04$^{a}$, \enspace58.91$^{a}$, \enspace74.89$^{a}$, \enspace75.77$^{a}$, \enspace82.0$^{}$, \enspace90.01$^{a}$ & \cite{Gromadzki2021}\\
2022aawb&0.13&-11.83$^{}$&34.51$^{}$ & \cite{Poidevin2022a}\\
2022abdu&0.13&0.23$^{a}$, \enspace1.06$^{a}$, \enspace1.09$^{a}$, \enspace17.92$^{a}$, \enspace25.86$^{e}$, \enspace38.25$^{a}$, \enspace52.35$^{ac}$&18.23$^{a}$, \enspace19.07$^{a}$, \enspace19.1$^{a}$, \enspace35.93$^{a}$, \enspace43.87$^{e}$, \enspace56.25$^{a}$, \enspace70.36$^{ac}$ & \cite{Grzesiak2022}\\
2022ful&0.15&-4.5$^{ab}$, \enspace31.67$^{}$&35.57$^{ab}$, \enspace71.74$^{}$ & \cite{Chu2022, Sollerman2022}\\
2022gyv&0.39&5.47$^{}$&64.6$^{}$ & \cite{Poidevin2022b}\\
2022le&0.2491&-49.0$^{a}$, \enspace-47.97$^{a}$, \enspace-36.91$^{ac}$&174.09$^{a}$, \enspace175.12$^{a}$, \enspace186.18$^{ac}$ & \cite{Gomez2022b}\\
2022ljr&0.2&13.2$^{}$&36.15$^{}$ & \cite{Davis2022}\\
2022lxd&0.54&-8.06$^{a}$, \enspace1.93$^{a}$&30.36$^{a}$, \enspace40.34$^{a}$ & \cite{Angus2022}\\
2022npq&0.26&-4.1$^{a}$&57.94$^{a}$ & \cite{Ayala2022}\\
2022ojm&0.47&-5.03$^{}$&41.6$^{}$ & \cite{Perez-Fournon2022}\\
2022pjq&0.17&-0.93$^{}$, \enspace27.69$^{ac}$&18.5$^{}$, \enspace47.12$^{ac}$ & \cite{Fulton2022}\\
2022ued&0.1087&84.04$^{}$, \enspace35.91$^{ad}$&127.03$^{}$, \enspace78.9$^{ad}$ & \cite{Perley2022a}\\
2022vxc&0.1&-26.47$^{ac}$, \enspace-23.75$^{}$, \enspace-54.61$^{}$, \enspace-51.94$^{a}$&117.15$^{ac}$, \enspace119.87$^{}$, \enspace89.01$^{}$, \enspace91.67$^{a}$ & \\
CSS140925&0.46&8.51$^{}$, \enspace16.79$^{}$, \enspace16.81$^{}$, \enspace16.83$^{}$&32.75$^{}$, \enspace41.03$^{}$, \enspace41.06$^{}$, \enspace41.08$^{}$ & \cite{Campbell2014}\\
DES14C1fi&1.302&10.02$^{e}$&43.51$^{e}$ & \cite{Yaron2023}\\
DES14C1rhg&0.481&4.54$^{}$&24.44$^{}$ & \\
DES14E2slp&0.57&-15.52$^{a}$&12.28$^{a}$ & \\
DES14S2qri&1.5&9.98$^{}$&27.66$^{}$ & \\
DES14X2byo&0.868&-2.13$^{ac}$&19.09$^{ac}$ & \\
DES14X3taz&0.608&-21.7$^{}$, \enspace-15.39$^{}$&10.71$^{}$, \enspace17.02$^{}$ & \\
DES15C3hav&0.392&-1.1$^{ac}$&25.19$^{ac}$ & \cite{Smith2016}\\
DES15E2mlf&1.861&-2.79$^{}$, \enspace0.32$^{}$, \enspace10.83$^{}$&12.9$^{}$, \enspace16.01$^{}$, \enspace26.52$^{}$ & \cite{Pan2017}\\
DES15S1nog&0.565&32.79$^{ac}$&55.01$^{ac}$ & \\
DES15X1noe&1.188&-9.82$^{}$&36.04$^{}$ & \\
DES15X3hm&0.86&19.96$^{a}$&46.07$^{a}$ & \\
DES16C2aix&1.068&-10.01$^{}$&38.11$^{}$ & \\
DES16C2nm&1.998&15.97$^{}$, \enspace16.3$^{}$, \enspace21.06$^{}$, \enspace21.45$^{}$, \enspace30.72$^{}$&41.66$^{}$, \enspace41.99$^{}$, \enspace46.75$^{}$, \enspace47.14$^{}$, \enspace56.41$^{}$ & \cite{Smith2018}\\
DES16C3cv&0.727&-3.43$^{}$&53.79$^{}$ & \\
DES16C3dmp&0.562&4.46$^{ac}$&27.06$^{ac}$ & \\
DES16C3ggu&0.949&-24.81$^{}$&31.22$^{}$ & \\
DES17C3gyp&0.47&-7.54$^{}$&24.69$^{}$ & \\
DES17X1amf&0.92&-0.42$^{}$&28.71$^{}$ & \\
DES17X1blv&0.69&2.33$^{ac}$&28.23$^{ac}$ & \\
LSQ12dlf&0.255&7.24$^{}$, \enspace8.85$^{}$, \enspace9.91$^{}$, \enspace10.73$^{}$, \enspace11.45$^{ac}$, \enspace15.99$^{}$, \enspace19.3$^{e}$, \enspace20.75$^{a}$, \enspace33.53$^{}$, \enspace34.86$^{}$, \enspace43.88$^{}$&38.81$^{}$, \enspace40.42$^{}$, \enspace41.49$^{}$, \enspace42.31$^{}$, \enspace43.03$^{ac}$, \enspace47.56$^{}$, \enspace50.88$^{e}$, \enspace52.33$^{a}$, \enspace65.11$^{}$, \enspace66.44$^{}$, \enspace75.46$^{}$ & \cite{Nicholl2014, Childress2016, Shivvers2019}\\
LSQ14an&0.1637&48.92$^{}$, \enspace53.2$^{}$, \enspace54.09$^{}$, \enspace72.14$^{}$, \enspace91.75$^{}$, \enspace92.58$^{}$, \enspace104.66$^{}$&132.98$^{}$, \enspace137.26$^{}$, \enspace138.14$^{}$, \enspace156.19$^{}$, \enspace175.8$^{}$, \enspace176.63$^{}$, \enspace188.71$^{}$ & \cite{Inserra2017}\\
LSQ14bdq&0.345&-41.16$^{}$, \enspace-40.35$^{}$, \enspace-39.64$^{}$&36.61$^{}$, \enspace37.42$^{}$, \enspace38.13$^{}$ & \cite{Nicholl2015b} \\
LSQ14mo&0.256&-5.34$^{}$, \enspace-0.54$^{}$, \enspace10.47$^{}$, \enspace16.91$^{}$, \enspace23.34$^{}$, \enspace58.27$^{e}$&19.58$^{}$, \enspace24.38$^{}$, \enspace35.39$^{}$, \enspace41.83$^{}$, \enspace48.27$^{}$, \enspace83.19$^{e}$ & \cite{Chen2017}\\
OGLE15xl&0.198&102.67$^{}$, \enspace105.18$^{a}$, \enspace106.0$^{e}$, \enspace7.75$^{}$, \enspace8.63$^{}$, \enspace19.44$^{}$, \enspace25.3$^{}$, \enspace42.68$^{a}$, \enspace58.57$^{}$&139.77$^{}$, \enspace142.27$^{a}$, \enspace143.1$^{e}$, \enspace44.85$^{}$, \enspace45.73$^{}$, \enspace56.53$^{}$, \enspace62.4$^{}$, \enspace79.77$^{a}$, \enspace95.67$^{}$ & \cite{Yaron2023}\\
PS110ahf&1.1&-79.19$^{}$&5.43$^{}$ & \\
PS110awh&0.9084&-16.79$^{}$, \enspace7.32$^{}$, \enspace13.6$^{}$&11.67$^{}$, \enspace35.77$^{}$, \enspace42.06$^{}$ & \cite{Chomiuk2011}\\
PS110bzj&0.65&8.6$^{}$, \enspace8.6$^{}$, \enspace15.87$^{}$, \enspace17.69$^{a}$, \enspace56.48$^{e}$&26.81$^{}$, \enspace26.81$^{}$, \enspace34.08$^{}$, \enspace35.9$^{a}$, \enspace74.68$^{e}$ & \cite{Lunnan2013}\\
PS110ky&0.9558&0.08$^{}$, \enspace2.12$^{}$, \enspace16.44$^{}$, \enspace27.69$^{ac}$&13.61$^{}$, \enspace15.65$^{}$, \enspace29.97$^{}$, \enspace41.22$^{ac}$ & \cite{Chomiuk2011}\\
PS110pm&1.206&13.92$^{}$&41.09$^{}$ & \\
PS111afv&1.407&4.75$^{}$&32.75$^{}$ & \cite{Lunnan2018b}\\
PS111aib&0.997&26.77$^{}$&92.95$^{}$ & \cite{Lunnan2018b}\\
PS111ap&0.524&0.98$^{}$&48.56$^{}$ & \\
PS111bam&1.565&8.91$^{}$&30.05$^{}$ & \\
PS111bdn&0.738&-1.23$^{}$, \enspace-1.17$^{}$&18.5$^{}$, \enspace18.56$^{}$ & \cite{Lunnan2018b}\\
PS111tt&1.283&19.01$^{}$&63.74$^{}$ & \cite{Lunnan2018b}\\
PS112bmy&1.572&41.21$^{}$, \enspace41.25$^{}$&72.88$^{}$, \enspace72.92$^{}$ & \cite{Lunnan2018b}\\
PS112bqf&0.522&1.71$^{}$&41.16$^{}$ & \cite{Lunnan2018b}\\
PS112cil&0.32&2.8$^{}$&31.45$^{}$ & \cite{Lunnan2018b}\\
PS113gt&0.884&26.13$^{}$&53.56$^{}$ & \cite{Lunnan2018b}\\
PS113or&1.52&8.21$^{}$, \enspace8.21$^{}$&44.34$^{}$, \enspace44.34$^{}$ & \cite{Lunnan2018b}\\
PS114bj&0.5215&-4.94$^{}$, \enspace-4.38$^{}$, \enspace-3.85$^{}$, \enspace-0.57$^{}$, \enspace34.26$^{ac}$, \enspace35.58$^{ac}$, \enspace-37.82$^{}$, \enspace-24.65$^{}$&118.43$^{}$, \enspace118.99$^{}$, \enspace119.52$^{}$, \enspace122.8$^{}$, \enspace157.63$^{ac}$, \enspace158.95$^{ac}$, \enspace85.55$^{}$, \enspace98.72$^{}$ & \cite{Lunnan2016}\\
PS15cjz&0.22&-29.11$^{}$&25.12$^{}$ & \\
PTF09atu&0.5015&34.51$^{}$, \enspace94.45$^{}$, \enspace113.09$^{}$, \enspace-24.1$^{}$, \enspace-22.77$^{}$, \enspace-0.13$^{}$, \enspace-0.13$^{}$, \enspace19.19$^{}$&107.96$^{}$, \enspace167.9$^{}$, \enspace186.55$^{}$, \enspace49.35$^{}$, \enspace50.68$^{}$, \enspace73.33$^{}$, \enspace73.33$^{}$, \enspace92.64$^{}$ & \cite{Quimby2011, Quimby2018}\\
PTF09cnd&0.2584&97.66$^{}$, \enspace119.91$^{}$, \enspace-25.51$^{}$, \enspace-22.33$^{}$, \enspace-15.18$^{}$, \enspace1.51$^{}$, \enspace7.87$^{}$, \enspace22.97$^{}$, \enspace26.94$^{}$, \enspace31.71$^{}$&153.29$^{}$, \enspace175.54$^{}$, \enspace30.12$^{}$, \enspace33.3$^{}$, \enspace40.45$^{}$, \enspace57.14$^{}$, \enspace63.49$^{}$, \enspace78.59$^{}$, \enspace82.57$^{}$, \enspace87.33$^{}$ & \cite{Quimby2011, Quimby2018, Shivvers2019}\\
PTF10aagc&0.206&1.86$^{}$, \enspace2.69$^{}$, \enspace119.6$^{}$, \enspace4.34$^{}$, \enspace13.47$^{}$, \enspace23.42$^{}$, \enspace30.05$^{}$, \enspace75.66$^{}$&11.88$^{}$, \enspace12.71$^{}$, \enspace129.63$^{}$, \enspace14.37$^{}$, \enspace23.49$^{}$, \enspace33.44$^{}$, \enspace40.07$^{}$, \enspace85.68$^{}$ & \cite{Quimby2018} \\
PTF10bfz&0.1701&1.97$^{}$, \enspace5.39$^{}$, \enspace13.94$^{}$, \enspace25.9$^{}$&22.99$^{}$, \enspace26.41$^{}$, \enspace34.95$^{}$, \enspace46.92$^{}$ & \cite{Quimby2018}\\
PTF10bjp&0.3584&11.62$^{}$, \enspace16.78$^{}$&55.21$^{}$, \enspace60.37$^{}$ & \cite{Quimby2018}\\
PTF10iam&0.109&1.36$^{a}$, \enspace27.84$^{ad}$, \enspace37.43$^{ad}$&15.04$^{a}$, \enspace41.52$^{ad}$, \enspace51.11$^{ad}$ & \cite{Arcavi2016}\\
PTF10nmn&0.1237&51.69$^{}$, \enspace57.92$^{}$, \enspace58.81$^{}$, \enspace81.05$^{}$&113.43$^{}$, \enspace119.66$^{}$, \enspace120.55$^{}$, \enspace142.8$^{}$ & \cite{Quimby2018}\\
PTF10uhf&0.2882&-3.03$^{}$, \enspace16.38$^{}$&18.23$^{}$, \enspace37.63$^{}$ & \cite{Quimby2018}\\
PTF10vqv&0.4518&85.02$^{}$, \enspace85.71$^{}$, \enspace7.87$^{}$, \enspace12.01$^{}$, \enspace22.34$^{}$, \enspace46.45$^{}$, \enspace47.82$^{}$&115.06$^{}$, \enspace115.75$^{}$, \enspace37.91$^{}$, \enspace42.04$^{}$, \enspace52.38$^{}$, \enspace76.48$^{}$, \enspace77.86$^{}$ & \cite{Quimby2018}\\
PTF12dam&0.1073&40.43$^{ac}$, \enspace50.37$^{}$, \enspace50.37$^{}$, \enspace50.37$^{}$, \enspace80.17$^{a}$, \enspace80.17$^{a}$, \enspace-32.72$^{}$, \enspace-31.82$^{}$, \enspace-30.91$^{}$, \enspace-30.01$^{ac}$, \enspace-29.11$^{}$, \enspace-28.2$^{}$, \enspace-28.2$^{}$, \enspace-28.2$^{}$, \enspace-27.3$^{}$, \enspace-21.88$^{a}$, \enspace-15.56$^{a}$, \enspace-10.14$^{}$, \enspace-6.53$^{}$, \enspace-3.82$^{a}$, \enspace-0.21$^{}$, \enspace-0.21$^{}$, \enspace3.4$^{}$, \enspace9.73$^{}$, \enspace14.24$^{}$, \enspace15.14$^{}$, \enspace19.66$^{}$&115.54$^{ac}$, \enspace125.48$^{}$, \enspace125.48$^{}$, \enspace125.48$^{}$, \enspace155.28$^{a}$, \enspace155.28$^{a}$, \enspace42.39$^{}$, \enspace43.29$^{}$, \enspace44.2$^{}$, \enspace45.1$^{ac}$, \enspace46.0$^{}$, \enspace46.91$^{}$, \enspace46.91$^{}$, \enspace46.91$^{}$, \enspace47.81$^{}$, \enspace53.23$^{a}$, \enspace59.55$^{a}$, \enspace64.97$^{}$, \enspace68.58$^{}$, \enspace71.29$^{a}$, \enspace74.9$^{}$, \enspace74.9$^{}$, \enspace78.52$^{}$, \enspace84.84$^{}$, \enspace89.35$^{}$, \enspace90.26$^{}$, \enspace94.77$^{}$ & \cite{Nicholl2013, Quimby2018, Shivvers2019}\\
PTF12gty&0.1768&-11.2$^{}$, \enspace-6.95$^{}$&50.14$^{}$, \enspace54.39$^{}$ & \cite{Quimby2018}\\
PTF12hni&0.1056&-6.54$^{}$, \enspace2.51$^{}$&20.35$^{}$, \enspace29.4$^{}$ & \cite{Quimby2018}\\
PTF12mxx&0.3296&-7.58$^{}$, \enspace8.21$^{}$&37.89$^{}$, \enspace53.69$^{}$ & \cite{Quimby2018}\\
SCP06F6&1.189&-1.85$^{e}$, \enspace10.02$^{}$, \enspace14.59$^{}$&39.86$^{e}$, \enspace51.74$^{}$, \enspace56.31$^{}$ & \cite{Barbary2009}\\
SNLS06D4eu&1.5881&-7.93$^{}$&11.85$^{}$ & \cite{Howell2013}\\
SNLS07D2bv&1.5&-4.2$^{}$&15.48$^{}$ & \cite{Howell2013}\\
SSS120810&0.156&6.52$^{}$, \enspace7.42$^{}$, \enspace12.54$^{acd}$, \enspace23.81$^{}$, \enspace31.61$^{}$, \enspace38.55$^{}$, \enspace40.05$^{a}$, \enspace56.72$^{}$&31.58$^{}$, \enspace32.48$^{}$, \enspace37.6$^{acd}$, \enspace48.88$^{}$, \enspace56.67$^{}$, \enspace63.61$^{}$, \enspace65.11$^{a}$, \enspace81.78$^{}$ & \cite{Nicholl2014}\\
iPTF13ajg&0.7403&82.81$^{}$, \enspace-5.68$^{}$, \enspace-5.34$^{}$, \enspace-5.11$^{}$, \enspace-0.51$^{ac}$, \enspace11.56$^{a}$, \enspace12.7$^{}$, \enspace13.01$^{}$, \enspace28.22$^{}$, \enspace30.52$^{a}$, \enspace30.52$^{a}$, \enspace48.91$^{}$&111.41$^{}$, \enspace22.92$^{}$, \enspace23.27$^{}$, \enspace23.5$^{}$, \enspace28.09$^{ac}$, \enspace40.16$^{a}$, \enspace41.31$^{}$, \enspace41.62$^{}$, \enspace56.82$^{}$, \enspace59.12$^{a}$, \enspace59.12$^{a}$, \enspace77.51$^{}$ & \cite{Vreeswijk2014}\\
iPTF13bdl&0.403&-33.09$^{}$, \enspace-31.82$^{}$, \enspace-30.76$^{}$&112.4$^{}$, \enspace113.66$^{}$, \enspace114.73$^{}$ & \cite{DeCia2018, Schulze2021}\\
iPTF13bjz&0.2712&-0.19$^{}$, \enspace4.53$^{}$&27.5$^{}$, \enspace32.22$^{}$ & \cite{DeCia2018, Schulze2021} \\
iPTF13cjq&0.3962&-3.13$^{}$&13.43$^{}$ & \cite{Schulze2021}\\
iPTF13dcc&0.4305&37.7$^{a}$&96.92$^{a}$ & \cite{Schulze2021}\\
iPTF13ehe&0.3434&-9.83$^{}$, \enspace-6.1$^{}$, \enspace13.57$^{}$&68.59$^{}$, \enspace72.32$^{}$, \enspace91.99$^{}$ & \cite{Yan2015, Schulze2021}\\
iPTF15eov&0.0535&-10.91$^{}$, \enspace-9.9$^{}$, \enspace-8.86$^{}$, \enspace-5.2$^{}$, \enspace4.23$^{e}$, \enspace11.91$^{}$, \enspace14.34$^{}$, \enspace27.05$^{}$, \enspace-16.98$^{}$, \enspace34.66$^{}$, \enspace42.12$^{}$, \enspace-14.6$^{}$, \enspace49.78$^{}$, \enspace53.48$^{}$, \enspace-13.74$^{}$, \enspace-12.24$^{}$, \enspace68.71$^{}$&11.19$^{}$, \enspace12.2$^{}$, \enspace13.24$^{}$, \enspace16.9$^{}$, \enspace26.33$^{e}$, \enspace34.01$^{}$, \enspace36.44$^{}$, \enspace49.15$^{}$, \enspace5.12$^{}$, \enspace56.75$^{}$, \enspace64.21$^{}$, \enspace7.5$^{}$, \enspace71.87$^{}$, \enspace75.58$^{}$, \enspace8.36$^{}$, \enspace9.86$^{}$, \enspace90.81$^{}$ & \cite{Taddia2019}\\
iPTF16asu&0.187&7.86$^{}$, \enspace9.57$^{}$, \enspace16.63$^{}$, \enspace19.13$^{}$, \enspace21.36$^{}$, \enspace-1.38$^{}$, \enspace0.35$^{}$&17.46$^{}$, \enspace19.17$^{}$, \enspace26.23$^{}$, \enspace28.74$^{}$, \enspace30.97$^{}$, \enspace8.22$^{}$, \enspace9.96$^{}$ & \cite{Taddia2019}\\
iPTF16bad&0.2467&94.89$^{}$, \enspace0.44$^{}$, \enspace2.87$^{}$&116.73$^{}$, \enspace22.28$^{}$, \enspace24.71$^{}$ & \cite{Yan2017a}\\
iPTF16eh&0.427&54.75$^{}$, \enspace77.81$^{}$, \enspace79.95$^{}$, \enspace98.13$^{ac}$, \enspace0.49$^{}$, \enspace6.8$^{}$, \enspace12.83$^{}$, \enspace37.17$^{}$&100.94$^{}$, \enspace124.0$^{}$, \enspace126.14$^{}$, \enspace144.32$^{ac}$, \enspace46.68$^{}$, \enspace52.99$^{}$, \enspace59.02$^{}$, \enspace83.36$^{}$ & \cite{Lunnan2018a}\\
\hline \hline \\\end{longtable}